\documentclass[a4paper, 10pt,reqno]{amsart}
\usepackage{amsmath}
\usepackage{amsfonts}
\usepackage{amsthm}
\usepackage{amstext}
\usepackage{enumerate}
     \newcommand{\supp}{\operatorname{supp}}

      \newcommand{\tr}{{\operatorname{tr}}}
      \newcommand{\dist}{{\operatorname{dist}}}

      \newcommand{\Ran}{{\operatorname{Ran}}}

      \newcommand{\Op}{{\operatorname{Op\!^w\!}}}
      \newcommand{\N}{{\mathbb{N}}}
      \newcommand{\R}{{\mathbb{R}}}
      \newcommand{\Z}{{\mathbb{Z}}}
      \newcommand{\C}{{\mathbb{C}}}
      \newcommand{\T}{{\mathbb{T}}}
 \newcommand{\p}{\mbox{\boldmath $p$}}
      \newcommand{\A}{\mbox{\boldmath $a$}}
\newcommand{\e}{{\rm e}}

\newcommand{\unif}{{\rm unif}}

\newcommand{\crt}{{\rm crt}}
\newcommand{\so}{{\rm so}}
\newcommand{\si}{{\rm si}}
\newcommand{\sa}{{\rm sa}}
\renewcommand{\sp}{{\rm sp}}
\newcommand{\col}{{\rm col}}
\newcommand{\exc}{{\rm exc}}
\renewcommand{\tr}{{\rm tr}}
 \newcommand\inp[2][]{#1 \langle #2#1\rangle}
     \theoremstyle{plain}%default
     \newtheorem{thm}{Theorem}[section]
     \newtheorem{prop}[thm]{Proposition}
     \newtheorem{lemma}[thm]{Lemma}
     \newtheorem{cor}[thm]{Corollary}
     \theoremstyle{definition}

     \newtheorem{example}[thm]{Example}
     
     \newtheorem{cond}[thm]{Condition}

     \newtheorem{remark}[thm]{Remark}
     \newtheorem{remarks}[thm]{Remarks}

     \numberwithin{equation}{section}
\title[Zero]{Classical and quantum dynamics for $2D$-electromagnetic potentials
asymptotically homogeneous of degree zero}

\author{H.  Cornean}
\address[H.  Cornean]{Department of Mathematical Sciences, Aalborg University\\Fredrik Bajers Vej 7G, 9220 Aalborg, Denmark}
\email{cornean@math.auc.dk}
\thanks{H.  Cornean and E. Skibsted are (partially) supported by the Danish 
F.N.U. grant {\it  Mathematical Physics and Partial Differential
  Equations}}

\author{I. Herbst}
\address[I. Herbst]{Dept. of Math., University of Virginia\\ 
Charlottesville, Virginia
  22903, USA} 
\email{iwh@virginia.edu}

\author{E. Skibsted}
\address[E. Skibsted]{Institut for  Matematiske
Fag \\
Aarhus Universitet\\ Ny Munkegade  8000 Aarhus C, 
Denmark}
\email{skibsted@imf.au.dk}

\begin{document}

\begin{abstract} We consider a charged particle moving in  the plane
  subject to electromagnetic potentials with non-vanishing radial
  limits. We analyse the classical and the quantum dynamics for large
  time in the   case the angular part of the (limiting) Lorentz force (defined for velocities that are purely
  radial) 
  has a finite number of zeros at fixed energy. Any such  zero
  defines a channel, and to the
  ``stable'' ones we associate  quantum  wave
  operators. Their completeness is studied in the case of zero as well
  as nonzero 
  magnetic flux. In the latter case one needs possibly to incorporate
  a channel of spiraling states. These states are similar to those  studied recently in the sign-definite
  case in \cite {CHS}.
\end{abstract}
\maketitle
\tableofcontents
\section{Introduction and results} \label{Introduction}
In this paper we continue the  study initiated in \cite{CHS} of the
dynamics of a charge $1$ particle moving in the plane subject to a
very long-range
magnetic field $\bf B$ perpendicular to the plane. More precisely we
assume ${\bf
B}=(0,0,r^{-1}b(x))$ where $x$ is the position in the plane and  $r=|x|$,
and that $b(\theta):=\lim_{r\to \infty} b(x)$ is a non-vanishing smooth
function on the unit-circle. 

The classical Hamiltonian may be put in the form
\begin{equation}
\label{eq:h1}
  h=h(x,\xi)=2^{-1}(\xi-\A)^2,
\end{equation}
with $\A(x)=\int_0^1r^{-1}b(sx)(-x_2,x_1)ds$ where
$x=(x_1,x_2)$. Here the mass is taken to be $1$.

In the bulk of this paper we shall
consider the more general symbol
\begin{equation}
\label{eq:h2}
  h=h(x, \xi)=2^{-1}(\xi -\A)^2+ V,
\end{equation} where $V(\theta)=\lim_{r\to \infty} V(x)$ similarly
defines a smooth
function on the unit-circle. 
 From this point of view the present paper may be seen as a natural
continuation/generalization  (in the $2$-dimensional framework) of the
previous works  \cite{CHS}, \cite{He}, \cite{HMV}, \cite{HS1} and
\cite{HS2}.

To simplify the presentation let us assume throughout this introduction
 that $V=0$, and that (using polar coordinates) $\A(x)=\A(\theta)=(-\sin
\theta{},\cos\theta{})b(\theta{})$. In the context of quantum
mechanics discussed below we take $\A(x)$ to be regularized  at the (singular)
 origin, but keep the above  form outside a compact set. Let us
denote
by $H$ the corresponding quantization which is a self-adjoint operator
on $L^2(\R^2_{x})$. It is a general fact that the spectrum is given by
$\sigma(H)=\sigma_{ess}(H)=[0,\infty)$; see \cite[Theorem 6.1]{CFKS}.

In \cite{CHS} we studied the case where $b(\theta)<0$ for all
angles. In outline we showed that above a certain energy $E_d>0$
all classical scattering orbits will go to infinity along logarithmic
 spirals, and an analogous result was proved for scattering quantum
 states localized in $(E_d,\infty)$.
  In this energy regime  the singular continuous spectrum and the
 pure point spectrum are empty and discrete, respectively.
At the particular energy $E_d$ there exists a one-parameter family of closed
classical orbits. Below  $E_d$ the classical orbits have infinitely
many ``loops''. The nature of the spectrum of $H$ in this regime
is not known
except in the constant $b$ case where  it is pure point,
cf. \cite[Theorem 6.2]{CFKS}. 

Now we shall study the case where $b(\theta)$  has a finite number of
non-degenerate 
zeros. We group those zeros into the ``stable'' ones where
$b'(\theta_j)>0$ and the ``unstable'' ones where
$b'(\theta_j)<0$. In either case there is a channel associated to
$\theta_j$. 
More precisely there are classical scattering orbits with
$\theta(t)\to \theta_j$ as $t\to \infty$. (Here and henceforth
$t\to\infty$ means $t\to+\infty$.) However 
they are of a  different nature in the following sense: There is a reduced
phase space in which the  stable $\theta_j$'s
correspond to  stable fixed points, while  the unstable $\theta_j$'s
correspond to  unstable fixed points (motivating their names). This
means that the orbits asssociated to an unstable $\theta_j$ are
``rare''
(by the stable manifold theorem  in the theory of dynamical systems),
while the orbits associated with a stable $\theta_j$ fill out a
continuum. The unstable channels do not exist in quantum mechanics.

For other works on quantum scattering theory for magnetic Hamiltonians we
refer to \cite {Ho2}, \cite{LT1}, \cite{LT2}, \cite {E}, \cite {NR},
\cite{R} and \cite {RY}. All these works require decay of the magnetic
vector potential (see \cite{CHS} for a more detailed account). In the
case of periodic fields it was shown in \cite{BS} and \cite {S} that 
the spectrum of the Hamiltonian is purely absolutely continuous. There are
other works on $2D$-magnetic Hamiltonians with continuous spectrum,
\cite{Iw}, \cite {MP}, \cite{BP} and  \cite {FGW}. Those do in fact
deal with more ``long-range''  vector potentials than considered in this
paper  
(although strictly disjoint from our class). The books \cite{CFKS},
\cite{DG2} and \cite{RS} contain further background information.

\subsection{Results in classical mechanics}
\label{Results in classical mechanics} For simplicity we take henceforth $V=0$. 
 We distinguish between the two cases 1) the flux
 $\int^{2\pi}_0b(\theta) d\theta\\  =0$, and 2) the flux
 $\int^{2\pi}_0b(\theta) d\theta<0.$ Partly motivated by the fact that
  there are collapsing
 orbits, $r(t)=|x(t)|\to 0$ in finite time, let us define a ``classical scattering
 orbit'' to be a solution to Hamilton's (or Newton's) equations with
 $r(t)\to \infty$ for $t\to \infty$.

\subsubsection{Zero flux}
In this case {\it any} scattering
 orbit approaches one of the zeros, $b(\theta_j)=0$, meaning
 $\theta(t)\to \theta_j$. Moreover there exists the limit $\lim_{t\to
 \infty}t^{-1}r(t)=\sqrt{2E}$, where $E>0$ denotes the energy of
 the orbit.
\subsubsection{Negative flux}
We are interested in 
$2\pi$-periodic 
solutions $\rho=\rho_E$ to the system 
\begin{equation}\label{eq:periodikkk}
\begin{cases}\frac{d\rho}{d\theta}=b+\eta
\\\eta=\sqrt{2E-\rho^2}>0,\;\int^{2\pi}_0{\rho\over
\eta}\;d\theta>0
\end{cases}\;.
\end{equation}
 Here $\rho$ and $\eta$ represent the radial and angular part of the
 velocity, respectively.
We know from \cite {CHS} that if $b<0$,  then there exists exactly one such solution for all
high enough energies and it defines a logarithmic spiral

We show in the present case where $b$ has zeros: 

\noindent i) There is at most one positive
$2\pi$-periodic solution $\rho_E$ to \eqref{eq:periodikkk} for
fixed $E$. If the set $\mathcal E$ of energies where the solution exists
is non-empty, then it is an open bounded interval $(E_d,E_e)$. 

\noindent ii)  If $\rho_E$ exists, then for 
{\it any} scattering
 orbit with energy $E$, either it behaves as described in the zero flux
 case,
or
\begin{equation}\label{eq:perieqq}
\lim_{t\to\infty}|\rho(t)-\rho_E(\theta(t))|=\lim_{t\to\infty}|\eta(t)-\eta_E(\theta(t))|=0.
\end{equation} 
The interpretation of \eqref{eq:perieqq} is attraction to
a spiral. 

\noindent iii) Above $E_e$, or for high
enough  energies if   $\mathcal E=\emptyset$, {\it all}  scattering
 orbits behave as described in the zero flux 
 case.
 
Our knowledge about the low-energy  region, $E<E_d$ if $\mathcal E\neq
\emptyset$, is rather sparse. (This is also true for the case  $b<0$ of
\cite {CHS}.) 

The positive flux case may be  reduced to the negative flux case by
permutation symmetry.

\subsection{Wave operators for a stable zero in quantum  mechanics}
\label{Results in quantum  mechanics}
Let us fix a stable zero, $b(\theta_j)=0$, $b'(\theta_j)>0$. We construct two types of
wave operators 1) one for high energies, and 2) one for small
energies. To define the relevant energy regimes we introduce the
quantities
\begin{equation}
  \label{eq:betass}
\beta=-
\frac{1}{2}+\frac{1}{2}\sqrt{1-4\frac{b'(\theta_j)}{\sqrt{2E}}},\;\tilde \beta =
- \frac{1}{2}-\frac{1}{2}\sqrt{1-4\frac{b'(\theta_j)}{\sqrt{2E}}}.
\end{equation}
Those numbers arise as eigenvalues for a linearized reduced flow,
cf. \cite{HS2}. Now 
``high'' means $4b'(\theta_j)<\sqrt{2E}$, while ``small'' means
$\Re\beta<-3^{-1}$; in particular the two regimes fixed by these
requirements, say $I_j=(E_j,\infty)$ and
$\bar I_j=(0, \bar E_j)$ respectively,  overlap.

\subsubsection{High energy regime}
 We consider the equation for $\eta$ from 
\eqref{eq:periodikkk} (assuming $\rho>0$)
\begin{equation*}
 \frac{d\eta}{d\theta}=-\frac{b + \eta 
  }{\eta}\sqrt{2E-\eta^2}, 
\end{equation*}
now as a singular equation with $\theta$ nearby
$\theta_j$ and the initial value $\eta(\theta_j)=0$. Away from a
discrete set of ``resonances'' there are  smooth solutions $\eta_E$ with the
asymptotics $\eta_E\asymp \sqrt{2E}\beta(E)(\theta-\theta_j)$ as
$\theta\to \theta_j$. Based
on these functions one may construct local solutions
$S=S(t,r,\theta)=S(t,x)$ to  the Hamilton--Jacobi equation
\begin{equation*}
  h(x,{\nabla}_{x}S)=-\partial_tS.
\end{equation*} By ``local'' we mean here that $|\theta-\theta_j|$ is
small and that $2^{-1}(t^{-1}r)^2$ (or equivalently $E=-\partial_tS$)
is well-localized (in particular bounded away from resonances) on the
support of any such $S$.

We also need a quantity $w=w(t,r,\theta)$ which is
defined as follows: First we notice that for any classical scattering orbit attracted
to $\theta_j$ there exists the limit $$w=\lim_{t\to\infty}
r(t)^{-\beta}(\theta(t)-\theta_j).$$ We use this formula for the orbit
that at time $t$ goes through 
$(r,\theta)$ on
the Lagrangian manifold associated with  an $S$ (in particular its energy is $E=-\partial_tS$).

Now the wave operator is given in terms of the family of local comparison
dynamics $$U_j(t):
L^2(I\times{\R})\to L^2({\R}_{x}^2)=L^2({\R}_+\times{\T};\;rdrd\theta)$$
given as
\begin{align}\label{valimold-88}
[U_j(t)\phi](r,\theta)=&\e^{iS(t,r,\theta)} r^{-1/2}J_t^{1/2}(r,\theta)1_{\{\cdot\}}(r,\theta)\\&\phi
\left (E(t, r,\theta),w(t,r,\theta)\right ).\nonumber
\end{align}
 Here  $I$ ranges over  small intervals free from resonances, $I\ni E(t, r,\theta)=-\partial_tS(t, r,\theta)$, $J_t$ is the Jacobian determinant arising 
from the  change of variables which makes $U_j(t)$ 
isometric and $1_{\{\cdot\}}$ projects to the region where $S(t,\cdot)$
  is defined. In terms of a suitable (regularized) quantization $H$ of the
  symbol $h$ the wave operator is specified by its value at given 
  $\phi\in L^2(I\times{\R})$ as 
  \begin{equation*}
    \Omega_j\phi=\lim_{t\to\infty}\e^{itH}U_j(t)\phi.
  \end{equation*}
We may view $\Omega_j$ as a uniquely defined (global) operator $\Omega_j:
L^2(I_j\times{\R})\to L^2({\R}_{x}^2)$, see Remark \ref{canonical} for
details.  Interpreted this way
$$H\Omega_j=\Omega_j M(E),\;\Ran
(\Omega_j)\subseteq1_{I_j}(H)L^2({\R}_{x}^2);$$ here $M(E)$
denotes multiplication by $E$.

The  above wave operator is closely related to the one at high
energies for the potential model constructed
recently in \cite{HS1}. For earlier works on a somewhat similar
``$x$-space modifier'' we refer to \cite{Y} and \cite{DG1}.
\subsubsection{Small energy regime}
We may assume that
$\theta_j=0$. We use rectangular coordinates $x=(x_1,x_2)$, and
similarly for the dual variable $\xi=(\xi_1,\xi_2)$. We
substitute in the expression \eqref{eq:h1}, Taylor expand up to
second order in $x_2/x_1$ and $\xi_2$, and replace $x_1$ by
$t\xi_1$. The result reads
\begin{equation*}
 h(t)=2^{-1}\xi^2-\frac {b'(\theta_j)}{\xi_1}\frac {x_2}{t}\xi_2+\frac {b'(\theta_j)}{\xi_1}\Big(1+\frac
{b'(\theta_j)}{2\xi_1}\Big)\Big(\frac
{x_2}{t}\Big)^2.
\end{equation*}

Next we quantize this expression and obtain a family of self-adjoint
operators $H(t),\;t>0,$ which generates an explicit propagator 
\begin{equation*}i\partial_t\bar U(t)=H(t)\bar U(t);\;U(1)=I.
\end{equation*}

Let  $\bar J_j=\left\{\xi_1: \xi_1 > 0, \frac{\xi^2_1}{2} \in \bar
  I_j\right\}$ and $p_1=-i{\partial\over \partial x_1}$. We prove the existence of the limit
  \begin{equation*}
 \bar\Omega_j = s-\lim_{t \to  \infty} \e^{itH} \bar
 U(t):1_{\bar J_j}(p_1)L^2(\mathbb{R}^2_{x}) \to  L^2({\R}_{x}^2).  
  \end{equation*}

Again the wave operator diagonalizes the free energy
$$H\bar\Omega_j=\bar\Omega_j 2^{-1}p_1^2,\;\Ran
(\bar\Omega_j)\subseteq1_{\bar I_j}(H)L^2({\R}_{x}^2).$$

A similar ``Dollard-type'' wave operator was considered at low
energies for the potential model in \cite{HS1}. 

\subsection{Wave operator for spirals}
\label{Wave operators for spirals}
Here  we recall the construction of \cite{CHS} adapted to the 
present case where the interval $\mathcal E=(E_d,E_e)$ of energies
possessing solutions to (\ref{eq:periodikkk}) is bounded (assuming
$\mathcal E\neq \emptyset$). Let
$$f(\theta)=\lim _{E\uparrow E_e}(\partial _E\rho_E(\theta))^{-1},$$
and 
\begin{equation}
  \label{eq:Areg}\mathcal D =\cup_{t>0}\{t\}\times \mathcal D _t;\;\mathcal D _t=\{(r,\theta)\in \R_+\times
  \T|\;0<{r\over t}<f(\theta)\}.
\end{equation}

We define on $\mathcal D$
\begin{equation}
  \label{eq:S}S(t,r,\theta)=r\rho_{E(t,r,\theta)}(\theta)-tE(t,r,\theta),
\end{equation} where the ``energy function''  $E(t,\cdot,\theta)$ is
the inverse of the function $\mathcal E\ni E\to r=
t/(\partial_E\rho_E)(\theta)\in (0,tf(\theta))$. 

Now we take as approximate dynamics the expression (\ref{valimold-88})
modified as follows: Take $\phi\in L^2(\mathcal E\times \T)$. Replace on the right hand side $\{\cdot\}$ by
$\mathcal D _t$, and take $w(t,r,\theta)$ to be the uniquely defined initial angle, say
$\theta_1$, for the orbit
that at time $t$ goes through 
$(r,\theta)\in \mathcal D _t$ on
the Lagrangian manifold associated with   $S$. More precisely given
such pair $(r,\theta)$ there
is a uniquely defined  pair $(r_1,\theta_1)\in \mathcal D _1$ such that the direct
flow (see 
(\ref{domnyca}))  starting at $(r_1,\theta_1)$ at  time $t=1$ ends
up at $(r,\theta)$ at time $t$.
 
Let us denote this expression by $[U_{\sp}(t)\phi](r,\theta)$. We can then
define $\Omega_{\sp}\phi=\lim_{t\to\infty}\e^{itH}U_{\sp}(t)\phi$. The
range of the 
wave operator $\Omega_{\sp}$ is a closed subspace of $L^2(\R^2)$ of states
whose large time behaviour is ``spiraling'', see \cite {CHS} for
further discussion Let $P_{\sp}$ denote  the orthogonal projection onto
this subspace. The wave operator diagonalizes the free energy
$$H\Omega_{\sp}=\Omega_{\sp} M(E),\;\Ran
(\Omega_{\sp})\subseteq1_{\mathcal E}(H)L^2({\R}_{x}^2).$$

\subsection{Asymptotic completeness in quantum mechanics}
\label{Asymptotic completeness in quantum mechanic}
We would like to associate a quantum channel to each zero,
$b(\theta_j)=0$.
Suppose $\mathcal C\subseteq(0,\infty)$ is an open  set containing no
eigenvalues of $H$. Then we would like to make sense of an expression
like
\begin{equation}
  \label{eq:pro22}
 P_j=P_{j,\mathcal C}=\lim _{f\uparrow
  1_{\mathcal C}}\lim _{\chi_j\downarrow 1_{\{\theta_j\}}}\lim _{t\to \infty}\e^{itH}\chi_j\e^{-itH}f(H),
\end{equation} where $f$ denotes a net of $C_0^\infty$-functions, and
  similarly $\chi_j$ is  a net of 
  operators given by multiplication by functions in $C^{\infty}(\T)$. All limits are taken in
  the strong sense. If it exists, $P_j$
  should project to a subspace $P_jL^2({\R}_{x}^2)\subseteq
  1_{\mathcal C}(H)L^2({\R}_{x}^2)$. In the case of a stable zero,
  one would 
  then ask for asymptotic completeness of the wave operators considered in
  Subsection \ref{Results in quantum  mechanics}, which for example
  for  the entire high energy regime would mean that 
  \begin{equation}
\label{eq:acomega}
    \Ran(\Omega_j)=P_{j,I_j\setminus \sigma_{pp}(H)}L^2({\R}_{x}^2).
  \end{equation}

We shall  justify the definition \eqref{eq:pro22} and  results like 
\eqref{eq:acomega} in various  cases.

\subsubsection{Completeness for the flux zero case}

Indeed if $\mathcal C=\mathcal C'\setminus \sigma_{pp}(H)\subseteq (0,\infty)$ is any open
set of energies the definition
\eqref{eq:pro22} can be justified, and we show  the decomposition into
channels formula
\begin{equation}
  \label{eq:chan0}
 1_{\mathcal C}(H)=\sum_{j}\oplus P_{j,\mathcal C}, 
\end{equation} cf. \cite{He}. 
In \eqref{eq:chan0} only those projections which correspond
to classically stable zeros are nonzero, cf. \cite{HS2}.

Moreover, high energy asymptotic completeness holds in the sense that for each stable zero 
 \begin{equation}\label{eq:acomega2}
    1_{\mathcal C'}(H)\Ran(\Omega_j)=P_{j,\mathcal
      C}L^2({\R}_{x}^2);\;\mathcal C'\subseteq I_j.
  \end{equation}
Thus in particular \eqref{eq:acomega} follows. 

Finally we show that 
  the singular continuous spectrum is empty,  and that  the
 pure point spectrum is discrete in $(0,\infty)\setminus\mathcal
  E_{\exc}$. Here $\mathcal
  E_{\exc}$ is either empty, finite  or at most  discrete in
  $(0,\infty)$; at those ``exceptional'' energies 
  there  exists a certain heteroclinic orbit for a
  reduced classical dynamics. We show 
\begin{equation*}
 \sup { \mathcal E_{\exc}}<8^{-1}\Big(\sup
\int_{\theta_1}^{\theta_2} bd\theta\Big)^2; 
\end{equation*}
 whence $0$ is actually the only possible accumulation point.
%
%
%\subsection{Open problems}
%\label{Open problems}
%
%\subsection{Organization of paper}
%\label{Organization of paper}
The (local) absence of ``exceptional'' 
orbits (see Condition
\ref{cond:noexcep} for  definition, and Example \ref{example:noexcep}
and Remark \ref{rem:excep} for 
discussion) is used in our proof of the limiting absorption
principle LAP in the zero flux case. Our proof  relies on a partial
differential equations 
scheme with a long history (and not Mourre theory, \cite{Mo}). It is
an  open problem whether there is
a Mourre estimate for  small energies so that 
LAP would follow without need of Condition
\ref{cond:noexcep}. 
  If  such an estimate exists, the conjugate operator
would need to be somewhat sophisticated, cf. \cite{ACH} and \cite{CHS}.

\subsubsection{Completeness for the general high energy region}
Without any condition on the flux nor existence of periodic solutions
(i.e. $\mathcal E\neq \emptyset$) the  results mentioned above for the
flux zero case hold provided $\mathcal C' =(E',\infty)\cap I_j$ with
$E'$ taken large enough. This number  may be  given in terms of a condition
on a reduced classical flow, see Condition \ref{cond: no large
  oscillation}. In the set $(E',\infty)$ 
  the singular continuous spectrum is empty  and the
 pure point spectrum is discrete.   For  the case $\mathcal E\neq \emptyset$ we
may take $E'=E_e$, cf. the discussion below.
\subsubsection{Completeness above  $ E_d$ for $\mathcal E\neq \emptyset$} We show that in $(E_d,\infty)$
  the singular continuous spectrum is empty and that the eigenvalues
 may only accumulate at $E_d$ or $\infty$.
  There are two regions 1)
above $E_e$ and 2) in  $\mathcal E$. As for 1) the previous results
hold with  $\mathcal C' =(E_e,\infty)\cap I_j$ (and similarly for
$\mathcal C' =(E_e,\infty)\cap \bar {I}_j$ if $E_e< \bar E_j$). As for
2) we take $\mathcal C'=\mathcal E$. Again we
can justify \eqref{eq:pro22}. As for \eqref{eq:chan0} we have the
substitute 
\begin{equation}
  \label{eq:chan000}
 1_{\mathcal C}(H)=P_{\sp}\oplus\sum_{j}\oplus P_{j,\mathcal C}, 
\end{equation}  where as above only stable zeros contribute to the
second  summation. In an obvious way the ranges of the
projections $P_{j,\mathcal C}$ of \eqref{eq:chan000} may be identified in terms of the ranges of the wave
operators, cf. \eqref{eq:acomega2} (hence  completeness holds). Here we need to use both of the wave
operators $\Omega_j$ and $\bar \Omega_j$ unless either $\mathcal
E\subseteq I_j$ or $\mathcal
E\subseteq \bar I_j$.

\subsection{Organization of paper}
\label{Organization of paper}

The paper is organized as follows: We include throughout the paper a possibly non-trivial
scalar potential depending only on the angle. We could also have included
faster decaying fields possibly with  local singularities, but for
simplicity of presentation we have put the discussion on such rather
trivial 
inclusions in a remark, see Remark \ref{remark: perturbations}. 

In
Section \ref{Classical mechanics} we introduce  notation,  basic
equations and 
assumptions, and two conditions. We state a basic result for
the classical system under
the zero flux condition, Condition \ref{cond:flux}. Further details and study of the classical  system with or
without zero flux are deferred to Section \ref{The general classical
 case}.

The wave operators  of Subsection \ref{Results in quantum  mechanics}
are explained in more detail in Section \ref{Wave operators}. As for
the one  in  Subsection \ref{Wave operators for spirals} we refer to
\cite {CHS} for a thorough account. 

In Section \ref{Propagation of
  singularities} we introduce the Hamiltonian, and we study various
  preliminaries for LAP in particular  a version of H\"ormander's
  propagation of singularities theorem \cite[Proposition
  3.5.1]{Ho2}. (To our knowledge Melrose was the first who applied
  this theorem in scattering theory, 
  see \cite{Melr} and \cite{HMV}.)

All of the sections, Sections \ref{Wave front set
  near}--\ref{Projection=0} and Section \ref{Asymptotic completeness},  are
  devoted to quantum mechanics under Condition \ref{cond:flux}. In  Sections \ref{Wave front set
  near}--\ref{LAP} 
 we prove 
  LAP. In Section \ref{Preliminary estimates} we use LAP to show
  various other basic estimates all of which have a clear classical
  interpretation. In Section \ref{Projections} we introduce a 
  channel projection $P_j$ for any given fixed point (stable or unstable), cf. \eqref{eq:pro22}. (It is slightly more
  complicated due to a possible energy-dependence of the fixed point
  angle in the general case.) We show some basic estimates for states associated to
  such a channel. In Section \ref{Projection=0}  we show that the
  projections for  the unstable fixed points do not contribute to the
  decomposition of channels formula, cf. \eqref{eq:chan0}. Finally,
  Section \ref{Asymptotic completeness} is devoted to showing
  asymptotic completeness for the wave operators of Section \ref{Wave
    operators}.

In Section \ref{Quantum mechanics in the  general setting} we study
LAP and completeness in the case of negative flux. We give an outline
of how the methods used in the zero flux case may be modified under
some conditions stated in this section. To keep this paper ``short''
we leave out most details, in particular those that are closely related to
 \cite {CHS}.

In Appendix  \ref{Collapsing orbits} we introduce collapsing classical 
orbits which correspond to orbits in  the configuration space with a
singularity at $x=0$. We investigate the smallness and general nature of the
set of collapsing orbits. Our results are not needed in our treatment
of the quantum model, however they are a natural supplement to Section
\ref{The general classical case} with some interest of their own. The
last part of Appendix  \ref{Collapsing orbits} is devoted to proving
discreteness of the set $\mathcal E _{\exc}$ of exceptional energies

In Section \ref{A similar model} we shall briefly discuss  another model exhibiting similar properties as the ones  for the model of the paper. This model is from Riemannian geometry.

\section{Assumptions, classical mechanics and the zero flux case} 

\label{Classical mechanics}
We shall study symbols of the form
\begin{equation}
\label{eq:h}
  h=2^{-1}(\xi-\A-\A_\delta)^2+V+V_{\delta},
\end{equation}
where $\A=\A(x)=\A(\theta)=(-\sin \theta{},\cos\theta{})b(\theta{})$, 
$V=V(x)=V(\theta)$,  and $\A_\delta=\A_\delta(x)$ and  
$V_{\delta}(x)$ have decay (measured by some $\delta>0$). For
 convenience we
assume below that $\A_\delta=0$ and $V_{\delta}=0$, and devote Remark \ref{remark:
 perturbations} to a  discussion on inclusion of  non-trivial  perturbations.
We assume that $b$ and $V$ are real-valued, smooth  and $2\pi$-periodic functions of the angle 
$\theta$.

We may write
\begin{equation*}
  h=2^{-1}(\rho^2+\eta^2)+ V,
\end{equation*}
where 
\begin{align}
  \label{eq:polar1}
 & \rho=\hat x \cdot \xi,\;\hat x=(\cos
\theta{},\sin\theta{})=x/r,\;r=|x|,\\
\label{eq:polar2}
&\eta=(-\sin
\theta{},\cos\theta{})\cdot {\xi} -b.
\end{align}

Let us fix an open interval $I\subseteq (\min V,\infty)$ 
free from critical values, i.e.
\begin{equation}
\label{eq:noncrit}
  V(\theta)=E\in I \Rightarrow V'(\theta)\neq 0.
\end{equation}

The equations of motion in the ``new time''
\begin{equation}\label{eq:time}
  \tau=\int r^{-1}dt,
\end{equation}
read
\begin{equation}\label{eq:equamotion}
\begin{cases}
\frac{d}{d\tau{}}\theta{}=\eta{}\\
\frac{d}{d\tau{}}\eta{}{}=-(\eta{}+b)\rho{}-V'\\
\frac{d}{d\tau{}}\rho{}=(\eta{}+b)\eta{}
\end{cases}\;.
\end{equation}
%\begin{align}
%\label{eq:equamotion}
%&\frac{d}{d\tau{}}\theta{}=\eta{}.\nonumber\\
%&\frac{d}{d\tau{}}\eta{}{}=-(\eta{}+b)\rho{}-V',\\
%&\frac{d}{d\tau{}}\rho{}=(\eta{}+b)\eta{},\nonumber
%\end{align}

We notice that this system is complete in the sense that the  maximal
solutions are defined for all $\tau \in \R$. Clearly the energy $h$ is
 a preserved observable. 

 Obviously  the fixed points are given by the two systems of equations 
\begin{align}
  \label{eq:fixed1}
 &\eta= b\rho+V'=0,\\
\label{eq:fixed2}
&\rho=+\sqrt{2(h-V)}\; {\rm or} \;\rho= -\sqrt{2(h-V)}.
\end{align}

The set of fixed points $z^+=z=(\theta,\eta,\rho)$ for which
$\rho>0$ and $h\in I$ 
will be denoted by $\mathcal{F}^+$. The set of fixed points
$z^-=(\theta,\eta,\rho)$ for which $\rho<0$ and $h\in I$ 
will be denoted by $\mathcal{F}^-$. Let $\mathcal{F}=
\mathcal{F}^+\cup \mathcal{F}^-$. For $E\in
I$ we put $\mathcal{F}^+(E)=\mathcal{F}^+\cap \{h=E\}$, and
introduce similarly
sets $\mathcal{F}^-(E)$ and $\mathcal{F}(E)$. With the conditions
\begin{equation}
  \label{eq:nondege1}
 \kappa^+:=\frac{d}{d\theta{}} (b\sqrt{2(E-V)}+V')\neq 0 \text { at
 all 
 }z^+\in\mathcal{F}^+(E),
\end{equation}
and
\begin{equation}
  \label{eq:nondege2}
 \kappa^-:=\frac{d}{d\theta{}} (-b\sqrt{2(E-V)}+V')\neq 0 \text { at
 all 
 }z^-\in\mathcal{F}^-(E),
\end{equation} 
there is at most  a  finite number of points in $\mathcal{F}(E)$. 
Moreover they depend
smoothly on $E$.

 For simplicity of presentation we shall tacitly assume
 \eqref{eq:noncrit}, \eqref{eq:nondege1}  and \eqref{eq:nondege2}
 throughout the paper unless otherwise stated (there are exceptions in
 Section \ref{The general classical case}).
 
 For any given $z^+=(\theta_j,0, \rho_j) \in\mathcal{F}^+(E)$ we may eliminate $\rho$ in the second equation of
\eqref{eq:equamotion} near  $z^+$ (by the energy relation) and obtain  an
 autonomous system whose linearization at the fixed point $(\theta,\eta)=(\theta_j,0)$ takes the following form: For
 $y=(\theta-\theta_j,\eta)^{\tr}$
\begin{displaymath} 
 \frac{d}{d\tau}y=A^+y;\;  A^+ =\left ( \begin{array}{cc}
0&1\\
-\kappa^+&-\sqrt{2(E-V)}
\end{array}\right ).
\end{displaymath}
Here $A^+$ is evaluated at $(E,\theta_j(E))$, that is 
$A^+=A^+(E,\theta_j(E))$.

The eigenvalues of $A^+$ are given by
\begin{equation}
  \label{eq:eigenv1}
 \lambda^+=- 2^{-1}\sqrt{2(E-V)}+(-)2^{-1}\sqrt{2(E-V)-4\kappa^+}.
\end{equation}

We proceed similarly 
at $z^-\in\mathcal{F}^-(E)$; the  eigenvalues of the corresponding matrix
$A^-$ are given by
\begin{equation}
  \label{eq:eigenv2}
 \lambda^-=2^{-1}\sqrt{2(E-V)}+(-)2^{-1}\sqrt{2(E-V)-4\kappa^-}.
\end{equation}

We assume for convenience that for $z\in\mathcal{F}(E)$
\begin{equation}
  \label{eq:assdif}
 2(E-V)\neq 4\kappa^{\sharp}. 
\end{equation}

With \eqref{eq:assdif} the matrix
$A^{\sharp}$ 
is diagonalized by the matrix
\begin{displaymath}
  T(E,z^{\sharp}(E))= \left( 
      \begin{array}{cc}
1 & 1\\
\lambda^{\sharp}_1 & \lambda^{\sharp}_2
      \end{array}\right),
\end{displaymath} where $\sharp$ refers to either plus or minus and
the subscript refers to an arbitrary numbering of the two eigenvalues
in question at $z^{\sharp}=z^{\sharp}(E)$. The observable
\begin{equation}
  \label{eq:liapunov}
 l=l_E=|T(E,z^{\sharp}(E))^{-1}\bar z|_{\C}^2;\;\bar z=(\theta-\theta^{\sharp}(E),\eta)^{\tr},
\end{equation}
 is a ``Liapunov-function'' near  $z^{\sharp}(E)$ if the fixed point
 is a sink  (where $l$ is decreasing) or a source (where $l$ is increasing); that
is, if the real part of the two eigenvalues has the same sign.
%\footnote {Cf. textbooks on stability near fixed
%points for dynamical systems??} 
If \eqref{eq:assdif} is violated one
 may still construct a similar smooth function $l=l_E(\theta, \eta
 )$, see \cite {HS2}, and this would be a substitute for
 \eqref{eq:liapunov} in our applications.

We denote by $\mathcal{F}^+_{\sa}$ the set of saddle points in
$\mathcal{F}^+$ (corresponding to having $\kappa^+<0$ in \eqref{eq:eigenv1}). Similarly $\mathcal{F}^+_{\si}=\mathcal{F}^+\setminus
\mathcal{F}^+_{\sa}$,  $\mathcal{F}^-_{\sa}$ is the set of saddle points in
$\mathcal{F}^-$ and $\mathcal{F}^-_{\so}=\mathcal{F}^-\setminus
\mathcal{F}^-_{\sa}$. Furthermore we put
$\mathcal{F}^+_{\sa}(E)=\mathcal{F}^+_{\sa}\cap \{h=E\}$, etc.

\begin{cond}[Zero flux condition]
\label{cond:flux}
\begin{equation}
  \label{eq:flux}
 \int^{2\pi}_0b(\theta) d\theta=0. 
\end{equation}
\end{cond}
We introduce 
\begin{equation}\label{fluks}
\tilde{b}(\theta):=\int_{0}^\theta b(\varphi)d\varphi,\quad
\theta\in \mathbb{R}.
\end{equation}
With \eqref{eq:flux} the function $\tilde b$ is
$2\pi$-periodic. 

\begin{lemma}\label{lemclasss} Suppose $E\in I$ obeys $E<\max V$, then
  $\mathcal{F}^+(E)$ and $\mathcal{F}^-(E)$ are non-empty.

Suppose Condition \ref{cond:flux} and $E\in I$ obeys $E>\max V$, then all of the  sets
$\mathcal{F}^+_{\sa}(E)$, $\mathcal{F}^+_{\si}(E)$,
$\mathcal{F}^-_{\sa}(E)$ and $\mathcal{F}^-_{\so}(E)$  are non-empty.
\end{lemma}
\begin{proof}  As for the  first part of
the lemma we pick $\theta_1$ such that $V(\theta_1)=\min V$ and a
maximal interval $J=( \theta_1^-,\theta_1^+)\ni \theta_1$ such that
$V(\theta)< E$ for $\theta\in J$. Look at $f:J\to \R$ given by
$f=b \sqrt{2(E-V)}+V'$. Since $\lim_{\theta\to
  \theta_1^-}f(\theta)<0$ and $\lim_{\theta\to
  \theta_1^+}f(\theta)>0$ there exists $\theta\in J$ such that
$f(\theta)=0$. This shows that  $\mathcal{F}^+(E)\neq \emptyset$. We may argue similarly for
  $\mathcal{F}^-(E)$. 

 As for the second part of the lemma  we prove that $\mathcal{F}^+(E)$
 has at least two elements for $E>\max V$.  We
need to show that $b(\theta)\sqrt{2(E-V(\theta))}+V'(\theta)$ has at
least two zeroes on the torus. But this is the same as showing that 
$$b+\frac{V'}{\sqrt{2(E-V(\theta))}}=\left
  (\tilde{b}(\cdot)-\sqrt{2(E-V(\theta))}\right )'$$
has at least two zeroes, which is implied by the fact that this
  function is periodic
and its
integral on the torus is zero. Clearly by \eqref{eq:nondege1} one of the zeros obeys $\kappa^+>0$ while
  another one obeys $\kappa^+<0$. 
We may argue similarly for
  $\mathcal{F}^-(E)$. 
\end{proof}

We notice that the second part of Lemma   \ref{lemclasss} is in agreement
with the Poinc\' are-Hopf theorem since for $E>\max
V$ the  corresponding energy surface is topological a torus. Below $\max
V$ the situation is different, in fact there might be no saddle
points  at all.  

\begin{prop}\label{lemdoi} Suppose Condition \ref{cond:flux}. Then
  every integral curve
  $\gamma(\tau)=(\theta(\tau),\eta(\tau),\rho(\tau))$  of
  the system \eqref{eq:equamotion} with energy $h=E\in I$ has the  property that there
  exist $z_1, z_2 \in \mathcal {F} (E)$ such that 
\begin{equation}\label{atrak}
\lim_{\tau\to +\infty}\gamma(\tau)=z_1.
\end{equation} and \begin{equation}\label{atrak40}
\lim_{\tau\to -\infty}\gamma(\tau)=z_2.
\end{equation} 
\end{prop}
\begin{proof} 
Let ${y}=(
b\rho+V',\eta)^{\tr}$. 
 Using
\eqref{eq:equamotion} we compute
for any classical orbit ${y}={y}(\tau)$
\begin{equation}
\label{eq:Bclas}
\frac{d}{d\tau}y=A{y}+\mathcal O(\eta^2),
\end{equation} where 
\begin{displaymath} 
A=\left ( \begin{array}{cc}
0&b'\rho+b^2+V''\\
-1&-\rho
\end{array}\right ).
\end{displaymath} 

Also we   introduce
\begin{align}\label{atrak2}
a_1(\tau)&=\rho(\tau)-\tilde{b}(\theta(\tau)) \nonumber \\ 
a_2(\tau)&=Ca_1(\tau)-2y_1(\tau)y_2(\tau). 
\end{align}
We differentiate $a_1$ and $a_2$ using 
(\ref{eq:equamotion}) and \eqref{eq:Bclas} and get

\begin{equation}\label{atrak3}
a_1'=\eta^2,\quad
(a_2-Ca_1)'\geq q+\mathcal{O}(\eta^2);\;q=y^2.
\end{equation}

Taking $C>0$ large enough we deduce that 
\begin{equation}
  \label{eq:b^2+}
 \frac{d}{d\tau{}}a_2\geq q.
\end{equation}
 
Notice that $a_1$ is always bounded due to the
zero flux condition.  This implies
that $a_2$ is also bounded. Whence by integrating $a_2'$ we infer that 
\begin{equation}\label{atrak4}
\int_0^\infty q(\tau)d\tau <\infty.
\end{equation}

By the finiteness of the integral in (\ref{atrak4}) there exists a sequence 
$\tau_n \to +\infty$ such that $q(\tau_n)\to 0$. Hence if we prove that
$\lim_{\tau\to+\infty}q(\tau)$ 
exists, it must be zero. In order to
achive this, we employ a Cook-type argument, that is, we prove that
$q'(\cdot)\in L^1([0,\infty))$. The verification is obvious since from
\eqref{eq:Bclas} we get
$$|q'(\tau)|\leq {\rm const} \cdot q(\tau),\quad \tau\geq0,$$  
so using again (\ref{atrak4}) we get the result. 

We can therefore conclude that
\begin{equation}\label{forke2}
\lim_{\tau\to +\infty}\left
  [b(\theta(\tau))\rho(\tau)+V'(\theta(\tau))\right ]=0 \text{ and }
  \lim_{\tau\to +\infty}\eta(\tau)=0.
\end{equation} 
 From this and compactness we conclude that along some sequence
$\tau_n\to +\infty$, 
\begin{equation}
  \label{eq:conv+}
  \theta\to \theta_1, \;\eta\to 0,\;\rho\to
\rho_1.
\end{equation}
 Here either,  $\rho_1=\sqrt{2(E-V(\theta_1))}$ or 
  $\rho_1=-\sqrt{2(E-V(\theta_1))}$. In particular $z_1=(\theta_1,0,
    \rho_1)\in \mathcal {F}(E)$. Using \eqref{eq:nondege1} and
    \eqref{eq:nondege2} we see that the convergence $z\to z_1$ of
    \eqref{eq:conv+} in fact holds for  $\tau\to +\infty$ rather than
    along  the
    sequence $\tau_n \to +\infty$.  

We may argue similarly for $\tau\to -\infty$.
\end{proof}

For a generalization of Proposition
\ref{lemdoi} we refer to Proposition \ref{prop:clas-gen}.
%\footnote{By topological reasons: index theory, Euler characteristic etc., see
%the book by
%Irwin (p.141), the situation is different below $\max
%V(\theta)$.}

To deal with quantum mechanis we shall need the following condition. 

\begin{cond}[No exceptional orbits condition] \label{cond:noexcep}An
  energy $E\in \R$ is said to
  be  exceptional if  there  exists an (exceptional)
integral curve $\gamma(\cdot)$  of the system \eqref{eq:equamotion} with
 $h=E$ such that 
simultaneously 
\begin{equation}\label{eq:unstable}
  \lim_{\tau \to +\infty} \gamma(\tau) \in \mathcal{F}^-_{\sa} \text{
  and }\lim_{\tau \to -\infty} \gamma(\tau) \in \mathcal{F}^+_{\sa}.
\end{equation}
Denoting by
  $\mathcal E _{\exc}$ the set of exceptional energies,  $\mathcal E
  _{\exc}\cap I=\emptyset$. 
\end{cond}

We remark that Condition  \ref{cond:noexcep} is fulfilled if $b=0$,
since in this case the observable $\e^{\int \rho}\rho$ is increasing in
$\tau$.

\begin{example}[Criterion at high energies] \label{example:noexcep}
Suppose $V=0$. 
  For all integral curves  $\gamma$ with energy $E>0$ we may pick  
  $\epsilon=\epsilon(E)>0$ such that along
  $\gamma$  
  \begin{equation}
    \label{eq:crit1}
  \frac{d}{d\tau{}}a_2\geq \frac {\epsilon}{2}(\eta^2+b^2),
  \end{equation} cf. \eqref{atrak2} and \eqref{eq:b^2+}. 
Let $C(E)=\epsilon^{-1}C$. We claim that \eqref{eq:unstable} is
not possible for 
\begin{equation}
  \label{eq:crit2}
  E > 8^{-1}\left ((1-C(E)^{-1})\sup [ \tilde
  b]\right )^2; \;[ \tilde
  b]_{\theta_1}^{\theta_2}=\int _{\theta_1}^{\theta_2} b(\theta) d\theta.
\end{equation}
To see this let us assume that there is an exceptional orbit starting
at $\theta=\theta_1$ and ending at $\theta=\theta_2$. Then we may use
\begin{equation*}
-[ \tilde
  b]_{\theta_1}^{\theta_2}=  -\int _{-\infty}^{\infty}b\eta d\tau
\end{equation*}
and \eqref{eq:crit1} to estimate 
\begin{equation*}
-[ \tilde
  b]_{\theta_1}^{\theta_2} \leq \epsilon^{-1}[ a_2
  ]_{\theta_1}^{\theta_2}  =C(E)(-2\sqrt {2E}
-[ \tilde
  b]_{\theta_1}^{\theta_2}),
\end{equation*} yielding
\begin{equation*}
2\sqrt {2E}\leq -(1-C(E)^{-1})[ \tilde
  b]_{\theta_1}^{\theta_2}.
\end{equation*} 

We remark that the observable $a=r(\rho-\tilde b+c)$, $c=2^{-1}(\max
\tilde b+\min \tilde b)$, provides a  Mourre estimate for energies $E>8^{-1}(\sup [ \tilde
  b])^2$ 
in this case. Thus the bound \eqref{eq:crit2} is a slight
improvement.
\end{example}

\begin{remark} \label{rem:excep} One obtains ``$r$'' from the system
  \eqref{eq:equamotion} by the formula $r=r_0\exp (\int
  \rho d\tau)$ In particular any exceptional orbit
  defines a one-parameter family of classical orbits in the
  full phase space. This may also be seen from the invariance of Newton's equation, $x(t)\to
  cx(c^{-1}t)$ for $c>0$. The configuration part $x(t)$ of any
  solution associated with an exceptional orbit has the peculiar
  property that it collapses at the origin both backwards and forwards in
  finite ``physical''  time $t$. (In other words any  maximal solution
  is only defined on a bounded interval.) In Appendix \ref{Collapsing
  orbits} we show that $\mathcal E_{\exc}\cap(\max V, \infty)$ is
  discrete in $(\max V, \infty)$.

For a more general
  discussion of collapsing orbits we also refer the reader to Appendix \ref{Collapsing orbits}. 
\end{remark}

Throughout the paper the notation $F(x>\epsilon)$ denotes 
a smooth increasing function $=1$ for $x>\frac {3}{4}\epsilon$ and
 $=0$ for $x<\frac {1}{2}\epsilon$;
 $F(\cdot<\epsilon):=1-F(\cdot>\epsilon)$. 
\section{Propagation of singularities} \label{Propagation of
  singularities}
For the  results of this section, Conditions
\ref{cond:flux} and \ref{cond:noexcep}
 are not used.

Let $L^{2,s}=X^{-s}L^2({\mathbb
R}^2_x)$, $X=\langle x\rangle =(1+|x|^2)^{1/2}$. Introduce 
$L^{2,-\infty}=\cup_{s\in \mathbb
R}L^{2,s}$ and $L^{2,\infty}=\cap_{s\in \mathbb
R}L^{2,s}$. We define the  wave front set $WF^s(u)$ of  a distribution
$u\in L^{2,-\infty}$  to be  the
subset of $\mathbb
T\times {\mathbb
R}^2_{\xi}$ whose complement is given by 
\begin{align}
   \label{eq:WF^s}
   &(\theta_0,\xi_0)=z_0\notin WF^s(u) \nonumber\\
&\Leftrightarrow\\
   &\exists\;{\rm neighborhoods }\;\mathcal{N}_{\theta_0}\ni
   {\theta_0},\;\mathcal{N}_{\xi_0}\ni {\xi_0}
\forall \chi_{\theta_0}\in
   C^{\infty}(\mathcal{N}_{\theta_0}),\;\chi_{\xi_0}\in
   C^{\infty}_0(\mathcal{N}_{\xi_0}): \nonumber\\
&\chi_{\theta_0}({\theta})\chi_{\xi_0}({p})u\in L^{2,s}. \nonumber
 \end{align}
 Here $p=\p=-i\nabla$.

For $z\in \mathbb
T\times {\mathbb
R}^2$
\begin{equation*}
L^{2,s}_z=\{u \in L^{2,-\infty}| z\notin  WF^s(u)\} 
\end{equation*}

Obviously 
\begin{equation*}
 WF^s(u)=\emptyset \Leftrightarrow \forall z\in\mathbb
T\times {\mathbb
R}^2:\;u \in L^{2,s}_z,
\end{equation*}
and 
\begin{equation*}
 u\in L^{2,s}  \Rightarrow WF^s(u)=\emptyset.
\end{equation*}
Conversely (by a compactness argument), if for some $\chi\in C^{\infty}_0(\mathbb{R}^2)$
\begin{equation*}
  u-\chi(p)u\in L^{2,s},
\end{equation*}
then 
\begin{equation*}
WF^s(u)=\emptyset \Rightarrow u\in L^{2,s}.  
\end{equation*}

We recall the continuous inclusions for Besov spaces
\begin{equation}
  \label{eq:besov}
L^{2,-1/2} \subset B^*_0\subset B^*\subset
L^{2,-1/2-\epsilon};\;\epsilon>0, 
\end{equation}
cf. \cite[ Section 14.1]{Ho1}. The norm on $B^*$ may be taken as
\begin{equation*}
  \|u\|_{B^*}^2=\sup _{R>1}R^{-1}\int_{|x|<R}|u|^2 dx,
\end{equation*}
and $u\in B^*_0$ if and only if  also 
\begin{equation*}
  \lim _{R\to \infty}R^{-1}\int_{|x|<R}|u|^2 dx =0.
\end{equation*} 
 
We shall use  the standard metric
$$g = \langle x \rangle^{-2}dx^2
+ d\xi^2,$$
and standard Weyl quantization $\Op (a)$ of symbols $a\in S(m,g)$.
Here we shall use $m$ of the form $m=\langle x \rangle^s \langle \xi \rangle^t$.

Now let us look at the  quantization $H$ of \eqref{eq:h}. We shall
 essentially assume that 
  $\A_{\delta}=0$ and $V_{\delta}=0$ (see Remark \ref{remark:
 perturbations} for an 
 extension), but for convenience we cut off the singularity at the
 origin. Whence we consider the symbol $\bar h$ obtained by
 replacing $\A$ and $V$  in the expression $h$ by
 $F(r>1)\A$ and $F(r>1)V$, respectively. A small  computation shows
 that 
 \begin{equation}
   \label{eq:weylh}
H:={\Op}(\bar h)=2^{-1}(\p-F(r>1)\A)^2+F(r>1)V.   
 \end{equation}

We shall prove a propagation of singularity result using
 \eqref{eq:polar1} and \eqref{eq:polar2} to  identify $\mathbb
T\times {\mathbb
R}^2_{\xi}$ as 
\begin{equation*}
  {\T}^*:=\mathbb
T\times {\mathbb
  R}^2_{(\eta, \rho)}.
\end{equation*}
Thus ${\T}^*$ is the phase space of the system
\eqref{eq:equamotion} (and not the cotangent bundle of ${\T}$ nor the
phase space of the full Hamiltonian system $T^*\R^2$).

If $(H-E)u=v$ for $E\in
I$ and  $v\in L^{2,s}$, then (by ellipticity)
\begin{equation}
  \label{eq:Char}
WF^s(u)\subseteq \T^*_E:=\{z\in
\T^* |\;h(z)=E\}.  
\end{equation}

The maximal solution of the system
\eqref{eq:equamotion} that 
passes $z$ at $\tau=0$ is denoted by $\gamma(\tau,
z)$ or $\phi_\tau (z)$. 

\begin{prop}
  \label{prop:propa} Suppose $u\in L^{2,s-1}$, $v\in L^{2,s+1}$, $E\in
I$ and  $(H-E)u=v$. Then 
 \begin{equation}
   \label{eq:bich} \gamma(\mathbb{R}\times (\mathbb{T}^*\setminus WF^s(u)))\subseteq
\mathbb{T}^*\setminus WF^s(u).
 \end{equation}
\end{prop}
\begin{proof}  Our proof is a modification of that of
\cite[Proposition 3.5.1]{Ho2}, see also \cite{Melr} and \cite{HMV}. Suppose $u\in L^{2,s}_z$ for some $z \in
  {\T}^*$. Then we need to show that $u\in L^{2,s}_{\gamma 
(\tau)}$  along  the integral curve  
$\gamma (\cdot)=\gamma(\cdot,z)$. Consider 
\begin{equation}
  \label{eq:maxtau}
\tau_0 = \sup \{\tau >0 | u\in L^{2,s}_{\gamma
(\tilde \tau)}\text { for all }\tilde \tau \in [0,\tau]\}.  
\end{equation}
We need to show that $\tau_0 = \infty$. (Similar arguments will work
in the backward direction.)

Suppose on the contrary that $\tau_0$ is finite. Then $\gamma
(\tau_0)$ is not a fixed point. Consequently we may pick a slightly smaller
$\tau_0' <\tau_0$ and  a transversal 2-dimensional submanifold at $\gamma(\tau_0')$, say
$\mathcal{M}$, such  that with $I=(-\epsilon+\tau_0' , \tau_0+\epsilon)$  
for some $\epsilon>0$
the map
\begin{equation*}
 I\times \mathcal{M} \ni
 (\tau,m) \to  \Psi (\tau,m )= \gamma(\tau-\tau_0', m) \in {\mathbb
T}^*,
\end{equation*}
is a diffeomorphism onto its range.

We pick $\chi \in C^{\infty}_0(\mathcal{M})$ supported near
$\gamma(\tau_0')$  such that $$\chi (\gamma(\tau_0'))=1.$$
 Since
$u\in L^{2,s}_{\gamma (\tau_0')}$ we can find a  non-positive function
$f \in
C^{\infty}_0(I)$ such that $f'\geq 0$
on a neighborhood of $[ \tau_0' , \tau_0+\epsilon)$, $f( \tau_0)<  0$
 and the set 
\begin{equation}
\label{eq:nowavefront}
  \Psi((-\epsilon+\tau_0' , \tau_0'] \times \supp \chi) \cap WF^s(u) =\emptyset.
\end{equation}

Let $f_K(\tau)=\exp (-K\tau)f(\tau)$ for $K>0$, and $X_{\kappa}=
(1+\kappa r^2)^{1/2}$ for $\kappa \in (0,1)$. We consider the classical observable 
\begin{equation}
\label{eq:propobs}
  b_\kappa=X^{1/2}a_\kappa;\; a_\kappa=X^sX_{\kappa}^{-3/2}F(r>1)(f_K
  \otimes \chi)\circ \Psi^{-1}.
\end{equation} 
First we fix $K$: A part of the Poisson bracket with
 $b_\kappa^2$ is
\begin{equation}
\label{eq:Poi1}
  \{h,X^{2s+1}X_{\kappa}^{-3}\}= X^{-1}(Y_\kappa \rho )X^{2s+1}X_{\kappa}^{-3},
\end{equation}
where $Y_\kappa=Y_\kappa(r)$ is uniformly bounded in $\kappa$.
We fix $K$ such that $2K\geq |Y_\kappa \rho |+2$  on $\supp b_\kappa$.

By  the system \eqref{eq:equamotion} 
\begin{equation}
  \label{eq:Poi2}
 \{ h,(f_K \otimes \chi)\circ \Psi^{-1}\} =r^{-1}([\frac {d} {d\tau }f_K] \otimes \chi)\circ \Psi^{-1}. 
\end{equation} 

>From \eqref{eq:Poi1} and \eqref{eq:Poi2}, and by the choice of $f$ and
$K$,  we conclude that  
\begin{equation}
\label{eq:Poi3}
  \{h,b_\kappa^2\}\leq -2a_\kappa^2 \text{ at   }
   \mathcal P
\end{equation}
given by  $$\mathcal P =[1,\infty)\times \Psi(\{\tau | f'(\tau)\geq0\}\times \supp \chi\}),$$
and therefore in particular
\begin{equation}
\label{eq:Poi4}
   \{h,b_\kappa^2\}(r,\gamma (\tau_0))< 0.
\end{equation}

Next we introduce $A_\kappa={\Op}(a
_\kappa)$ and $B_\kappa={\Op}(b
_\kappa)$. We write 
\begin{equation}
  \label{eq:comm1}
 \langle i[H, B_\kappa^2]\rangle _u=-2\Im \langle v,B_\kappa^2 u\rangle, 
\end{equation} 
and estimate the right hand side using the calculus of
pseudodifferential operators  \cite[Theorems 18.5.4, 18.6.3]{Ho1} and the Cauchy-Schwarz inequality to
obtain the uniform bound 
\begin{equation}
  \label{eq:comm2}
 |\langle i[H, B_\kappa^2]\rangle _u|\leq C_1\|v\|_{s+1}\|A_\kappa
 u\|+C_2
\leq \|A_\kappa
 u\|^2 +C_3.
\end{equation}

 On the other hand using \eqref{eq:weylh}, \eqref{eq:nowavefront},
 \eqref{eq:Poi3}  and the calculus
 \cite[Theorems 18.5.4, 18.6.3, 18.6.8]{Ho1}  we infer that 
\begin{equation}
  \label{eq:comm3}
 \langle i[H, B_\kappa^2]\rangle _u\leq -2\|A_\kappa u\|^2+C_4.
\end{equation} 

Combining \eqref{eq:comm2} and \eqref{eq:comm3} yields 
\begin{equation*}
 \|A_\kappa u\|^2\leq C_5=C_3+C_4,
\end{equation*} 
which in combination with  \eqref{eq:Poi4} in
turn yields a uniform bound 
\begin{equation}
  \label{eq:comm4}
 \|X_{\kappa}^{-1}\chi_{\gamma (\tau_0)}u\|^2_s\leq C_6.
\end{equation}
Here $\chi_{\gamma (\tau_0)}$ signifies a non-trivial phase-space localization
factor
of the form entering in \eqref{eq:WF^s} centered at the point $\gamma
(\tau_0)$ (using  polar
coordinates).

We let $\kappa\to 0$ in \eqref{eq:comm4} and infer that $u \in
L^{2,s}_{\gamma (\tau_0)}$, which is a contradiction.
\end{proof}

\begin{remark} The assumption $u\in L^{2,s-1}$ of Proposition
  \ref{prop:propa}
is not needed. This may be shown by iterating the method of proof,
cf. \cite[Proposition 3.5.1]{Ho2}.
\end {remark}
 
An  elaboration of the above proof gives the following result,
cf. \cite[Proposition 3.5.1]{Ho2}. For $z\in {\T}^*$ we use the
notation $\mathcal{N}_{z}$ to denote a neighborhood of $z$, and for $\chi \in C^{\infty}_0(\mathcal{N}_{z})$  the
notation ${\Op}( \chi )$  for the operator ${\Op}(a)$ with symbol $a(x,\xi)=F(r>1)\chi$
 as defined by the identification \eqref{eq:polar1} and
 \eqref{eq:polar2}.
\begin{prop}
  \label{prop:propaz} Suppose $u\in L^{2,s-1}$, $v\in L^{2,s+1}$, $\Re
  \zeta\in
I$, $\Im \zeta \geq 0$  and  $(H-\zeta)u=v$. Suppose $z_1,z_2
\in {\T}^*$ are linked as $\gamma (\tau_0,z_1)=z_2$ for a positive
$\tau_0$. Then 
 \begin{align}
   \label{eq:bichz}
 &\forall \mathcal{N}_{z_1}\exists \mathcal{N}_{z_2}\forall \chi_2 \in
 C^{\infty}_0(\mathcal{N}_{z_2}) \exists \chi_1 \in
 C^{\infty}_0(\mathcal{N}_{z_1}):\\
& \|{\Op}( \chi_2 )u\|^2_s\leq C\left(\|u\|^2_{s-1}+\|{\Op}( \chi_1 )u\|^2_s+\|v\|^2_{s+1}\right).\nonumber
 \end{align}
The bound \eqref{eq:bichz} is uniform in $\zeta$ in a bounded set.
\end{prop}

\begin{remark}
\label{remark:propa2} Under the condition of Proposition
\ref{prop:propa} the sign condition in Proposition
\ref{prop:propaz} is redundant. Thus for $\zeta$ real the wave front
 bound \eqref{eq:bichz} is valid in both directions of the flow.
\end {remark}

\begin{prop}
  \label{prop:wave_upper} Suppose $u\in L^{2,-1}$,
  $E \in
I$ and  $(H-E)u=0$. Suppose $WF^{-\frac {1}{2}}(u)\subseteq
\T^*_+=\{z\in \T^*|\rho>0\}$. Then
 \begin{equation}
   \label{eq:wave_empty}
  u\in B^*_0.
 \end{equation}
\end{prop}
\begin{proof} We follow the proof of \cite[Theorem
  30.2.6]{Ho1}. Pick a real-valued decreasing $\psi\in
  C^{\infty}_0({[0,\infty)})$ such that  $\psi(r)=1$ in a small neighborhood of $0$ and $\psi'(r)=-1$ for $1/2\leq
r \leq 1$. Let
 $\psi_R(x)=\psi(r/R)$ for $R>1$.
\begin{equation}
  \label{eq:comWF}
 0=\langle i[H-E, \psi_R]\rangle _u=2^{-1}R^{-1}\Re \langle (\hat x \cdot p
 +p\cdot \hat x)\psi'(r/R)\rangle _u.
\end{equation}
 
We use \eqref{eq:besov} and the assumption of the proposition to bound the right hand
side of \eqref{eq:comWF} as
\begin{equation*}
  \cdots \leq \epsilon R^{-1}\langle \psi'(r/R)\rangle _u + T_R,
\end{equation*}
where
\begin{equation*}
  \lim_{R\to \infty}T_R=0.
\end{equation*} 

We conclude \eqref{eq:wave_empty}.
\end{proof}
\section{Microlocal bounds}
\label{Wave front set near}
In this section Condition \ref{cond:flux} (but not 
 Condition  \ref{cond:noexcep}) is imposed. 
Suppose we have given $u\in L^{2}$ and  $v\in L^{2,t}$ such that
 $(H-\zeta)u=v$,  where $t\in (1/2,1)$, 
  $\Re
  \zeta\in
I$ and  $\Im \zeta \geq 0$, 
  cf. Proposition \ref{prop:propaz}. 
\subsection{Bounds at $\mathcal{F}^-_{\so}$}
\label{Wave front set1}

We shall prove  microlocal bounds
  of $u$ at $\mathcal{F}^-_{\so}$ (the set of
  ``sources'') somewhat along the line of the proof of
  Proposition \ref{prop:propa}. Consider a  branch $z^-(\cdot )\in \mathcal{F}^-_{\so}$. Instead of the 
observable \eqref{eq:propobs} we now consider 
\begin{equation}
\label{eq:propobs2}
  b_\kappa=X^{1/2}a_\kappa;\; a_\kappa=(\frac
  {X}{X_{\kappa}})^{t-1}X_{\kappa}^{-1/2}F(r>1)f(h)F(l<\epsilon),
\end{equation} where $f\in C^{\infty}_0(I)$, $f$ real-valued,  and $l=l_h(\theta, \eta
 )$ is the 
``Liapunov-function''
as defined in \eqref{eq:liapunov}, in the present context in terms of $z^-$.

 Picking $\epsilon >0$ small enough in \eqref{eq:propobs2} we
may estimate part of  the Poisson bracket with $b_\kappa^2$ as 
\begin{equation}
  \label{eq:negpoiss1}
 \cdots \{h, F(l<\epsilon)\}\leq  \delta r^{-1}F'(l<\epsilon), 
\end{equation} 
for some positive $\delta$  (which essentially is determined 
as  a uniform lower bound of the real part of the eigenvalues
\eqref{eq:eigenv2}).

Obviously on the support of $b_\kappa^2$
\begin{equation}
  \label{eq:negpoiss2}
 \{h, (\frac {X}{X_{\kappa}})^{2t-1}\}=(1-\kappa)(2t-1)rX^{-1}X_{\kappa}^{-3}\rho(\frac {X}{X_{\kappa}})^{2t-2}<0.
\end{equation}

We quantize \eqref{eq:propobs2} writing $A_\kappa={\Op}(a
_\kappa)$ and $B_\kappa={\Op}(b
_\kappa)$. Using \eqref{eq:negpoiss1}  and \eqref{eq:negpoiss2} we may estimate
\begin{equation}
  \label{eq:comms1}
 \langle i[H, B_\kappa^2]\rangle _u\leq -2\sigma\langle
 X_{\kappa}^{-2}\rangle _{A_\kappa u}+C_1\| u\|_{-1}^2.
\end{equation}  

On the other hand since $(\Im \zeta) B_\kappa^2\geq 0$ 
\begin{align}
  \label{eq:comms2}
 &\langle i[H-\zeta, B_\kappa^2]\rangle _u\geq 2\Im \langle B_\kappa^2u,
 v\rangle \nonumber\\
&\geq -\sigma\|X_{\kappa}^{1/2-t}A_\kappa u\|^2-C_2\|u\|_{-1}^2-C_3\|v\|^2_t.
\end{align}

We combine \eqref{eq:comms1} and \eqref{eq:comms2} to obtain 
\begin{equation}
  \label{eq:bound122}
  2\sigma\|X_{\kappa}^{-1}A_\kappa u\|^2 -\sigma\|X_{\kappa}^{1/2-t}A_\kappa u\|^2\leq (C_1+C_2)\|u\|_{-1}^2+C_3\|v\|^2_t,
\end{equation}
from which we  conclude by 
letting $\kappa\to 0$ (and dividing by $\sigma$) that 
\begin{align}
  \label{eq:bound1}
  &\|\tilde Au\|_{t-1}^2 \leq C_4\|u\|_{-1}^2+C_5\|v\|^2_t;\\&\tilde
  A={\Op}(F(r>1)f(h)F(l<\epsilon)),\nonumber
\end{align}
with constants independent of $\zeta$ in a bounded  set.

We notice that a similar bound at $\mathcal{F}^-_{\sa}$ would require
an extra input. We will come back to that point in Section \ref{Saddlepo}.

\subsection{Bounds away from  $\mathcal{F}$}
\label{Wave front set2}
Consider the quantization $A_2={\Op}(\bar a_2)$ of the symbol $\bar a_2=F(r>1)f(h)a_2$ where $a_2$
is given  by \eqref{atrak2} and the other factors are as in
\eqref{eq:propobs2}. Introduce 
\begin{equation}
  \label{eq:e}
e=r^{-1}F(r>1)f(h) ((b\rho+V')^2+\eta^2)\;(\in S(\langle x \rangle^{-1},g)). 
\end{equation} 
We compute 
\begin{equation}
  \label{eq:com a_2}
  i[H,A_2]\geq {\Op}(e) +{\Op}(r_{-3}), 
\end{equation}
where 
$r_{-3}\in S(\langle x \rangle^{-3},g)$.
Since $A_2$ is bounded we may pick a constant
$C_2$ such that $C_2-A_2\geq 0$. Using the scheme of \eqref{eq:comms1}
and \eqref{eq:comms2} with $B_\kappa^2$ replaced by $C_2-A_2$ 
(and using  \eqref{eq:com a_2} of course) we obtain  the estimate
\begin{equation}
  \label{eq:awayfromF}
  \langle {\Op}(e) \rangle_{u}\leq
  C_1\|u\|_{-1}^2+\sigma \|u\|_{-t}^2 + C_2\sigma^{-1}\|v\|^2_t;\;\sigma>0.
\end{equation}

\subsection{Bounds at $\mathcal{F}^+$}
\label{Wave front set3}

Let $z^+(\cdot)\in \mathcal{F}^+$. We
  use the following construction, cf.  \eqref{eq:propobs2}, with $l$ defined in terms
  of $z^+$: 
\begin{equation*}
  b=X^{1/2}a;\; a=X^{-t}F(r>1)f(h)F(l<\epsilon).
\end{equation*}
Mimicking the proof of \eqref{eq:bound1} with the quantization of this
 symbol and using
 \eqref{eq:awayfromF} we derive the bound
\begin{align}
  \label{eq:bound11}
  &\|\tilde Au\|_{-t}^2 \leq
  C_1\|u\|_{-1}^2+C_2\|v\|^2_t+\sigma\|u\|_{-t}^2;\\&\tilde A={\Op}(F(r>1)f(h)F(l<\epsilon)),\;\sigma>0,\;C_j=C_j(\sigma).\nonumber
\end{align}

\section{Bounds at $\mathcal{F}^-_{\sa}$}
\label{Saddlepo}
In this and the following section we shall need  Conditions
\ref{cond:flux} and \ref{cond:noexcep}. Let $u$ and $v$ be given as in
Section \ref{Wave front set near}.
We shall prove 
\begin{align}
  \label{eq:bound4}
  &\|\tilde Au\|_{t-1}^2 \leq C_1\|u\|_{-1}^2+C_2\|v\|^2_t;\\&\tilde
  A={\Op}(F(r>1)f(h)F(l<\epsilon)),\nonumber
\end{align} 
where  $l=l_h$ is the function 
associated to a given branch $z^-(\cdot)\in \mathcal{F}^-_{\sa}$,
cf. \eqref{eq:liapunov}. 

Using the observable $a_2$ of \eqref{atrak2} we introduce the
following equivalence relation on the set $\mathcal
F^-_{\sa}(E)$: $z\sim \tilde z$ for $z, \tilde
z \in \mathcal
F^-_{\sa}(E)$ if and only if $a_2(z)= a_2(\tilde z)$. 
On the set $\mathcal
F^-_{\sa}(E)/\sim$ we have the following partial ordering:
$z\prec  \tilde z$ for $z, \tilde
z \in \mathcal
F^-_{\sa}(E)$ if and only if $a_2(z)\leq a_2(\tilde z)$. 

We have the important property that for $z \prec  \tilde z$ any
incoming orbit at the representative $z$ can not emanate from  the  representative $\tilde z$. In particular
if $z(E)\in \mathcal{F}^-_{\sa}(E)$ belongs to a   minimal class then by Condition
\ref{cond:noexcep} the two  incoming
orbits at the representative $z(E)$ will come from  $\mathcal{F}^-_{\so}(E)$,
cf Proposition \ref{lemdoi} and Condition
 \ref{cond:noexcep}. 

Suppose  first that $z(E) \in \mathcal
F^-_{\sa}(E)$   belongs to a   minimal class at a fixed energy
$E=E_0\in I$. As a preliminary step we are interested in
quantizing the symbols  \eqref{eq:propobs2}  with $f$  
supported in an interval $[E_0-\delta,E_0+\delta]$ with
$\delta<<\epsilon$. The unstable manifold at neighboring $E$'s is
denoted by $M^u_E$. We introduce $\T^*_-=\{z\in \T^*|\rho<0\}$, and define for $\epsilon>0$
\begin{equation*}
  K=\Big \{z\in \T^*_-|\;{\epsilon\over 4}\leq l_{E_0}\leq2\epsilon,\; h=E_0,\;
  \dist (z,M^u_{E_0})\geq{\epsilon\over 2}\Big  \}.
\end{equation*}
If $\epsilon$ is small enough,  then for all  $z\in K$ we have
$\phi_\tau(z)\to \tilde z \in \mathcal F^-_{\so}$ for $\tau\to -\infty$. This means that
$\phi_\tau(z)$ in the far past will belong to a set where we have a
good wave front set bound due to the result of Subsection \ref{Wave front set1}. Since $K$ is compact we may choose
$\tau_0<<-1$ such that we have a good wave front set bound of an open
neighborhood $U_{\tau_0}\subseteq \T^*$ of the
compact set $\phi_{\tau_0}(K)$. Define
$U_1=\phi_{-\tau_0}(U_{\tau_0})$. By Proposition \ref{prop:propaz}
there are good bounds on compact subsets of  $U_1$. Fix such
$\epsilon>0$ and such open set $U_1$.

Next we define 
\begin{equation*}
  U_2^\delta=\{z\in \T^*_-|\;{\epsilon\over 4}< l_{h}<2\epsilon,\; |h-E_0|<2\delta,\;
  \dist (z,M^u_{h})<{\epsilon}\};\;\delta>0,
\end{equation*} and also
\begin{equation*}
  K^\delta=\{z\in \T^*_-|\;{\epsilon\over 2}\leq l_{h}\leq \epsilon,\; |h-E_0|\leq\delta,\;
  \dist (z,M^u_{h})\geq{\epsilon}\};\;\delta\geq 0.
\end{equation*} 
 \begin{lemma}\label{lem:Stability2}
   For all $\delta>0$ small enough, $K^\delta\subseteq U_1$.
 \end{lemma}
 \begin{proof} Clearly $K^0\subseteq U_1$. Using that $\nabla
   h(z(E_0))\neq 0$ and the fact that $|h(z)-E_0|\leq\delta$ for any
   given $z\in K^\delta$ we may choose a neighboring $\tilde z$ with
   $h(\tilde z)=E_0$; $|z-\tilde z|\leq C\delta$. Asssuming 
   $\delta<<\epsilon$  we see that $\tilde z\in K$. We may assume that
   $\dist (K, \T^*\setminus U_1)>C\delta$. Whence $z\in U_1$.
 \end{proof}

Due to the lemma we may choose $\delta>0$ such that 
\begin{equation*}
  \tilde K :=\{z\in \T^*_-|\;{\epsilon\over 2}\leq l_{h}\leq\epsilon,\;
  |h-E_0|\leq \delta\}\subseteq U_1\cup  U_2^\delta.
\end{equation*}
Next pick non--negative $\psi_1,\psi_2\in C^\infty (\T^*)$ subordinate to this
covering, that is $\supp \psi_1 \subseteq U_1$, $\supp \psi_2
\subseteq U_2^\delta$ and $\psi_1+\psi_2=1$ on $\tilde K$.

For any $f$ 
supported in an interval $[E_0-\delta,E_0+\delta]$ we may decompose 
$F'(l_h<\epsilon)f(h)=(\psi_1+\psi_2)F'(l_h<\epsilon)f(h)$.
In agreement with the discussion  above the contribution
from the term involving $\psi_1$ may be  treated by \eqref{eq:bound1} and
Propositions \ref{lemdoi} and \ref{prop:propaz}  yielding   a bound of the type \eqref{eq:bound1}. The contribution from the term
involving $\psi_2$ has the right sign (more precisely,
$\psi_2\{h,l_h\}\geq 0$). 
We conclude
 \eqref{eq:bound4} 
for  $z^-$ for which $a_2(z^-(E_0))$ is minimal under the additional
condition that $f$ is supported near the fixed $E_0\in I$. 

By repeating these arguments for the other $z^-\in \mathcal{F}^-_{\sa}$ in
the order of increasing  $a_2$, measured at $E=E_0$,  we conclude
\eqref{eq:bound4} for all $z^-\in \mathcal{F}^-_{\sa}$.

Now by letting $E_0\in I$ vary, the bound for any
$f\in C_0^\infty(I)$ finally follows by  a compactness argument. 

\section{Limiting absorption
principle}
\label{LAP}
 
Let us define $u=R(\zeta)v=(H-\zeta)^{-1}v$ for $v\in L^{2,t}$, $t\in
  (1/2,1)$, and 
  $\zeta$ with 
  $\Re
  \zeta\in
[a,b] \subseteq I$ and  $\Im \zeta > 0$. 
  We aim at the  following limiting
  absorption principle bound,
 \begin{equation}
    \label{eq:lapbound1}
    \|u\|_{-t}\leq C \|v\|_{t}.
  \end{equation}

Suppose \eqref{eq:lapbound1} is not true. Then 
  \begin{equation}
    \label{eq:lapbound10}
    \|u_n\|_{-t}> n \|v_n\|_{t}
  \end{equation} for 
sequences $(u_n)$ and $(v_n)$ of functions of this type and with $\zeta_n \to E \in
I$.  
We normalize $u_n$ in $L^{2,-t}$. The resulting new sequence
(whose elements are also denoted
 by  $u_n$) has a weak limit, say $u$, in $L^{2,-t}$. Clearly
$(H-E)u=0$ in the distributional sense.

\subsection{$u$ not zero}
\label{u not zero}
We combine the bounds \eqref{eq:bound1}, 
\eqref{eq:awayfromF}, \eqref{eq:bound11} and \eqref{eq:bound4} 
to one estimate 
\begin{equation}\label{eq:usubn}
  \|u_n\|_{-t}^2 \leq
  C_1\|u_n\|_{-1}^2+C_2\|(H-\zeta_n)u_n\|^2_t.
\end{equation}
Letting $n\to \infty$  using a compactness argument yields 
\begin{equation*}
  1\leq C_1\|u\|_{-1}^2.
\end{equation*}

Clearly it follows that   
\begin{equation}
  \label{eq:unot0}
 u\not =0\;{\rm in}\; L^{2,-1}.
\end{equation}
 
\begin{remark} One may also derive \eqref{eq:usubn} using the observable
  \begin{equation*}
    a_3=C(g_\epsilon' \rho -\tilde b) -2(b\rho+V')\eta;\;g_\epsilon(r) =r-r^{1-\epsilon},\;\epsilon>0.
  \end{equation*}
This is done by showing that 
\begin{equation*}
  \{h,F(r>1)f(h)a_3\}\geq \sigma r^{-1-\epsilon}F(r>1)f(h) +O(r^{-1-2\epsilon})
\end{equation*}
for $F(r>1)f(h)$ as in \eqref{eq:propobs2} and for a small $\sigma>0$;
and similarly for a quantization, cf. Subsection \ref{Wave front set2}.

Here Condition  \ref{cond:noexcep} is not used. However we will need
Condition  \ref{cond:noexcep} for the bound \eqref{eq:wave_empty2}
stated below.
\end{remark} 

\subsection{Weak decay}
\label{Weak decay}

We shall show 
that
\begin{equation}
   \label{eq:wave_empty2}
  u\in B^*_0.
 \end{equation}

First we apply \eqref{eq:bound1}, \eqref{eq:awayfromF} and
\eqref{eq:bound4} to $u_n$ and let $n\to \infty$. The result reads
\begin{equation} 
\label{eq:bound44} 
WF^{t-1}(u)\cap \mathcal{F}^-= \emptyset, 
\end{equation}
and 
\begin{equation} \label{eq:bound44333}
WF^{-\frac {1}{2}}(u)\subseteq  \mathcal{F}. 
\end{equation}

In particular
\begin{equation*} 
WF^{-\frac {1}{2}}(u)\subseteq  \mathcal{F}^+, 
\end{equation*}
which in conjunction with Proposition \ref{prop:wave_upper}
yields \eqref{eq:wave_empty2}.

\subsection{Strong decay}
\label{Strong decay}
In this section we shall improve \eqref{eq:wave_empty2} along the line
of the proof of \cite[Theorem  30.2.9]{Ho1}.
\begin{prop} \label{prop:uinl2}The function $u$ is in $ L^{2,1}$ and obeys
  \begin{equation}
    \label{eq:ubound1}
    \|u\|_1\leq C\|u\|_{-1},
  \end{equation} where $C$ is independent of $E\in [a,b]$.
\end{prop}
\begin{proof} 
First we shall prove that 
\begin{equation}
  \label{eq:ubound5}
WF^{t-1}(u)=\emptyset.  
\end{equation}

For any  $z^+\in \mathcal{F}^+_{\si}$ we may 
  use the construction \eqref{eq:propobs2} with $l$ defined in terms
  of $z^+$ to show that indeed
  \begin{equation}
    \label{eq:ubound2}
   z^+\notin WF^{t-1}(u). 
  \end{equation}
 
In fact we may follow the scheme of Subsection \ref{Wave front set1} 
 with $\zeta$, $u$ and $v=(H-\zeta)u$ replaced by $E\in
I$,  $u_R=\psi_Ru$ and $v_R=(H-E)u_R$, respectively. Here
$\psi_R(x)=\psi(r/R)$ is given as in the proof of Proposition
\ref{prop:wave_upper}. Notice that for this choice of $b_\kappa$ the Poisson
bracket is now non-negative (this would not allow us to have a complex
$\zeta$). Using the estimate \eqref{eq:wave_empty2} allows us to take $R\to
\infty$ in the commutator of \eqref{eq:comms2}
\begin{equation*}
  \lim_{R\to\infty}\langle i[H-E, B_\kappa^2]\rangle
  _{u_R}=0.\end{equation*}
Consequently \eqref{eq:comms1} leads to  the following bound
(given in terms of the new $l$)
\begin{equation*}
  \|X_\kappa^{-1}A_\kappa u\|^2 \leq C\|u\|_{-1}^2,
\end{equation*} 
from which we in turn obtain \eqref{eq:ubound2} by letting $\kappa\to 0$.

Thus
 \begin{equation}
  \label{eq:ubound4}
WF^{t-1}(u)\cap \mathcal{F}^+_{\si}=\emptyset.  
\end{equation} 

Next we repeat the above arguments as well as those of Section
\ref{Saddlepo} for the fixed points   
$z^+\in \mathcal{F}^+_{\sa}$ now arranged in
the order of {\it decreasing}  $a_2$;
 here we use  the bound \eqref{eq:ubound4} and Proposition \ref{prop:propa}
 (for the state $u$) in  the backward direction of the flow,
 cf. Remark \ref{remark:propa2}.  
  This leads to 
\begin{equation}
  \label{eq:ubound444}
WF^{t-1}(u)\cap \mathcal{F}^+_{\sa}=\emptyset.  
\end{equation}
 
In combination with Proposition \ref{prop:propa} the estimates 
\eqref{eq:bound44}, \eqref{eq:ubound4} and
\eqref{eq:ubound444} yield \eqref{eq:ubound5}.
 
Next we notice that the above regularization procedure  in conjunction with
\eqref{eq:ubound5}  and Section \ref{Saddlepo} may work to improve
\eqref{eq:bound44} by one power,  $t-1\to t$. 
The arguments above yield an improvement of 
\eqref{eq:ubound4} and
\eqref{eq:ubound444}  by one power. Whence we obtain
\eqref{eq:ubound5} with $t-1\to t$. By bootstrapping we 
conclude  that in fact $u\in L^{2,\infty}$, in
particular we infer that  $ u\in L^{2,1}$. 

The estimate \eqref{eq:ubound1} follows by
keeping track of bounding constants,  cf. Remark \ref{remark:propa2}.
\end{proof}
\begin{remark}\label{remark:eigen}
  It follows from the above proof that in fact Proposition
  \ref{prop:uinl2} holds for all $u\in B_0^*$ such that  $(H-E)u=0$
  with  $E\in [a,b]$. 
\end{remark}
\subsection{Completing the proof of LAP}
\label{Completing the proof of LAP}
In this section we complete the proof of limiting absorption
principle.
We mimic the proof of \cite[Theorem 30.2.10]{Ho1}.

It follows from \eqref{eq:unot0} and Proposition \ref{prop:uinl2}  that $u$ is an
$L^2$--eigenfunction, and from the uniformity in $E\in [a,b]$ and a
compactness argument that the
set of $E$'s constructed as above must be finite  in $[a,b]$. Thus for $\Re
  \zeta$ 
staying away from a finite set (by a positive distance) we deduce \eqref{eq:lapbound1}  which in turn implies absence of
eigenvalues. Thus we have shown  
that $\sigma_{pp}(H)$ is discrete in $I$. (Alternatively we may invoke
Remark \ref{remark:eigen} to show this.)

If $[a,b]$ does  not contain  eigenvalues of $H$  the bound
\eqref{eq:lapbound10} lead to a contradiction and  therefore  \eqref{eq:lapbound1}  holds with a
constant independent of $\zeta$. We have shown:

\begin{thm} \label{thm:LAP}With the assumptions from Section \ref{Classical
    mechanics} (including Conditions \ref{cond:flux} and
    \ref{cond:noexcep}) $\sigma_{pp}(H)$ is discrete in
    $I$. Moreover for any subinterval $[a,b]\subset I\setminus
    \sigma_{pp}(H)$ and $t>1/2$ there is a  constant $C>0$ such that 
    \begin{equation}
      \label{eq:lapfinal}
      \|R(E+i\epsilon)\|_{\mathcal{B}(L^{2,t},L^{2,-t})}\leq C;\;E\in [a,b],\;\epsilon>0.
    \end{equation}
\end {thm}
\begin{cor} \label{cor:smoothbnd}Suppose the conditions of Theorem \ref{thm:LAP} and 
\begin{equation}
  \label{eq:emptypp}
 I\cap \sigma_{pp}(H)=\emptyset. 
\end{equation} Then for all $s>2^{-1}$ and $f\in C^{\infty}_0(I)$ there exists a
constant $C>0$ such that 
\begin{equation}
  \label{eq:smoothbnd}
 \int_{\infty}^{\infty} \|X^{-s}\e^{-itH}f(H)\psi\|^2  dt\leq
C \|\psi\|^2.  
\end{equation}
\end {cor}
\begin{remark}\label{remark: perturbations}
We introduce a class of perturbations, cf. \eqref{eq:h}, for
which small modifications of the methods of
this paper may yield 
generalizations of our results. In this sense it is  not an ``optimal'' class of
perturbations, but for convenience let us assume:

The scalar potential $V_\delta =V_\delta^1 +V^2$ where $V^2$ is real-valued,
 compactly supported and Laplacian
 bounded with norm less than $1$. The potentials $\A_\delta$ and
 $V_\delta^1$ are smooth (with values in $\R^2$ and $\R$,
 respectively) and obey the bounds 
 \begin{equation}
   \label{eq:1}
   \partial _{x}^\alpha \A_\delta=\mathcal
   O(|x|^{-\delta-|\alpha|})\;\text {and}\;\partial _{x}^\alpha V_\delta^1=\mathcal
   O(|x|^{-\delta-|\alpha|}),
 \end{equation} for some $\delta>0$.

 Upon adding such perturbations to the expression \eqref{eq:weylh},
  $\delta>0$ suffices for Sections \ref{Propagation of
  singularities}--\ref{Projections}. For 
  Sections \ref{Projection=0}--\ref{Asymptotic completeness} and
  Section \ref{Quantum mechanics in the  general setting} one would 
  need $\delta>1$.    
\end{remark} 
\section{Preliminary estimates} 
\label{Preliminary estimates}
We impose in this and in the following two sections Conditions
\ref{cond:flux} and \ref{cond:noexcep}. As in the previous sections
$I$ refers to an open interval obeying \eqref{eq:noncrit}.  We shall also assume \eqref{eq:emptypp}
in this and the following sections. 
Our bounds involve  functions $f,\tilde{f}\in C^{\infty}_0(I)$ such that $0\leq\tilde f\leq1$ and $\tilde f=1$ in
a neighborhood of $\supp f$.
Moreover $e$ refers to the symbol \eqref{eq:e} with $f$
replaced by $\tilde{f}$, and the subscript $t$ indicates ``expectation'' in the state
$$\psi(t)=\e^{-itH}f(H)\psi; \;{\rm viz.}\; \langle A\rangle _t :=\langle
\psi(t),A\psi(t)\rangle.$$

For convenience let us in the following assume that there exists
$\epsilon_0>0$ such that 
\begin{equation}
  \label{eq:eps1}
 V(\theta)\in I\Rightarrow |V'(\theta)|\geq 2\epsilon_0,
\end{equation} cf. \eqref{eq:noncrit}. (Obviously this may be achieved
possibly by a slight shrinking of $I$.)
Let $\epsilon_1>0$ obey
\begin{equation}
  \label{eq:eps2}
 2\epsilon_1\max |b|\leq\epsilon_0,\;2\epsilon_1^2\leq \dist  (\supp
 \tilde f, {\R}\setminus I).
\end{equation}

\begin{lemma}\label{lemma:1}For all functions $f,\tilde{f}$, constants
 $\epsilon_1>0$  and states $\psi(t)$ as above 
\begin{equation} 
  \label{eq:bounde}
\int_{1}^{\infty} |\inp[\big]{{\Op}(e)} _t| dt\leq
C \|\psi\|^2,  
\end{equation}
\begin{align} 
  \label{eq:boundrovert}
\int_{1}^{\infty} t^{-1}\inp[\big]{F'\big (\frac{r}{t}>&\bar C\big )}_t dt\leq
C \|\psi\|^2;\\&\bar C>2\sqrt{2(\sup (\supp \tilde f)-\min V)}, \nonumber 
\end{align}
\begin{align} 
  \label{eq:boundrho}
\int_{1}^{\infty} t^{-1}|\inp[\big]{{\Op}(&b_1) }_t| dt\leq
C \|\psi\|^2;\\&b_1=F(r>1)\tilde f (h)F\big (\frac{r}{t}<\bar C\big )F(\rho<\epsilon_1),\nonumber 
\end{align} 
\begin{align} 
  \label{eq:boundrovertmin}
\int_{1}^{\infty} t^{-1}|\inp[\big]{{\Op}(&b_2) }_t| dt\leq
C \|\psi\|^2;\\&b_2=F(r>1)\tilde f (h)F\big (\frac{r}{t}<\frac{\epsilon_1}{2}\big )F(\rho>\epsilon_1).\nonumber 
\end{align}
\end{lemma}
\begin{proof} The  bound \eqref{eq:bounde} follows from \eqref{eq:com
    a_2} and Corollary  \ref{cor:smoothbnd}.

The  maximal velocity bound \eqref{eq:boundrovert} is rather standard, see for example
\cite[Lemma 6.4]{CHS} or \cite [Lemma 8.1]{HS2}.

As for the  bound \eqref{eq:boundrho}  we first show that 
\begin{equation}
  \label{eq:boundrho2}
\int_{1}^{\infty} t^{-1}|\langle {\Op}(c_1)  \rangle _t |dt\leq
C \|\psi\|^2,  
\end{equation}
where 
\begin{equation*}
  c_1=F(r>1)\tilde f (h)F(\frac{r}{t}<\bar C)F(|\rho|<2\epsilon_1).
\end{equation*}
For that we notice that for $\epsilon'>0$ small enough 
\begin{equation}
  \label{eq:bnd}
 F(r>1)\tilde f (h)F(|\rho|<2\epsilon_1)  F((b\rho+V')^2+\eta^2 <\epsilon')=0.
\end{equation}
Here we used \eqref{eq:eps1} and \eqref{eq:eps2}. 

We infer \eqref{eq:boundrho2} from Corollary  \ref{cor:smoothbnd},
\eqref{eq:bounde}  and \eqref{eq:bnd}.

To show \eqref{eq:boundrho}  we introduce  the family of ``propagation observables'' 
\begin{equation*}
 \Phi(t)= {\Op}(c_2);\;c_2=\frac{r}{t}b_1.
\end{equation*}
We consider the expectation of the Heisenberg derivative ${\bf{D}}\Phi(t)=i[H, \Phi(t)]+\frac
{d}{dt}\Phi(t)$ in the state $\psi(t)$. We shall henceforth
denote by ${\bf d} a=\{h,a\}+\frac{d}{dt}a$ the classical
Heisenberg derivative of an observable $a$. For the first factor we
may use that 
\begin{equation}\label{eq:clas700}
  {\bf d}\big (t^{-1}r\big )=t^{-1}(\rho-\frac{r}{t}).
\end{equation}

Clearly
\begin{align}\label{clas}
&(\rho-\frac{r}{t}) F(\rho<\epsilon_1)\big (1-
F(|\rho|<2\epsilon_1)\big )\nonumber\\&\leq-\epsilon_1 F(\rho<\epsilon_1)\big (1-  F(|\rho|<2\epsilon_1)\big ).
\end{align}

 We obtain \eqref{eq:boundrho} using Corollary  \ref{cor:smoothbnd},
 \eqref{eq:boundrovert}, \eqref{eq:boundrho2}, \eqref{eq:clas700}, \eqref{clas}  and the fact that $\Phi(t)$
 is uniformly bounded.

The bound \eqref{eq:boundrovertmin} follows similarly, cf. \cite
[Lemma 8.1]{HS2}.

\end{proof}
\begin{lemma} \label{lemma:2}With functions $f,\tilde{f}$, constants
 $\epsilon_1, \bar
 C>0$, states $\psi(t)$  and symbols $b_1, b_2$ as in Lemma  \ref{lemma:1} 
\begin{align}
\label{eq:largevelo}
   \lim_{t\to\infty}\|F\big (\frac{r}{t}&>\bar C\big )\psi(t)\|=0,\\
\label{eq:rhopos}
   \lim_{t\to\infty}\langle {\Op}(&b_1) \rangle _t=0,\\
\label{eq:min2}
   \lim_{t\to\infty}\langle {\Op}(&b_2) \rangle _t=0,\\
   \label{eq:etabound}
   \lim_{t\to\infty}\langle {\Op}(&b_3) \rangle _t=0;\\
&b_3=re=F(r>1)\tilde f (h)((b\rho+V')^2+\eta^2).\nonumber
 \end{align}
\end{lemma}
\begin{proof}The maximal velocity bound \eqref{eq:largevelo} is 
  standard, see for example \cite[Proposition 6.8]{CHS}. The bound
  \eqref{eq:rhopos} follows from \eqref{eq:boundrho} and the fact that
  $\frac {d}{dt}\langle {\Op}(b_1) \rangle _t$ is integrable, cf. the proof of
  Lemma \ref{lemma:1}. We argue similarly for \eqref{eq:min2}. For
  \eqref{eq:etabound} we may proceed as follows:
  By \eqref{eq:bounde} and \eqref{eq:largevelo} it suffices to show
  that the time-derivative of  $\langle {\Op}(b_3)
  \rangle _t$ is integrable. For that we use Corollary
  \ref{cor:smoothbnd}, \eqref{eq:Bclas} and  \eqref{eq:bounde}.
\end{proof}
\section{Projections $P_j$} 
\label{Projections} 
Motivated by  Lemma \ref{lemma:2} we shall 
for all  branches $z_j=z_j(E)\in \mathcal{F}^+$, $E\in I$,  associate a projection
$P_j:{\Ran}(1_{I}(H))\to {\Ran}(1_{I}(H))$. It is characterized in terms of  the
strong limit
\begin{equation}
  \label{eq:pro}
 P_jf(H)=s-\lim _{t\to \infty}\e^{itH}{\Op}(\chi_j\tilde f) \e^{-itH}f(H).
\end{equation}
Here $f\in C^{\infty}_0(I)$ is arbitrary and the symbol
$\chi_j\tilde f$ is given in terms of any function  $\tilde f\in
C^{\infty}_0(I)$ such that  $0\leq\tilde f\leq1$ and $\tilde f=1$ in
a neighborhood of $\supp f$, and in terms of any small $\epsilon>0$, 
as
\begin{equation}
  \label{eq:chi_j}
 \chi_j\tilde f=F(r>1)\tilde f(h)F(|\theta-\theta_j(h)|<\epsilon).
\end{equation}
(We have suppressed a trivial periodization in the variable $\theta$.)
\begin{prop}\label{prop:projection}
Given \eqref{eq:emptypp}, the expression \eqref{eq:pro} is well-defined and independent of the
choices of $\tilde f$ and small $\epsilon$ in \eqref{eq:chi_j}. Taking
$f\uparrow 1_{I}$ the resulting limit $P_j=s-\lim _{f\uparrow
  1_{I}}P_jf(H)$ is indeed an orthogonal projection with
 $\Ran (P_j) \subseteq\Ran (1_{I}(H))$. Moreover, $P_j$ reduces $H$.
\end{prop}
 \begin{proof}
We  introduce an ``intermediate'' $f_1$, that is $f_1 =1$  in
a neighborhood of $\supp f$ and $\tilde f =1$ in
a neighborhood of $\supp f_1$. To show the existence of the limit
on the right hand side of \eqref{eq:pro} it suffices by commutation, Corollary
\ref{cor:smoothbnd} and Lemma \ref{lemma:2} to prove the existence  of 
$$P_{j,f}=s-\lim _{t\to \infty}f_1(H)\e^{itH}{\Op}\Big (\chi_j\tilde fF\big (\frac{r}{t}>\frac{\epsilon_1}{4}\big )F\big (\frac{r}{t}<\bar C\big )\Big )
\e^{-itH}f(H).$$
For that we differentiate and need to verify integrability. Clearly
the contribution to the Heisenberg derivative from the factor
$F(\frac{r}{t}>\frac{\epsilon_1}{4})F(\frac{r}{t}<\bar C)$ may be
treated by \eqref{eq:boundrovert}--\eqref{eq:boundrovertmin}. 

The
Poisson bracket with the factor $F(|\theta-\theta_j(h)|<\epsilon)$ is
obviously proportional to
$r^{-1}F'(|\theta-\theta_j(h)|<\epsilon)$. This term is treated as
follows: Let $\epsilon_1>0$ be given as in Lemma \ref{lemma:1}. Then
for a sufficiently small $\epsilon'>0$ (in particular $\epsilon'<<\epsilon$)
\begin{multline}
  \label{eq:decprime}
F'(|\theta-\theta_j(h)|<\epsilon)=F'(\cdot)F(\rho<\epsilon_1)\\+ F'(\cdot)F(\rho>\epsilon_1)F(\eta^2 +(b\rho+V')^2>\epsilon'). 
\end{multline}
Here we used \eqref{eq:nondege1}.

The contribution from the first term on the right hand side of
\eqref{eq:decprime} is treated using \eqref{eq:boundrho}; the second
term by \eqref{eq:bounde}.

 We conclude that  the limit on the right hand side of 
\eqref{eq:pro} exists.
 Clearly $P_{j,f}$  is  independent of the
choices of $f_1$, $\tilde f$ and $\epsilon$,
cf. \eqref{eq:decprime}. In particular there is a unique orthogonal
projection $P_j$ (given by extension by continuity using the spectral
theorem) such that 
\begin{equation*}
\begin{cases}
P_j\psi=0 &\text {for} \;\psi\in \Ran (1_{\R\setminus I}(H))\\
P_jf(H)\psi=P_{j,f}\psi &\text {for} \;f\in C^{\infty}_0(I) \;\text {and \;for\;arbitrary\;}\psi
\end{cases}\;.
\end{equation*} Notice  the formula $P_{j,g}f(H) = P_{j,f}$ which holds when $g = 1$ on the
support of $f$ for $f,g\in C_0^\infty(I)$.  This shows that
$P_jf(H)\psi = P_{j,g}f(H)\psi$ and hence in particular that  $P_j$ is well-defined.
Obviously $P_j$ reduces $H$
\end{proof}

It follows from Lemma \ref{lemma:2} and Proposition \ref{prop:projection} that
the projections span the spectral subspace for the interval $I$.
\begin{prop}\label{prop:projectioncomp} Given \eqref{eq:emptypp}, 
  \begin{equation}
    \label{eq:completejss}
  1_{I}(H)=\sum_jP_j.  
  \end{equation}
\end{prop}

We may supplement the list of ``global'' estimates in Lemma \ref{lemma:2} by the
following ones for  states associated to $P_j$. Let
\begin{equation}
  \label{eq:rhojE}
\rho_j(E)=\sqrt{2\big(E-V(\theta_j(E))\big)};\;E\in I.  
\end{equation}
\begin{lemma}\label{lemma:extrabounds}For all small $\bar \epsilon>0$, functions $f,\tilde{f}$ as
  above and   $\psi=f(H)\psi \in \Ran (P_j)$
  \begin{align}
    \label{eq:rho}
 \|{\Op}(&d_1)\psi(t)\| \to 0
 \;{\rm{as}}\;t\to \infty;\\&d_1=F(r>1)\tilde
 f(h)F(|\rho-\rho_j(h)|>\bar
 \epsilon),\nonumber\\
  \label{eq:r/t}
\|{\Op}(&d_2)\psi(t)\| \to 0
 \;{\rm{as}}\;t\to \infty;\\&d_2=F(r>1)\tilde f(h)F\big (\big |\frac{r}{t}
-\rho_j(h)\big | >\bar \epsilon\big ).\nonumber 
  \end{align}
\end{lemma}
\begin{proof} The bound \eqref{eq:rho} follows from the energy
 relation $\rho^2+\eta^2=2(h-V)$ and Lemma \ref{lemma:2}. 

As for the bound \eqref{eq:r/t} we introduce the observable
$\Phi(t)={\Op}(c_1)$ with 
\begin{multline*}
  c_1=F(r>1)\tilde
 f(h)F(|\theta-\theta_j(h)|<\epsilon)\times\\F\big (\frac{r}{t}>\frac{\epsilon_1}{4}\big )F\big (\frac{r}{t}<\bar C\big )F(|\rho-\rho_j(h)|<\epsilon')F((b\rho+V')^2+\eta^2<\epsilon'')\times\\\big |\frac{r}{t}
-\rho_j(h)\big |F\big (\big |\frac{r}{t}
-\rho_j(h)\big |>\bar \epsilon \big ).
\end{multline*} 
We choose $0<\epsilon,\epsilon''<<\epsilon'$. Then the
contribution to the  expectation of the Heisenberg derivative of
$\Phi(t)$ in the state $\psi(t)$ from  differentiating
the factor $F(|\rho-\rho_j(h)|<\epsilon')$ vanishes (by the energy 
relation). As for the contribution from the last factor we write, cf. \eqref{eq:clas700},
\begin{equation}\label{eq:boldd}
  {\bf d}\big (t^{-1}r\big )=t^{-1}\Big((\rho-\rho_j(h))-\Big (\frac{r}{t}-\rho_j(h)\Big )\Big).
\end{equation} 

The first term on the right hand side of
\eqref{eq:boldd} is bounded by $\epsilon'$. Hence (given
that $2\epsilon'$ is smaller that $\bar \epsilon$) the second term
``dominates''. This fact and previously used estimates yields an
integral estimate for a symbol $c_2$ given by omitting the outer factor $|\frac{r}{t}
-\rho_j(h)|$ in the expression $c_1$. By the standard method, see the
proof of \eqref{eq:rhopos}, we readily obtain from this bound that
indeed \eqref{eq:r/t} holds. 

\end{proof}
\section{$P_j=0$ at $\mathcal{F}^+_{\sa}$} \label{Projection=0} 
In this section we shall invoke \cite{HS2} to show that there are no
states associated to  any branch $z_j=z_j(\cdot)\in \mathcal{F}^+_{\sa}$. 

As in \cite{HS1} and \cite{HS2} we  linearize the equations of
motion nearby any $z_j(\cdot)\in \mathcal{F}^+$ as follows: Write for $E\in I$
\begin{equation*}
\omega_1(E)=(\cos \theta_j(E),\sin \theta_j(E)),\;\omega_2(E)=(-\sin \theta_j(E),\cos \theta_j(E)),  
\end{equation*}
and introduce new coordinates
\begin{equation*}
  x=x_1(\omega_1(E)+u\omega_2(E)),\;\xi=\xi(E)+\mu\omega_1(E)+v\omega_2(E),
\end{equation*}
where (with $\rho(E)=\rho_j(E)$ given by \eqref{eq:rhojE})
\begin{equation*}
 \xi(E)=\rho(E)\omega_1(E)+b(\theta_j(E)) \omega_2(E).
\end{equation*}

Notice that 
\begin{equation}\label{eq:coord}
  x_1=x\cdot \omega_1(E)\;{\rm and}\; u=\frac {x_2}{x_1};\;x_2=x\cdot \omega_2(E).
\end{equation}

 We may express $\mu$ as a function of $u,v$ and $E$ (for small
 $u$ and $v$) using the energy relation. 
Introducing the ``time'' $\bar\tau=\ln x_1(t)$ we obtain an
 autonomous system whose linearization at the fixed point $(u,v)=0$ takes the following form: For
 $w=(u,v)^{\tr}$
\begin{displaymath} 
 \frac{d}{d\bar\tau}w=Bw;\;  B= \rho^{-1}\left ( \begin{array}{cc}
-b'-\rho&1\\
-d&b'
\end{array}\right );\;d=b^2+b'^2+2b'\rho+V''.
\end{displaymath}
Here $B$ is evaluated at $(E,\theta_j(E))$. We used \cite[(1.7)]{HS2} and the leading
asymptotics
\begin{equation*}
  %\label{eq:leadingmu}
 \mu=- \rho^{-1}\left (\frac{1}{2}du^2-b'vu+\frac{1}{2}v^2\right
 )+\mathcal O(|w|^3).
\end{equation*}

The eigenvalues of $B$ are given by dividing those of \eqref{eq:eigenv1} by
$\rho$, that is 
\begin{equation}
  \label{eq:betas}
\beta=-
\frac{1}{2}+\frac{1}{2}\sqrt{1-4\frac{\kappa^+}{\rho^2}},\;\tilde \beta =
- \frac{1}{2}-\frac{1}{2}\sqrt{1-4\frac{\kappa^+}{\rho^2}}.
\end{equation}

To apply \cite{HS2} we need to verify various conditions. As for the
non-resonance condition \cite[(H8)]{HS2} we consider 
\begin{cond}[Few  resonances condition]
\label{cond:reson} At  the branch $z_j(\cdot)\in \mathcal{F}^+$ there
is a discrete set $\mathcal D \subset I$ such that  
\begin{equation}
  \label{eq:resnot0}
\frac{d}{dE}\left (\frac{\kappa^+}{\rho^2}\right )\neq
0\;\text{for}\;E\in I\setminus \mathcal D.  
\end{equation}
\end {cond}

With Condition \ref{cond:reson}  imposed for a branch 
$z_j(\cdot)\in \mathcal{F}^+$ it is easy to see that the resonances
of fixed order can  accumulate at most at  a discrete set of energies in $I$. We
recall that  $E\in I$ is resonance of order $m\geq2$ if we may write
\begin{equation}
  \label{eq:res2}
  \beta^{\natural}=m_1\beta+m_2\tilde\beta,
\end{equation} where $\beta^{\natural}$ signifies either $\beta$ or
$\tilde\beta$ and the coefficients to the right are non-negative integers with $m_1+m_2=m$.

For the case of $z_j(\cdot)\in \mathcal{F}^+_{\sa}$ that we address in this
section
it is enough to avoid \eqref{eq:res2} with only $\beta^{\natural}=\tilde \beta$ to
the left and for all orders $m\leq m_0$, where $ m_0$ is sufficiently large,
see \cite[(H8)]{HS2}. Under Condition \ref{cond:reson} the ``bad''
subset of  $I$ (where  \eqref{eq:res2} holds for some $m\leq m_0$) can be
``reached'' using  the spectral theorem. (Here we use the fact that its
set of accumulation points is at most discrete.)

For completeness of presentation we notice  that in the case of
$z_j(\cdot)\in \mathcal{F}^+_{\sa}$ the
union of resonances of all orders is dense.

Since the other conditions
\cite[(H1)--(H7)]{HS2} are readily verified using Lemmas
\ref{lemma:1}, \ref{lemma:2} and \ref{lemma:extrabounds} we conclude
from   \cite[Theorem 1.1]{HS2}:
\begin{thm}\label{thm:absence} Suppose Condition \ref{cond:reson} for
  a  branch 
  $z_j(\cdot)\in \mathcal{F}^+_{\sa}$. Then
  \begin{equation}
    \label{eq:00}
    P_j1_{I}(H)=0.
  \end{equation}
\end{thm}

We notice that Condition \ref{cond:reson} is trivially fulfilled in
the case $V=0$. 
\section{Wave operators at $\mathcal{F}^+_{\si}$} 
\label{Wave operators} 
In this section we shall consider two types of wave  operators at any
branch $z_j(\cdot)\in \mathcal{F}^+_{\si}$. We
construct one for high energies, cf. \cite[Theorem 1.2]{HS1}, and with a
further assumption one for small energies,  
cf. \cite[Theorem 11]{HS1}. With this 
extra  assumption there
will be an interval of overlap, so that for this energy regime both
wave operators exist. 
The constructions do not need Conditions \ref{cond:flux},
\ref{cond:noexcep} nor \eqref{eq:emptypp}.

\subsection{Wave operator at high energies}
\label{Wave operator at high energies}
We assume that for all $E\in I$
\begin{equation}
  \label{eq:conhigh}
 0<4\kappa^+<\rho^2\;{\rm at}\;z_j(E); 
\end{equation}
or  alternatively that both eigenvalues in \eqref{eq:betas} are
negative reals thoughout $E\in I$.

Let us look at the initial value problem
\begin{equation}\label{periodikkerik}
\frac{d\eta}{d\theta}=-\frac{1}{\eta}\left\{(\eta +
  b)\sqrt{2(E-V)-\eta^2}+V'\right \},\quad \eta(\theta_j)=0, 
\end{equation}
nearby $\theta_j$ where $z_j(\cdot)=(\theta_j(\cdot),0,\rho_j(\cdot))\in \mathcal{F}^+_{\si}$ Formally the differential equation arises
from eliminating $\rho$ and $\theta$ in the system
\eqref{eq:equamotion}. 
Obviously it is a {\it singular } ordinary
differential equation. We can solve it away from resonances as
follows. 

Let us denote by $\lambda$ the biggest of the two eigenvalues in
\eqref{eq:eigenv1} and $\tilde\lambda$ the smallest one (cf. the
notation \eqref{eq:betas}). Away from resonances the Sternberg
diffeomorphism provides  integral curves for the system
\eqref{eq:equamotion}
(with $\rho$ eliminated) of the form 
\begin{equation}
  \label{eq:sternb}
 (\theta(\tau)-\theta_j,\eta(\tau))=\Psi(+(-)\e^{\lambda^\natural \tau} (1,\lambda^\natural));\;\tau\;{\rm
 large}. 
\end{equation}

We use \eqref{eq:sternb} with $\lambda^\natural =\lambda$ and look at
the parametrized curve 
\begin{equation*}
 (\theta-\theta_j,\eta)=\Psi(s(1,\lambda));\;|s|\;{\rm
 small}.
\end{equation*} 
Using $\theta$ as a new parameter we obtain a
solution $\eta=\eta_E(\theta)$ to \eqref{periodikkerik} with 
\begin{equation}
  \label{eq:asyptoticseta}
\eta=\lambda( \theta-\theta_j)+{\mathcal O}(|\theta-\theta_j|^2). 
\end{equation}

We notice that this $\eta$ is smooth as a function of
$(E,\theta)$ (away from resonances), see \cite[Appendix]{HS1}. Also we remark that although
it is not unique as a solution to the initial value problem
\eqref{periodikkerik} and subject to \eqref{eq:asyptoticseta} (there
is a one-parameter family of $C^1$--solutions), all derivatives
$\eta^{(k)}_E(\theta_j)$ are uniquely determined by these
requirements; 
 for this assertion $E$ needs to be non-resonance.
Following \cite[Subsection 3.5]{CHS} we introduce
\begin{equation}
  \label{eq:rhodef}
  S_E=r\rho_E;\;\rho_E=\rho_E(\theta)=\sqrt{2(E-V)-\eta_E^2}.
\end{equation}

This function, $S_E=S_E(r,\theta)=S_E(x)$, solves the eikonal equation 
\begin{equation*}
  h(x,{\nabla}_{x}S_E)=0.
\end{equation*}

Using the formulas
\begin{equation}\label{Konvekspro}
  \partial_E\rho_E=\rho_\e^{-1},\;\partial^2_E\rho_E=-(1+(\rho_E\partial_E\eta_E)^2)\rho_\e^{-3};\;{\rm
  at}\;\theta=\theta_j,
\end{equation}
we may use a Legendre transformation to
obtain a smooth solution $S=S(t,r,\theta)=S(t,x)$ for $t>0$ to  the Hamilton--Jacobi equation
\begin{equation*}
  h(x,{\nabla}_{x}S)=-\partial_tS,
\end{equation*}
cf. \cite{CHS}.  Assuming that $I=(E^-,E^+)$ is small centered around,
say $E^0$, and does not
contain resonances, cf. Condition \ref{cond:reson}, $S$ is defined
 in a region $\mathcal
D_\epsilon=\cup_{t>0}\{t\}\times{\mathcal D}_{\epsilon,t}$ where $x\in {\mathcal D}_{\epsilon,t}$ (with $\epsilon>0$  taken
small) if and only if 
\begin{equation*}
   |\theta-\theta_j(E^0)|<\epsilon,\;(\partial_E\rho)^{-1}(E=E^-,\theta)<\frac{r}{t}<(\partial_E\rho)^{-1}(E=E^+,\theta).
 \end{equation*} 
It is given explicitly as $S=r\rho_E -tE$ where $E\in I$ is
determined by 
\begin{equation}
  \label{eq:legendr}
\frac{r}{t}=(\partial_E\rho_{E})^{-1}(\theta),  
\end{equation}
and 
possesses  the homogeneity property
\begin{equation*}
  S(t,r,\theta)=tS(1,\frac{r}{t},\theta).
\end{equation*}

Following \cite{CHS} we define the ``direct'' flow
\begin{equation}\label{domnyca}
\begin{cases}
\frac{d\tilde{\theta}}{dt}(t)=\tilde{r}^{-1}[\tilde{r}^{-1}\partial_{\theta}
S(t,\tilde{r},\tilde{\theta})-b(\tilde{\theta})]
 \\ \frac{d\tilde{r}}{dt}(t)=\partial_r S(t,\tilde{r},\tilde{\theta})
\end{cases}\;.
\end{equation}

The domain $\mathcal
D_\epsilon$ is preserved under the forward direct flow, and there is  energy conservation
\begin{equation*}
  -\partial_tS(t,\tilde{r}(t),\tilde{\theta}(t))={\rm const},
\end{equation*}
cf. \cite[Lemma 3.11]{CHS}.
 
Using the ``time'' $\tau$ of \eqref{eq:time} we may rewrite \eqref{domnyca} as the following system of equations
\begin{equation}\label{directflow}
\begin{cases}
\frac{d\tilde{\theta}}{d\tau}(\tau)=\eta_E(\tilde{\theta})
 \\ \frac{d\tilde{r}}{d\tau}(\tau)=\tilde{r}\rho_E(\tilde{\theta})
\end{cases}\;.
\end{equation}

Let us find the asymptotics of the quantity
$(\tilde\theta-\theta_j)\tilde r^{-\beta}$ (with $\beta$  given as in
\eqref{eq:betas}) for the solution to \eqref{directflow}
starting at time $\tau=0$ at $(\tilde\theta_0,\tilde r_0)$. By integrating the first
equation in \eqref{directflow} we obtain
\begin{equation}
  \label{eq:asytheta}
 \tilde\theta(\tau)-\theta_j=(\tilde\theta_0-\theta_j) \e^{\lambda\tau}(1+ o(1)).
\end{equation}
Clearly the solution to the second equation is 
\begin{equation}
  \label{eq:solr}
 \tilde r(\tau)= \tilde r_0\e^{\int_0^\tau \rho_E(\tilde\theta(\tau'))d\tau'}.
\end{equation} We write
\begin{equation}
  \label{eq:expanho}
 \rho_E(\theta) =\rho_E(\theta_j)(1+\zeta_E(\theta) \eta_E (\theta)),  
\end{equation}
where we have introduced the smooth function
\begin{equation*}
\zeta_E(\theta) =\frac {\sqrt{1+\frac{2V(\theta_j)-  2V(\theta)-\eta_E (\theta)^2}{\rho_E(\theta_j)^2}}-1}{\eta_E (\theta)}.
\end{equation*}
Using \eqref{eq:expanho} and the first equation of \eqref{directflow}
we may write \eqref{eq:solr} as 
\begin{equation}
  \label{eq:solr2}
 \tilde r(\tau)= \tilde r_0\e^{\rho_E(\theta_j)\left 
 (\tau+\int_{\tilde\theta_0}^{\tilde\theta}
 \zeta_E(\theta)d\theta\right )}=\tilde r_0\e^{\rho_E(\theta_j)\left 
 (\tau+\int_{\tilde\theta_0}^{\theta_j}
 \zeta_E(\theta)d\theta+o(1)\right )}.
\end{equation} 
 Clearly we obtain from \eqref{eq:asytheta} and \eqref{eq:solr2}  the asymptotics 
 \begin{equation}
   \label{eq:asyquant}
   (\tilde\theta-\theta_j)\tilde r^{-\beta}\to
 (\tilde\theta_0-\theta_j)\tilde r_0^{-\beta}\e^{-\lambda \int_{\tilde\theta_0}^{\theta_j}
 \zeta_E(\theta)d\theta}\;{\rm for}\;\tau\to \infty.
 \end{equation}

Motivated by these considerations  we introduce a smooth observable $w=w(t,r,\theta)$ on $\mathcal
D_\epsilon$ as follows:

\noindent{\bf Step I}. Define for given $(t,r,\theta)\in\mathcal
D_\epsilon$ 
\begin{equation}
  \label{eq:energycons3}
  E=-\partial_tS(t,{r},{\theta})\;(\in I).
\end{equation}
\noindent{\bf Step II}. Introduce 
\begin{equation}
  \label{eq:param}
 \theta_j=\theta_j(E),\;\beta=\beta(E),\; \lambda=\lambda(E)=\rho_E(\theta_j)\beta.
\end{equation}
{\bf Step III}. Put
\begin{equation}
  \label{eq:wdefined}
 w= (\theta-\theta_j) r^{-\beta}\e^{-\lambda \int_{\theta}^{\theta_j}
 \zeta_E(\theta')d\theta'}.
\end{equation}
>From the very construction we see that 
\begin{equation}
  \label{eq:hom}
  w=w(t,r,\theta)=t^{-\beta}w(1,\frac{r}{t},\theta),
\end{equation}
and that $w$ specifies asymptotics of the solution to \eqref{domnyca}
that at  time $t$ passes through $(r,\theta)$.

We can now define a comparison dynamics $$U(t)=U_j(t):
L^2(I\times{\R})\to L^2({\R}_{x}^2)=L^2({\R}_+\times{\T};\;rdrd\theta)$$
as follows:
\begin{align}\label{valimold-8}
[U(t)\phi](r,\theta)=&\e^{iS(t,r,\theta)} r^{-1/2}J_t^{1/2}(r,\theta)1_{{\mathcal D}_{\epsilon,t}}(r,\theta)\\&\phi
\left (E(t, r,\theta),w(t,r,\theta)\right ),\nonumber
\end{align}
where $E(t, r,\theta)=-\partial_tS(t, r,\theta)$, and $J_t=t^{-\beta-1}J(\frac{r}{t},\theta)$ is the Jacobian determinant arising 
from the  change of variables which makes $U(t)$ asymptotically 
isometric. Notice that we may use  \eqref{Konvekspro} and
\eqref{eq:wdefined} to show that indeed the map ${\mathcal
  D}_{\epsilon,1}\ni (\frac{r}{t},\theta)\to (E,w)(1,\frac{r}{t},\theta)$
is a diffeomorphism onto its range. Hence by also using \eqref{eq:hom}
we have $\lim _{t\to \infty}\|U(t)\phi\| =\|\phi\|$ for all $\phi\in L^2(I\times{\R})$.

Formally the ``generator'' of  $U(t)$ is 
\begin{equation}\label{eq: generator}
H-2^{-1}\gamma^2;\;\gamma=p-\nabla_{x}S.  
\end{equation}

If $\phi\in C_0^\infty(I\times{\R})$ the right hand side of
\eqref{valimold-8} is smooth for $t$ large enough, and we may use the
Cook  method as in \cite{HS1} to obtain the following wave
operator. 
%For convenience we identify below 
% $L^2(\R^2)$ with $L^2(\R_+\times \mathbb{T})$ 
%through the 
%unitary transformation 
%\begin{equation}\label{copil1}
%L^2(\R^2)\ni f(x)\rightarrow r^{1/2}f(r,\theta)
%\in L^2(\R_+\times \mathbb{T}),
%\end{equation} and consider the Hamiltonian $H$ of \eqref{eq:weylh} as
%an operator on $L^2(\R_+\times \mathbb{T})$.
Let $\mathcal{H}_{I} = 1_{I}(H)L^2({\R}_{x}^2)$, and let $M(E)$
denote multiplication by $E$ on $ L^2(I\times\mathbb{R})$.
\begin{thm}\label{thm:-1}
Suppose there are no resonances (of any order) in $I$. Suppose
\eqref{eq:conhigh}. Then there exists
\begin{eqnarray}\label{valimoldo3}
\Omega_j = s-\lim_{t\rightarrow\infty}
\e^{itH}U_j(t):L^2(I\times\mathbb{R})\mapsto L^2({\R}_{x}^2),
\end{eqnarray}
with  $\Ran (\Omega_j) \subseteq \mathcal{H}_{I}.$ Moreover $H\Omega_j=\Omega_j M(E)$.
\end{thm}
\begin{remark} \label{canonical}The smallness assumption on $I$ and the choice of
   $E\in
  I$ for defining the domain of $S$ above and hence the comparison
  dynamics $U_j(t)$ are not important. In fact our
  construction is asymptotically 
 canonical. This means that if we cover a given ``large'' $I$ free
  from resonances by
  ``smaller'' pieces say $I=\cup I_{k}$ and choose arbitrarily
  energies  $E^0_{k}\in I_{k}$, then the corresponding dynamics
  $U_{j,k}(t)$ 
glue in the following sense: Decompose  any given $\phi\in
  L^2(I\times{\R})$ as $\phi=\sum \phi_k$ where $\phi_k\in
  L^2(I_{k}\times{\R})$. Then asymptotically as $t\to \infty$ the
  state  $U_{j}(t)\phi:=\sum U_{j,k}(t)\phi_k$ is independent of the
  covering and cutting procedures. In particular the definition  $\Omega_j\phi=\lim_{t\rightarrow\infty}
\e^{itH}U_{j}(t)\phi=\sum \lim_{t\rightarrow\infty}
\e^{itH}U_{j,k}(t)\phi_k=\sum \Omega_{j,k}\phi_k$ is canonical. 

This procedure may be pushed further. Suppose a given open set
$\mathcal C \subseteq \R$ has a countable covering $\mathcal C =\cup
I_{k}$ where each interval meets the requirements of Theorem
\ref{thm:-1}. Then we decompose  any $\phi\in
  L^2(\mathcal C \times{\R})$ as above  and define $\Omega_j\phi=\sum \Omega_{j,k}\phi_k$.

An alternative procedure of defining a ``global'' wave operator would
be the one of \cite{HS1}. This amounts to introducing  the wave operator in
terms
of an $S$ obtained by gluing the various local
pieces considered above together, cf.  \cite[Appendix]{HS1}; the
domain of this $S$ is complicated. 
\end{remark}

\subsection{Wave operator at small energies}
\label{Wave operator at small energies}
We need the condition 
\begin{equation}
  \label{eq:condb=0}
  b(\theta_j)=V'(\theta_j)=0.
\end{equation}
Notice that this means that $\theta_j$ does not depend on $E\in I$.
Clearly the case $b= 0$
 covered by \cite[Theorem 1.1]{HS1} is a particular example. The 
 case $V= 0$ is an example that has not been treated before.

With \eqref{eq:condb=0} the coordinates $x_1, x_2$ of \eqref{eq:coord}
are independent of $E$ and similarly for the dual coordinates $\xi_1,
\xi_2$ given by writing $\xi=\xi_1\omega_1+\xi_2\omega_2$. We
substitute in the expression for the symbol $h$, Taylor expand up to
second order in $x_2/x_1$ and $\xi_2$, and replace $x_1$ by
$t\xi_1$. The result is the expression
\begin{align}
  \label{eq:compa1}
 &h(t)=2^{-1}\xi_1^2+ h_2(t)+V(\theta_j);\\
&h_2(t)=h_{2,t,\xi_1}\big(\frac {x_1}{t}, \xi_2\big)=2^{-1}\xi_2^2+
\alpha_1\frac {x_2}{t}\xi_2+\alpha_2\Big(\frac
{x_2}{t}\Big)^2;\nonumber\\
&\alpha_1=-\frac {b'(\theta_j)}{\xi_1},\nonumber\\
&2\alpha_2=\frac {b'(\theta_j)}{\xi_1}\Big(2+\frac
{b'(\theta_j)}{\xi_1}\Big) +\frac {V''(\theta_j)}{\xi_1^2}.\nonumber
\end{align}

We denote for fixed $\xi_1>0$ the Weyl-quantization  of $h_{t,2,\xi_1}$
by $H_2(t)$, and by $U_2(t)$ the corresponding dynamics
\begin{equation*}
 i\partial_tU_2(t)= H_2(t)U_2(t);\;U_2(1)=I.
\end{equation*}
It way be written explicitly in terms of 
\begin{equation*}
  H_{2,\xi_1}:=2^{-1}p_2^2+\alpha_1\frac {x_2p_2+p_2x_2}{2}+\Big(\frac
{\alpha_1}{2}-\frac
{1}{8}+\alpha_2\Big)x_2^2,
\end{equation*}
as $$
U_2(t) = U_{2,\xi_1}(t) =S_{t^{-\frac{1}{2}}} \e^{\frac{ix_2^2}{4}} \e^{-i 
(\ln t) H_{2,\xi_1}} \e^{-\frac{ix_2^2}{4}},
$$
where 
$S_{t^{-\frac{1}{2}}}$ is the  scale transformation
$$
S_{t^{-\frac{1}{2}}} g(x_1,x_2) = t^{- \frac{1}{4}}g\left(x_1, \frac{x_2}{\sqrt{t}}\right).
$$

The quantum dynamics corresponding to the expression \eqref{eq:compa1}
is given on the space $L^2(\mathbb{R}^2_{x})$ as 
\begin{equation}
  \label{eq:quandyn}
 \bar U(t)= \e^{-i(t-1)V(\theta_j)}\e^{-\frac{itp^2_1}{2}} U_{2,p_1}(t)\e^{\frac{ip^2_1}{2}}. 
\end{equation}

To get a wave operator with this comparison dynamics we need a small
energy assumption. This may be  given in terms of the $\beta$ of
\eqref{eq:betas}; recall that the quantity $\rho^{-2}\kappa^+$
depends on $E$. We shall need the condition
\begin{equation}
  \label{eq:condsmall}
  \Re \beta <-\frac {1}{3}\;\text{for}\;E\in I.
\end{equation} 
Clearly the inequality \eqref{eq:condsmall} is fulfilled for $4\kappa^+>\rho^2$ in
which case $\beta $ and $\tilde\beta$ are complex with real part equal
to $-2^{-1}$. 

In Subsection \ref{Completeness  at low energies} we shall need the condition that 
\begin{equation}
  \label{eq:discD}
  \beta \neq \tilde\beta\;\text{for}\;E\in I\setminus  \mathcal D, \;\text{where}\;\mathcal D
\subset I\;\text{is\;discrete},
\end{equation}
 cf. Condition \ref{cond:reson}. For convenience (although not needed)
 we shall also impose
 \eqref{eq:discD} in the theorem stated below.

Let $J=\left\{\xi_1: \xi_1 > 0, \frac{\xi^2_1}{2} \in I\right\}$, and let 
  $\mathcal{H}_{I}$ be
  given as in Theorem \ref{thm:-1}.
\begin{thm}\label{thm:-2} Suppose \eqref{eq:condsmall} and \eqref{eq:discD}. Then there
  exists
  \begin{equation}
    \label{eq:lowwaveoperator}
 \bar\Omega_j = s-\lim_{t \to  \infty} \e^{itH} \bar
 U(t):1_{J}(p_1)L^2(\mathbb{R}^2_{x}) \to  L^2({\R}_{x}^2),  
  \end{equation}
with $\Ran (\bar\Omega_j) \subseteq \mathcal{H}_{I}.$ Moreover $H\bar\Omega_j=\bar\Omega_j 2^{-1}p_1^2$.
\end{thm}

One may prove Theorem \ref{thm:-2} along the lines of the proof for
the potential case of \cite {HS1}. For reference in
Subsection \ref{Completeness  at low energies} let us notice the following
ingredient of the proof: Let 
\begin{equation}
  \label{eq:gam1}
 \gamma=p_2-(\beta+\alpha_1)\frac {x_2}{t}\;\text{and}\; \tilde\gamma=p_2-(\tilde\beta+\alpha_1)\frac {x_2}{t},
\end{equation}
where $\beta$, $\tilde\beta$ and $\alpha_1$ are evaluated at
$2^{-1}p_1^2$ and $p_1$, respectively. Then 
\begin{equation}
  \label{eq:gam2}
  \gamma(t):=\bar
 U(t)^*\gamma\bar
 U(t)=t^{-\beta} \gamma\;\text{and}\; \tilde\gamma(t):=\bar
 U(t)^*\tilde\gamma\bar
 U(t)=t^{-\tilde\beta} \tilde\gamma.
\end{equation}  Due to
 \eqref{eq:discD} the equations \eqref{eq:gam2} can trivially be solved
 for the quantities $p_2(t)$ and $t^{-1}x_2(t)$ for energies
 $2^{-1}p^2_1 \notin \mathcal D$. 
\begin{remarks} In Theorem \ref{thm:-2} the interval $I$ is not
 needed to be ``small'' (except for the requirement \eqref{eq:condsmall}). With \eqref{eq:condsmall} the condition of no
 resonances of Theorem \ref{thm:-1} is fulfilled. Under the conditions
 \eqref{eq:conhigh}, \eqref{eq:condb=0},  \eqref{eq:condsmall} and
 \eqref{eq:discD} one
 can compute explicitly 
 \begin{equation*}
 s-\lim_{t \to  \infty} U(t) ^*\bar
 U(t): 1_{J}(p_1)L^2(\mathbb{R}^2_{x}) \to L^2(I\times\mathbb{R}),
 \end{equation*} linking the  wave operators of Theorems
 \ref{thm:-1} and \ref{thm:-2}.
\end{remarks} 
\section{Asymptotic completeness} 
\label{Asymptotic completeness}
In this section we discuss asymptotic completeness of the wave
operators introduced in Section \ref{Wave operators}. Now we impose
Conditions \ref{cond:flux} and 
\ref{cond:noexcep}, and \eqref{eq:emptypp}.

\subsection{Completeness at high energies}
\label{Completeness  at high energies}
With the above assumption we have the following result.
\begin{thm}\label{thm:-11}
Under the conditions of Theorem \ref{thm:-1}
\begin{eqnarray}\label{valimoldo2}
\Ran (\Omega_j) =P_j\mathcal{H}_{I}.
\end{eqnarray}
\end{thm}
We shall outline a proof of this theorem using weak propagation
  estimates similar in spirit to \cite{CHS} and \cite{HS2}. (The proof in the
  potential case given in \cite{HS1} uses strong propagation
  estimates.) 

Clearly  $\Ran (\Omega_j) \subseteq P_j\mathcal{H}_{I}.$ 
We split the proof of the opposite inclusion into four steps:

I) Preliminary preparation of a state $\psi(t)=\e^{-itH}f(H)\psi;\;f\in
C_0^\infty(I)$ and  $\psi \in P_j\mathcal{H}_{I}$.

II) Strong localization for  the observable $\gamma^2$.

III) Propagation estimates for $U_{j}(t)\phi$.

IV) Verification of  the Cauchy condition.

\noindent{\bf Step I}. We may
summarize the estimates of  Lemmas \ref{lemma:2} and
\ref{lemma:extrabounds} as follows: Let $L_1(t)$ be the operator with
symbol $l_1(t)$ given in terms of small $\epsilon_1,\dots,\epsilon_4>0$ as 
\begin{align*}l_1(t)&=\tilde f (h)F_1\cdots F_4\\
  &=\tilde f (h)F\Big (\Big |\frac{r}{t}
-\rho_j(h)\Big |<\epsilon_1\Big )F(|\rho-\rho_j(h)|<\epsilon_2)\\&F(|\theta-\theta_j(h)|<\epsilon_3)F((b\rho+V')^2+\eta^2 <\epsilon_4).
\end{align*} Then 
\begin{equation}
  \label{eq:local1}
  \psi(t)\approx L_1(t)\psi(t),
\end{equation} meaning that $\|\psi(t)- L_1(t)\psi(t)\|\to 0$
for $t\to \infty$.

It will be important to control the Heisenberg derivative of
$L_1$. For that purpose we need to replace the last factor $F_4$ by an
equivalent localization described as follows. Let  ${y}=(
b\rho+V',\eta)^{\tr}$. Instead of taking $F_4=F_4(y^2)$ as above we choose
$F_4=F_4(|T_1^{-1}y|^2)$, where $T_1$ is the matrix 
 \begin{displaymath} 
T_1 =\left ( \begin{array}{cc}
\kappa^+&\kappa^+\\
\lambda_1&\lambda_2
\end{array}\right )
\end{displaymath} given in terms of $\kappa^+=\kappa^+(h)$ and the two
eigenvalues $\lambda_1$ and $\lambda_2$ in \eqref{eq:eigenv1}
(arbitrarily  labelled and with
$E=h$). 
This matrix diagonalizes
\begin{displaymath} 
A_1 :=\left ( \begin{array}{cc}
0&\kappa^+\\
-1&-\rho
\end{array}\right ),
\end{displaymath} which arises in the equation 
\begin{equation*}
\frac{d}{d\tau}y=\left (A_1 +o(1)\right ){y}
\end{equation*} valid for any orbit approaching ${z}_j(h)$ (in
particular having 
energy $E=h$). We
used \eqref{eq:Bclas}. See \eqref{eq:liapunov} for a similar construction.

Since the eigenvalues of this $A_1$ coincides with those of
\eqref{eq:eigenv1}, we see that classical   
\begin{equation}\label{eq:F_4decre}
\frac{d}{d\tau}|T_1^{-1}y|^2\leq 2\left (\rho\beta +o(1)\right )|T_1^{-1}y|^2;
\end{equation} in particular $|T_1^{-1}y|^2$ is decreasing.

Clearly there is an analogue of \eqref{eq:F_4decre} for the 
Poisson bracket. We have the estimate
\begin{equation}
  \label{eq:poisF_4}
  \big \{h,F_4\big (|T_1^{-1}y|^2<\epsilon_4\big ) \big \}\geq -\epsilon
  F_4'(\cdot)\rho/r\geq 0
\end{equation}
for a small $\epsilon>0$.

Next, we fix the order of scale
\begin{equation}
  \label{eq:epsilons}
  \epsilon_4<<\epsilon_3<< \epsilon_2<<\epsilon_1,
\end{equation}
for the ``new''  $L_1(t)$ with
symbol
\begin{align}\label{eq:l_111}l_1(t)&=\tilde f (h)F_1\cdots F_4\\
  &=\tilde f (h)F\Big (\Big |\frac{r}{t}
-\rho_j(h)\Big |<\epsilon_1\Big
)F(|\rho-\rho_j(h)|<\epsilon_2)\nonumber \\&F(|\theta-\theta_j(h)|<\epsilon_3)F\big
(|T_1^{-1}y|^2<\epsilon_4\big).\nonumber 
\end{align} 

To see how \eqref{eq:epsilons} comes in we look at the contribution
to  the classical Heisenberg derivative from the factor $ F_1$. From the proof of \eqref{eq:r/t} we learn
that 
\begin{align}
  \label{eq:poisF_1}
  &\tilde f (h)\left ({\bf d} F_1\Big(\Big |\frac{r}{t}
-\rho_j(h)\Big |<\epsilon_1\Big)\right ) F_2\cdots F_4\nonumber\\&\geq -\epsilon t^{-1}\tilde f (h)F_1'(\cdot)F_2\cdots F_4\geq 0
\end{align} for $\epsilon_2 <\epsilon_1/2$  and for $\epsilon>0$
sufficently small.

We also learn from the proof of \eqref{eq:r/t} that the contribution from the Poisson
bracket from the factor $ F_2$ is negligible
\begin{align}
  \label{eq:poisF_2}
  \tilde f (h)F_1\left ({\bf d}
  F_2(|\rho-\rho_j(h)|<\epsilon_2|)\right ) F_3 F_4 =\mathcal O(t^{-2})
\end{align} for $\epsilon_3,\epsilon_4 <<\epsilon_2$.

Finally, the contribution from the Poisson
bracket from the factor $ F_3$ is negligible
\begin{align}
  \label{eq:poisF_3}
  \tilde f (h)F_1F_2\left ({\bf d}
  F_3(|\theta-\theta_j(h)|<\epsilon_3)\right )F_4 =\mathcal O(t^{-2})
\end{align} for $\epsilon_4 <<\epsilon_3$, see \eqref{eq:decprime}. 

We conclude that 
\begin{equation}
  \label{eq:DL_1}
{\bf D}L_1(t)\geq  \mathcal O(t^{-2}),
\end{equation} and the estimates
\begin{equation} 
  \label{eq:boundb11}
\int_1^{\infty} t^{-1}|\big\langle {\Op}(b_1) \big\rangle _t| dt\leq
C \|\psi\|^2;\;  b_1=-\tilde f (h)F_1'(\cdot)F_2\cdots F_4,
\end{equation} and 
\begin{equation} 
  \label{eq:boundb12}
\int_1^{\infty} t^{-1}|\big\langle {\Op}(b_2) \big\rangle _t| dt\leq
C \|\psi\|^2;\;  b_2=-\tilde f (h)F_1\cdots F_3F_4'(\cdot).
\end{equation} We used \eqref{eq:poisF_1} an \eqref{eq:poisF_4}.

\noindent{\bf Step II}.  Classically $\gamma= \xi-\nabla_{x} S$ vanishes
along  any orbit  $x (t)$ associated with the channel we study. Here
$S=S(t,r,\theta)$ is defined in a  domain $\mathcal D_\epsilon$ as 
described in Subsection \ref{Wave operator at high energies}. We
compute
\begin{equation*}
 \frac {d}{dt}\gamma=-\left (\partial _{x} \partial_{\xi} h+\nabla^2_{x}
 S+o(t^{-1})\right )\gamma. 
\end{equation*}
Notice that this follows from a classical analogue of
\eqref{eq:local1} since (recall that ${\bf d}$ is the classical
Heisenberg derivative)
\begin{equation}
  \label{eq:difgamma}
 l_1(t){\bf d}\gamma=-l_1(t)\left (\partial _{x} \partial_{\xi}h +\nabla^2_{x}
 S+t^{-1}\mathcal O(\epsilon_1)\right )\gamma. 
\end{equation} Here $|\mathcal O(\epsilon_1)|\leq C\epsilon_1$
 uniformly in $t$.

We are motivated to calculate the 
matrix
\begin{equation}
  \label{eq:Ahes}
  A_2=-t\left (\partial _{x} \partial_{\xi}h+\nabla^2_{x}S_E\right )
\end{equation} at the particular energy
\begin{equation*}
  I\ni E=h=-\partial_tS(t,r,\theta).
\end{equation*}
Represented in  the coordinates $x_1, x_2$ of \eqref{eq:coord} with $E=h$ we
may compute using \eqref{eq:asyptoticseta} and \eqref{eq:legendr}
\begin{displaymath} 
 A_2= -\left ( \begin{array}{cc}
(1+\sigma^2)^{-1}&\sigma(1+\sigma^2)^{-1}+b\rho^{-1}\\
\sigma(1+\sigma^2)^{-1}&-\tilde\beta+\sigma^2(1+\sigma^2)^{-1}
\end{array}\right );\;\sigma=\rho \beta b (\kappa^+)^{-1}.
\end{displaymath}  Here the various quantities are computed at 
$(E,\theta)=(h,\theta_j(h))$.

The eigenvalues of $A_2$ are $-1\;{\rm and}\;\tilde \beta.$ 
 Let $T_2$ denote a real $2\times 2$--matrix that diagonalizes $A_2$.  We obtain 
\begin{equation}
  \label{eq:difgamma2}
 l_1(t){\bf d}|T_2^{-1}\gamma|^2\leq 2 t^{-1} \big (\tilde \beta
 +\mathcal O(\epsilon_1)\big )l_1(t)|T_2^{-1}\gamma|^2.
\end{equation}
 
Clearly \eqref{eq:difgamma2} yields the decay
\begin{equation}
  \label{eq:gmadecay}
  \gamma(t) =\mathcal O(t^{\tilde\beta +\epsilon'});\;\epsilon'>0,
\end{equation} along the orbit  $x (t)$.

We aim at a similar statement in quantum mechanics. (Notice
that by \eqref{eq:gmadecay} $\gamma^2$ is integrable.) For that we
pick $\bar f\in C^{\infty}_0(I)$ such that  $0\leq\bar f\leq1$ and
$\bar f=1$ in
a neighborhood of $\supp \tilde f$. Let $l_2(t)$ be given as in
\eqref{eq:l_111}  with  $\tilde f$ and $\epsilon_k$
replaced by $\bar  f$ and $2\epsilon_k;\;k=1,\dots, 4$,
respectively. We
consider localization in terms of  the operator $P=P(t)={\Op}(p(t))$ where
\begin{equation}\label{eq:p}
 p(t)= l_2(t)(T_2^{-1}\gamma)^2 l_2(t).
\end{equation} Notice that, for example $(1-l_2(t))l_1(t)=0.$

Using \eqref{eq:local1} and the above calculations for the classical
case we may mimic the proof of \cite[Proposition  5.1] {HS2} (see also the
one of 
\cite [Proposition 7.1]{CHS}). We obtain the following strong
localization
statement. Fix $\epsilon'>0$ such that $\epsilon'<
-2\tilde\beta(E)-1$ for all $E\in \supp {\bar f}
$. Then for $\epsilon_1$ small enough
\begin{equation}
  \label{eq:local2}
  \psi(t)\approx F\big(t^{1+\epsilon'}P(t)<1\big)L_1(t)\psi(t).
\end{equation} 

The proof yields the integral estimate 
\begin{equation} 
  \label{eq:boundb13}
\int_1^{\infty} t^{-1}\Big\langle L_1(-F')\big(t^{1+\epsilon'}P<1\big) L_1\Big\rangle _t dt\leq
C \|\psi\|^2,
\end{equation} in agreement with \eqref{eq:difgamma2}.

 \noindent{\bf Step III}. Recall that $U_{j}(t)$ is defined in terms
 of (a 
 small)  
 $\epsilon>0$, $I$ and the corresponding 
 domain $\mathcal D_{\epsilon,t}$. We let $\epsilon_1, \dots, \epsilon_4
 $ and $\epsilon'>0$ be given in agreement with Step II. 

The integral estimates \eqref{eq:boundb11},
\eqref{eq:boundb12} and \eqref{eq:boundb13} remain true upon
replacing  $\psi$ and $\psi(t)=\e^{-itH}f(H)\psi$ by $\phi\in
C_0^{\infty}(I\times {\R})$ and $U_{j}(t)\phi$, respectively.

To see this we introduce  a smooth function $g\in
C_0^{\infty}(\mathcal D _{\epsilon,1})$ such that with
$g_t(r,\theta)=g({r\over t},\theta)$ the following support property 
 holds  for  all  multi-indices  $|\alpha|\geq0$:
\begin{equation}\label{eq:L_1 g_t}
  l_1(t)\partial^\alpha_{t,x}(1-g_t)=0.
\end{equation} 

We consider the expectation of the family of uniformly bounded
operators
\begin{equation}\label{eq:locF}
  \Phi(t)=g_tL_1(t)F\big(P(t)t^{1+\epsilon'}<1\big)L_1(t)g_t
\end{equation} in the states  $U_{j}(t)\phi$. 
 We compute using  \eqref{eq:L_1 g_t}
and the expression \eqref{eq: generator} for the ``generator'' (notice
that $g_t \in
C_0^{\infty}(\mathcal D _{\epsilon,t})$)
\begin{equation*}
  L_1(t)i\partial_tg_tU_{j}(t)\phi=L_1(t)\left (H-2^{-1}P(t)\right
  )g_tU_{j}(t)\phi+O(t^{-\infty}),
\end{equation*}
and from the derivation of the bounds for $\psi$ we see that indeed
\begin{equation*}
  \frac {d}{dt}\Phi(t)+i[H-2^{-1}P(t),\Phi(t)]\geq \mathcal
  O(t^{-1-\epsilon'
}),
\end{equation*} expressing the ``attractiveness'' of all
localizations. Since the right hand side is integrable we obtain the integral estimates  for $g_tU_{j}(t)\phi$ 
 and hence for $U_{j}(t)\phi$.

\noindent{\bf Step IV}. We need to verify the existence of the limit 
\begin{equation*}
  \lim_{t\to \infty} U_{j}(t)^*\e^{-itH}\psi;\;\psi(t)=\e^{-itH}f(H)\psi.
\end{equation*} 

As in Step III we consider  $\phi\in
C_0^{\infty}(I\times {\R})$ and its   approximate evolution
$U_{j}(t)\phi$. We may compute the derivative
\begin{equation*}
  \frac{d}{dt}\big\langle U_{j}(t)\phi,\Phi(t)\psi(t)\big\rangle,
\end{equation*} where $\Phi(t)$ is given by \eqref{eq:locF}, as
before. Using the integral estimates \eqref{eq:boundb11},
\eqref{eq:boundb12} and \eqref{eq:boundb13} for $\psi(t)$ and the
similar  bounds for $U_{j}(t)\phi$, we obtain the estimate
\begin{equation*}
  \int _T^{\infty}|\frac{d}{dt}\big\langle
  U_{j}(t)\phi,\Phi(t)\psi(t)\big\rangle|dt\leq o(T^0)\|\phi\|,
\end{equation*} as $T\to \infty$. 

The proof is complete.
  
\subsection{Completeness at low  energies}
\label{Completeness  at low energies}
With the same assumptions as those of Subsection \ref{Wave operator at
  small energies} we have the following result.

\begin{thm}\label{thm:-111}
Under the conditions of Theorem \ref{thm:-2}
\begin{eqnarray}\label{valimoldo20}
\Ran (\bar\Omega_j) =P_j\mathcal{H}_{I}.
\end{eqnarray}
\end{thm}

We shall outline a proof of this result using weak propagation
  estimates. (The one for the potential case in \cite{HS1} uses strong propagation
  estimates.) Since we may follow the scheme of the proof of Theorem
  \ref{thm:-11} we shall be  brief.
 
Due to the discreteness assumption of the $\mathcal D$ in
\eqref{eq:discD} we may assume that $I\cap \mathcal D=\emptyset$.

Clearly  $\Ran (\bar\Omega_j) \subseteq P_j\mathcal{H}_{I}.$ For the
opposite inclusion, we follow Step I literally; the function $l_1(t)$
given by \eqref{eq:l_111} will be used below. Step II is
different. (Notice
that the $\gamma$ of \eqref{eq:gam1} differs from the $\gamma$ of the
proof of Theorem \ref{thm:-11}.) We introduce the symbol $l_2(t)$ as
before, but instead of \eqref{eq:p} we define
\begin{equation}
  \label{eq:newp}
  p(t)=l_2(t)\Big(\gamma^2+\tilde \gamma^2+(\xi_1-t^{-1}x_1)^2\Big )l_2(t),
\end{equation} where $\gamma$ and $\tilde \gamma$ are given by \eqref{eq:gam1}.

Fix $\epsilon'\in (0, \tfrac{1}{2})$ such that 
\begin{equation}
  \label{eq:BETAINE}
  3\Re\beta(E)<-1-\epsilon'\;\text{for }\;E\in \supp \tilde f.
\end{equation}

We notice that 
\begin{equation}
  \label{eq:symbp}
  t^{\frac{2}{3}(1+\epsilon')}p(t)\in S_{\unif}(1, t^{\frac{2}{3}\epsilon'-\frac{4}{3}}dx^2+t^{\frac{2}{3}(1+\epsilon')}d\xi^2),
\end{equation} 
using here a uniform symbol class; see \cite[Definition
3.1]{HS2} and the discussion after the definition.

The ``Planck's constant''  $t^{-\frac{1}{3}+\frac{2}{3}\epsilon'}\to 0$.

We may mimic the proof of  \cite[Lemma 4.4]{HS2} to obtain the
following analogue of \eqref{eq:local2}:
\begin{equation}
  \label{eq:local222}
  \psi(t)\approx \Op\big(F(t^{\frac{2}{3}(1+\epsilon')}p(t)<1)\big)L_1(t)\psi(t).
\end{equation} 

The analogue of \eqref{eq:boundb13} reads:

\begin{equation} 
  \label{eq:boundb133}
\int_1^{\infty} t^{-1}|\big\langle L_1\Op\big(F'\big(t^{\frac{2}{3}(1+\epsilon')}p(t)<1\big)\big)L_1\big\rangle _t| dt\leq
C \|\psi\|^2.
\end{equation} 

For completeness of presentation we remark that one may also derive
statements like \eqref{eq:local222} and \eqref{eq:boundb133} using
the functional calculus as done for  \eqref{eq:local2} and
\eqref{eq:boundb13}. We find the pseudodifferential approach somewhat
simpler; the reader might not. In any case a pseudodifferential
approach does not work for \eqref{eq:local2} and
\eqref{eq:boundb13}, cf. a discussion in \cite{HS2}.

Now returning to the remaining steps, we need in  Step III to show
similar 
integral estimates as those from  Step I and \eqref{eq:boundb133} for $\bar
 U(t)\phi$ where $\phi\in 1_{J}(p_1)L^2(\mathbb{R}^2_{x})$. In fact
 it suffices to have these estimates for $\phi\in g(p_1)\mathcal
 S(\mathbb{R}^2_{x})$, where $g\in C_0^\infty (J)$ and $\mathcal
 S(\cdot)$ denotes the Schwartz space. For such states we have $i\partial _t\bar
 U(t)\phi =\bar H(t)\bar
 U(t)\phi $ and we may use that the difference between the generators,
 $\bar H(t)-H$, is integrable in combination with the localization  
\eqref{eq:local222}.

For Step IV we proceed as  before. 

The outline of proof is complete.
\section{The general classical case} 
\label{The general classical case} 
We shall look at a periodic $b$ with
$\int_0^{2\pi}b\;d\theta{\leq{0}}$. (Possibly by replacing $b(\theta)\to
-b({\pi\over 2}-\theta)$ this assumption is for free.)

We  impose the following condition for a given open interval $I$ of energies.

\begin{cond}[Torus]
\label{cond:torus} 
\begin{equation}
  \label{eq:x1}
  \inf I\geq \max V.
\end{equation}
\end{cond}

With this assumption the reduced phase space $\T^*_E:=\{z\in
\T^* |\;h(z)=E\}$ is a torus. The case of
energies $E<\max V$ is somewhat easier to treat, see Remark \ref{remark:belowener}.

We have the following result on the classical dynamics for $\tau\to
+\infty$. A similar result holds for $\tau\to
-\infty$.

Let for $E\in I$
\begin{equation}
    \label{eq:resimpphase}
    \T^{*,+}(E)=\Big \{z\in \T^{*}_E|\;\rho> -\sqrt  {2(E-V(\theta))}\Big \},
  \end{equation} and 
\begin{equation}
    \label{eq:resimpphase2}
    \T^{*,-}(E)=\Big \{z\in \T^{*}_E|\;\rho< \sqrt
 {2(E-V(\theta))}\Big \}.
  \end{equation} 

Clearly the topological structure of both of the sets $\T^{*,+}(E)$ and $\T^{*,-}(E)$ is an annulus.

The following result involves the notion of {\it regular solutions}
  $\rho=\rho_r(\theta)$ of the equation $\rho'=b+\eta$ which may exist
  for  $\int_0^{2\pi}b\;d\theta{<{0}}$. For later purposes let us also 
  introduce the set 
\begin{align}\label{eq: E}
  \mathcal E =\big \{&E>\max V|\text{ a }\;2\pi-\text{periodic solution } \rho_E \text { exists},\nonumber \\&\;2(E-V)>\rho_E^2,\;\int ^{2\pi}_0{\rho_E\over \sqrt {2(E-V)-\rho_E^2}}\;d\theta>0\big \}.
\end{align}
 Here  by definition   the function $\rho_E$ solves
 \begin{equation}
   \label{eq:peri}
 \rho_E'=b+\eta_E;\;\eta_E=\sqrt {2(E-V)-\rho_E^2};  
 \end{equation} it is called an {\it upper regular} solution. If we
 drop the sign condition of the integral of \eqref{eq: E} and only
 require that $2(E-V)>\rho_E^2$ for  a solution of  \eqref{eq:peri},
 then we say that it is a {\it regular} solution.

The result stated below also involves the notion of {\it singular
  solutions/cycles} $\rho=\rho_r(\theta)$ of the equation $\rho'=b+\eta$.
Let us introduce the set 
$\mathcal{F}_1 (E)=\{\theta\in \T | \exists z\in \mathcal {F}(E):z=(\theta,0,\rho)\}$.

We shall discuss two types of singular 
cycles  which we call upper and lower singular cycles, respectively.

A $2\pi$--periodic $C^1$--solution  $\rho{_E}$  to \eqref{eq:peri} 
 is called {\it an upper singular cycle}
 if  $S^+:=\{\theta\in \mathcal{F}_1 (E)|\;\eta_E(\theta)=0\}\neq
 \emptyset$ and this set $S^+=\{\theta\in \mathcal {F}_1 (E)|\;\rho_E(\theta)=\sqrt {2(E-V(\theta)}\}$. With the latter
    assumption and \eqref{eq:nondege1}, $\kappa^+(\theta_0)<0$  for
 all $\theta_0\in S^+$ (thus 
    saddles). 

A $2\pi$--periodic $C^1$--solution  $\rho{_E}$ of
\eqref{eq:peri} is called {\it a lower singular cycle}
 if  $S^-:=\{\theta\in \mathcal{F}_1 (E)|\;\eta_E(\theta)=0\}\neq \emptyset$ and $S^-=\{\theta\in \mathcal
    {F}_1 (E)|\;\rho_E(\theta)=-\sqrt {2(E-V(\theta)}\}$. Clearly with
    \eqref{eq:nondege2}, $\kappa^-(\theta_0)<0$ for  all $\theta_0\in
    S^-$ (thus saddles again).

If $S^+$ (or $S^-$) only consists of one point for a given
singular cycle, then the cycle  corresponds to a homoclinic orbit for
the dynamics \eqref{eq:equamotion}.

\begin{prop} \label{prop:clas-gen}Suppose Condition \ref{cond:torus},
  \eqref{eq:nondege1}  and \eqref{eq:nondege2} at some $E\in I$, and that $\gamma=(\theta,\eta,\rho)$ is an
   arbitrary classical orbit with energy $E$. Then  one of the following 
  cases occurs:
\begin {enumerate}[\normalfont  i)]
  \item \label{it:cl1}There exist  sequences $\tau_n^- \to +\infty$
    and $\tau_n^+ \to +\infty$
    such that 
    \begin{equation}
      \label{eq:rhoneg}
    \rho(\tau_n^-) =-\sqrt {2(E-V(\theta(\tau_n^-)))}, 
    \end{equation} and 
\begin{equation}
      \label{eq:rhoneg2}
    \rho(\tau_n^+) =+\sqrt {2(E-V(\theta(\tau_n^+)))}, 
    \end{equation}
\item \label{it:cl2}The set $\mathcal F(E) \neq \emptyset$, and there exists $z\in \mathcal F(E)$ such that
  $\gamma(\tau)\to z$ for $\tau \to +\infty$. 
\item \label{it:cl3} There exists a regular solution
  $\rho=\rho_r(\theta)$ to the equation $\rho'=b+\eta$,
  such that  
  \begin{equation}
    \label{eq:rhor}
    \lim_{\tau\to + \infty}|\rho(\tau)-\rho_r(\theta(\tau))|=0;\; \lim_{\tau\to + \infty}|\eta(\tau)-\eta_r(\theta(\tau))|=0.
  \end{equation}
\item \label{it:cl4}There exists a singular cycle $\rho_s$ such that
   ($\eta_s=\sqrt{2(E-V)-\rho_s^2}$) 
  \begin{equation}
    \label{eq:rhos}
    \lim_{\tau\to + \infty}|\rho(\tau)-\rho_s(\theta(\tau))|=0;\;\lim_{\tau\to + \infty}|\eta(\tau)-\eta_s(\theta(\tau))|=0.
\end{equation}
\end {enumerate}
\end {prop}
\begin{proof} Using that ${d\over d\tau} (\rho-\tilde
b)=\eta^2$ (cf. the proof of Proposition \ref{lemdoi}) we have two possibilities, either 1) $\lim _{\tau\to
  +\infty}\theta(\tau) =+\infty$, or 2)  $\theta(\tau)$ stays bounded
  near $+\infty$. (Notice that 1) occurs precisely if
  $\int_0^{\infty} \eta^2\;d\tau=\infty$. To show these assertions it is
  convenient to distinguish between the case where the flux is zero
  and the case where it is nonzero.) In the case of 2) we may show
  that \ref{it:cl2})  occurs by mimicking  the proof of Proposition \ref{lemdoi}. So we can assume 1).

Suppose also that \eqref{eq:rhoneg} is false for all sequences $\tau_n^- \to +\infty$. Then
  $\gamma(\tau)$ takes values in the set $\T^{*,+}(E)$ given by
  \eqref{eq:resimpphase} for all large $\tau$'s. The topological
  structure of $\T^{*,+}(E)$ is an annulus and therefore we deduce
  (using 1)) that
  $\eta> 0$ eventually. Next we may write $\tau=\tau(\theta)$ using the
  equation $\theta'=\eta$, and thus $\rho=\rho(\theta)$. Let
  \begin{equation}
    \label{eq:rhope}
   \rho_p(\theta)= \lim _{n\to \infty}\rho(\theta+2\pi n);
  \end{equation}
 notice that the sequence is monotone. Clearly $\rho=\rho_p$ is a periodic
solution to $\rho'=b+\eta_p;\;\eta_p=\sqrt{2(E-V)-\rho^2}$. If $\eta_p$
does not have zeroes, \ref{it:cl3}) occurs. If $\eta_p$ has a zero,
then \ref{it:cl4}) occurs.

If \eqref{eq:rhoneg2} is false for all sequences $\tau_n^+ \to
 +\infty$ we look at the ``annulus'' $\T^{*,-}(E)$ given by
 \eqref{eq:resimpphase2}
 and argue similarly.
\end{proof}

\begin{remark}\label{remark:negversion} There is a version of
 Proposition \ref{prop:clas-gen} for negative times; it is given  by
 replacing $+\infty$ by $-\infty$ at all occurrences. The proof is
 similar. 
\end{remark}

\begin{remark}\label{remark:flux00}
 In the case of zero flux, Proposition \ref{prop:clas-gen}
 \ref{it:cl3}) and  \ref{it:cl4}) do not occur (seen by integrating
 \eqref{eq:peri}).  
Moreover it follows from   
 Proposition \ref{lemdoi} that \ref{it:cl1}) does not occur in this
 case. Whence    \ref{it:cl2}) and only \ref{it:cl2}) occurs  in this
 case. 
\end{remark}

\begin{remark}\label{remark:belowener} Suppose the non-critical
 condition \eqref{eq:noncrit}, \eqref{eq:nondege1} and
 \eqref{eq:nondege2} for an 
 energy  $E<\max V$. Then for all  classical orbits  with this energy
 Proposition \ref{prop:clas-gen} \ref{it:cl2})
 occurs. This follows readily from the proof of Proposition \ref{prop:clas-gen}.
\end{remark}
\subsection{The set $\mathcal {E}$}
\label{The set mathcal E} 

Motivated by Proposition \ref{prop:clas-gen}
 \ref{it:cl3}) we shall in this subsection study the set $
\mathcal E$. The analysis does not use \eqref{eq:nondege1}  nor \eqref{eq:nondege2}.

Let us show that the periodic orbit $\rho_E$ in the definition of $
\mathcal E$ is unique: Suppose $\rho_1$ and $\rho_2$ are two solutions
obeying the above conditions for the same $E$.
Let $\theta_2$ be
a  solution to $\theta_2'=\eta_2(\theta_2)$. Let $r_2({\theta})=\exp
(\int \rho_2/\eta_2\;d\theta)$. Look at
$A_2(\theta_2)=r_2(\theta_2)(\rho_1-\rho_2)({\theta_2})$. We compute and
estimate 
$A_2'=r_2\big (\rho_2(\rho_1-\rho_2)+\eta_2(\eta_1-\eta_2)\big )\leq
0$. Thus 
$A_2$ is bounded from above. Since $r_2(\theta_2)\to +\infty$ and the
$\rho$'s are periodic we conclude that $\rho_1\leq \rho_2$. By
symmetry we see that $\rho_1=\rho_2$.

 \begin{prop}\label{prop:interm}Suppose $V=0$ and $\mathcal E \neq \emptyset$. We have \begin {enumerate}[\normalfont  1)]
  \item \label{it:131}$\mathcal E$ is an open interval $\mathcal
    E=(E_d,E_e),\;E_d>0$.
\item \label{it:132}$\lim_{E\to E_d}\partial_E \rho{_E}(\theta) =+\infty.$
\item \label{it:133}$\lim_{E\to E_d}\int ^{2\pi}_0{\rho_E\over \eta_E}\;d\theta=0.$
\item \label{it:134}$\partial_E \rho{_E}(\theta) >1/\sqrt {2E}.$
\item \label{it:1343}$\partial_E^2 \rho{_E}(\theta) <0.$
\item \label{it:1344}$\partial_E {\rho{_E}\over \eta_E} >0.$
\item \label{it:1345} The function $E\to \rho{_E}(\theta)- \sqrt {2E}$
  is increasing.
\item \label{it:135}$\lim_{E\to E_e}\int ^{2\pi}_0{\rho_E\over
    \eta_E}\;d\theta=+\infty.$
\item \label{it:136} $\eta_E(\theta)\leq \max |b|$.
\item \label{it:137} Suppose $2\tilde E:= 2E-\max b^2>0$. Then $\rho_E
  \geq \sqrt {2\tilde E}$ and\\ 
  $\partial_E \rho{_E}(\theta) \leq 1/\sqrt {2\tilde E}.$ In
  particular $\lim_{E\to E_e}\partial_E \rho{_E}(\theta) =0$ if $E_e=\infty$.
\end{enumerate}
 \end{prop}
 \begin{proof} As in the proof of [CHS, Proposition 2.4] we see that
   $\mathcal E $ is open. Take for the moment a maximal subinterval
   $(E_d,E_e)\subseteq \mathcal E$.  By integrating \eqref{eq:peri} we
   see that $\sqrt {2E_d}\geq (2\pi)^{-1}\int ^{2\pi}_0 -b(\theta)
   \;d\theta>0$; in particular $E_d>0$. 
%Let us notice another lower
%   bound, also obtained by integrating \eqref{eq:peri},
%   \begin{equation*}
%     2\sqrt {2E_d}\geq \Big (\max_{\theta_1,\theta_2} \int ^{\theta_2}_{\theta_1} -b(\theta)
%   \;d\theta\Big )+\int ^{2\pi}_0 b(\theta)
%   \;d\theta.
%   \end{equation*}

We have the formula
   \begin{equation}\label{eq:formparE}
    \partial_E \rho (\theta)=\int ^\theta_ {-\infty}{1\over
    \eta(\theta')}\e^{\int ^{\theta'}_\theta {\rho\over
    \eta}\;d\bar \theta}\;d\theta'. 
   \end{equation} In particular
   \begin{equation}\label{eq:lowb}
   \partial_E \rho (\theta)> \int ^\theta_ {-\infty}{1\over
    \eta(\theta')}\e^{\int ^{\theta'}_\theta {\sqrt {2E}\over
    \eta}\;d\bar \theta}\;d\theta'= {1\over \sqrt {2E}},
   \end{equation} 
 showing  \ref{it:134}).

Rewriting \ref{it:134}) as $\partial_E (\rho_E(\theta) -\sqrt {2E})>0$
 yields \ref{it:1345}). 

As for \ref{it:1344}) we compute $\partial_E {\rho{_E}\over \eta_E}
=(2E\partial_E \rho{_E}-\rho_E)/\eta^3$. Using \eqref{eq:lowb} again
this leads to $\partial_E {\rho{_E}\over \eta_E}\geq (\sqrt{2E}-\rho_E)/\eta^3>0$.

As for \ref{it:1343}) we refer to the proof of [CHS, (2.19)].
 
By monotonicity, cf.  \eqref{eq:lowb},  there exist 
\begin{equation*}
 \rho_d(\theta)=\lim _{E\to E_d}\rho_E(\theta),\;\eta_d(\theta)=\lim _{E\to E_d}\eta_E(\theta), 
\end{equation*} and if $E_e<\infty$
\begin{equation*}
 \rho_e(\theta)=\lim _{E\to E_e}\rho_E(\theta),\;\eta_e(\theta)=\lim _{E\to E_e}\eta_E(\theta). 
\end{equation*}
Let $\eta_j$ denote either $\eta_d$ or $\eta_e$ (assuming $E_e<\infty$
for the latter). Suppose
$\eta_j(\theta_0)=0$ for some $\theta_0$. Then by \eqref{eq:peri},
$b(\theta_0)=0$. We claim that 
\begin{equation}
  \label{eq:infiover eta}
  \int ^{2\pi}_0{1\over \eta_j}\;d\theta=\infty.
\end{equation} Suppose not, then we learn from writing
$\eta^2=(\sqrt {2E}-\rho)(\sqrt {2E}+\rho)$ near
$\theta=\theta_0$ (with $\eta=\eta_j$ and $E=E_j$) and using \eqref{eq:peri} that 
\begin{equation}
  \label{eq:bnd8} 
 \eta^2\leq C\Big ((\theta-\theta_0)^2 +\Big |\int _{\theta_0}^\theta
 \eta(\theta')\;d\theta'\Big |\Big )=:g(\theta).
\end{equation}
Let us look at $g(\theta)$ to the right of $\theta_0$. We have 
\begin{equation*}
  g'\leq 2C(\theta-\theta_0) +C\sqrt g. 
\end{equation*} Let  $f(\theta)=K(\theta-\theta_0)^2$ with $K>0$ taken 
such that \begin{equation*}
  f'=  2C(\theta-\theta_0) +C\sqrt f. 
\end{equation*} Then 
\begin{equation*}
  (f-g)'\geq C(\sqrt f-\sqrt g)={C\over \sqrt f+\sqrt g}(f-g). 
\end{equation*} Since $\int {C\over \sqrt f+\sqrt g}\;d\theta<\infty$
we conclude that $\sqrt K |\theta-\theta_0|\geq \sqrt f\geq \sqrt
g\geq \eta$.  This contradicts the
finiteness assumption, and therefore \eqref{eq:infiover eta} holds.

Let us show \ref{it:132}) and \ref{it:133}). We first show that for
$\rho_d (\theta)\neq \sqrt {2E_d}$ for all $\theta$. Indeed if $\rho_d
(\theta)= \sqrt {2E_d}$ then $b(\theta)=0$, and by \ref{it:1345}) for $E'<E$
\begin{equation*}
 0>\rho_E(\theta)-\sqrt{2E} \geq \rho_{E'}(\theta)-\sqrt{2E'},
\end{equation*} yielding a contradiction by letting $E'\to E_d$.

If $\rho_d (\theta)= -\sqrt {2E_d}$ for some $\theta$, then $\lim_{E\to
  E_d}\int ^{2\pi}_0{\rho_E\over \eta_E}\;d\theta=- \infty$ by
  \eqref{eq:infiover eta} and  Fatou's Lemma,  which conflicts  the fact that  $\int ^{2\pi}_0{\rho_E\over \eta_E}\;d\theta>0$.

But then 
\begin{equation}
  \label{eq:int0}
  \int ^{2\pi}_0{\rho_d\over \eta_d}\;d\theta = 0, 
\end{equation}
since
the integral exists and $E_d$ is a positive  endpoint of the maximal interval. 
 This shows \ref{it:133}), and \ref{it:132}) follows from
 \eqref{eq:formparE},  \ref{it:133}) and Fatou's Lemma.

We can now prove that $
\mathcal E$ is an interval. Let $E_d^1$ and $E_d^2$ be the left
endpoints of two maximal intervals. Similar notation is used for the
corresponding periodic functions $\rho$ and $\eta$. Let $\theta_2$ be
a  solution to $\theta_2'=\eta_2(\theta_2)$. Let $r_2({\theta})=\exp
(\int \rho_2/\eta_2\;d\theta)$. Look at
$A_2(\theta_2)=r_2(\theta_2)(\rho_1-\rho_2)({\theta_2})$. We compute
$A_2'=r_2\big (\rho_2(\rho_1-\rho_2)+\eta_2(\eta_1-\eta_2)\big )$. If
$E_d^1<E_d^2$ we see that  $A_2'\leq -\epsilon$ contradicting that
$A_2$ is bounded. Thus (by symmetry) $E_d^1=E_d^2$. This shows that  $
\mathcal E$ is an interval.

The statement  \ref{it:136}) follows readily using \eqref{eq:peri} and the
periodicity, and \ref{it:137}) from \ref{it:136}) and by estimating
\eqref{eq:formparE}, cf. \eqref{eq:lowb}.

Finally,   to show \ref{it:135}) we can assume that $E_e$ is finite,
cf. \ref{it:136})  and \ref{it:137}). By \ref{it:1344}) the limit
\begin{equation*}
 \lim_{E\to E_e}\int ^{2\pi}_0{\rho_E\over \eta_E}\;d\theta \in
(0,\infty] 
\end{equation*}
 exists. Suppose it is finite. Then by the Lebesgue
monotone convergence theorem and 
\eqref{eq:infiover eta} we see that $\eta_e(\theta)>0$ for all
$\theta$, contradicting the maximality of $E_e$.
\end{proof}

Small modifications of the  above proof yield the following results in the general case.
 
\begin{prop}\label{prop:intermgen}Suppose $\mathcal E \neq \emptyset$. We have \begin {enumerate}[\normalfont  1)]
  \item \label{it:131g}$\mathcal E$ is an open set with $E_d := \inf \mathcal
    E>\min V$.
%\item \label{it:132g}$\lim_{E\to E_d}\partial_E \rho{_E}(\theta) =+\infty.$
%\item \label{it:133g}$\lim_{E\to E_d}\int ^{2\pi}_0{\rho_E\over \eta_E}\;d\theta=0.$
\item \label{it:134g}$\partial_E \rho{_E}(\theta) >1/\sqrt {2(E-\min V)}.$
\item \label{it:1343g}$\partial_E^2 \rho{_E}(\theta) <0.$
%\item \label{it:1344g}$\partial_E {\rho{_E}\over \eta_E} >0.$
\item \label{it:1345g} The function $E\to \rho{_E}(\theta)- \sqrt {2(E-\min V)}$
  is increasing.
\item \label{it:135g}Suppose $\hat I=(\hat E_d, \hat E_e)$ is a
  maximal subinterval of $\mathcal E$. Then $$\lim_{E\to (\hat E_e)^-}\int ^{2\pi}_0{\rho_E\over
   \eta_E}\;d\theta=+\infty.$$
\item \label{it:136g} $\eta_E(\theta)\leq \max |b|$.
\item \label{it:137g} Suppose $2\tilde E:= 2(E-\max V)-\max b^2>0$. Then $\rho_E
  \geq \sqrt {2\tilde E}$ and 
  $\partial_E \rho{_E}(\theta) \leq 1/\sqrt {2\tilde E}.$ 
%In
 % particular $\lim_{E\to E_e}\partial_E \rho{_E}(\theta) =0$ if $E_e=\infty$.
\end{enumerate}
 \end{prop}
 
Under the conditions of Proposition \ref{prop:intermgen} it  is an
open problem whether $\mathcal {E}$ is an interval.
 
We notice that in the case $b<0$ and $V=0$ studied in \cite {CHS} the set $\mathcal E =(E_d, \infty)$. As the following example shows
there are other cases where $\mathcal E \neq \emptyset$.
\begin{example} Take $V=0$, $b(\theta):=-c<0$, if 
$0\leq \theta\leq 2\pi-\epsilon$, and
$-c\leq b(\theta)\leq c$. Whence we do not have a sign assumption on the interval
$(2\pi-\epsilon, 2\pi)$. In particular $\mathcal{F}(E)\neq
\emptyset$ is allowed.
For this class of examples:

For all $E>c^2/2$ we can find
$\epsilon=\epsilon_E>0$ small enough such that $E\in \mathcal E$. 

This may be proved by elementary differential
inequality techniques.
\end{example}

On the other hand, if $V=0$ and  $b(\theta) \geq 0$ in an interval $I=[\theta_1,\theta_2]$ of length
$|I|=\pi$ (no sign assumption outside $I$), then indeed  $\mathcal E
= \emptyset$: Suppose on the contrary that $E\in \mathcal E$ exists. Then the
corresponding periodic solution obeys $\rho'\geq \eta=\sqrt{2E-\rho^2}$ on
$I$. Hence 
\begin{align}
  \label{eq:bnega}
  &\pi=|I|\leq\int _I{\rho'\over \sqrt{2E-\rho^2}}d\theta=\int
  _{\rho(\theta_1)}^{\rho(\theta_2)}{1\over
  \sqrt{2E-\rho^2}}d\rho \nonumber
  \\&=\sin ^{-1}\Big ({\rho(\theta_1)\over \sqrt{2E}}\Big )-\sin ^{-1}\Big ({\rho(\theta_2)\over \sqrt{2E}}\Big )<\pi. 
\end{align}

Furthermore in the more general situation, $V=0$ and  $b(\theta) \geq 0$ in an interval
$I=[\theta_1,\theta_2];\;\theta_1< \theta_2$,  the same
estimation as above leads to

\begin{equation}
  \label{eq:bnega2}
  |I|\leq \sin ^{-1}\Big ({\rho(\theta_1)\over \sqrt{2E}}\Big )-\sin ^{-1}\Big ({\rho(\theta_2)\over \sqrt{2E}}\Big ). 
\end{equation}
Since $ \rho(\theta_j)\geq \sqrt{2E-\max b^2}$, cf. Proposition
\ref{prop:interm} \ref{it:137}), the right hand side is close to ${\pi\over
  2}-{\pi\over 2}=0$ for $E$ large, in particular $<|I|$. So $\rho$
does not exist for $E$ large.

\begin{remark}\label{rem:e>enedred}
  For $E>E_d$ in the case where $\mathcal E\neq \emptyset$ one easily
  verify that Proposition \ref{prop:clas-gen} \ref{it:cl1}) 
  does not occur by 
checking that  $\e^{\int \rho}(\rho-\rho_{E'})$ is increasing in
$\tau$  for $E'<E$ in $\mathcal E$.
\end{remark}

\subsection{Singular cycles }
\label{Singular cycles}
Motivated by Proposition \ref{prop:clas-gen}
 \ref{it:cl4}) and Proposition \ref{prop:interm} \ref{it:135}) (notice
 that $\rho_e=\lim_{E\to E_e}\rho_E$ indeed is an upper singular
 cycle) we shall in this subsection study singular cycles under
 slightly more
 restrictive conditions than in Proposition \ref{prop:interm}.

 \subsubsection{Uniqueness of the  upper singular cycle}    
\label{Uniqueness of upper singular cycle}
\begin{prop}
  \label{prop:unique} Suppose $V=0$, Condition \ref{cond:torus} and 
  \eqref{eq:nondege1} for all $E\in I$.  Then there is at most one
  energy in $I$ for which an upper singular cycle exists, and
  this solution is uniquely determined.
\end{prop}
\begin{proof} Let $\rho_1=\rho_{E_1}$ and $\rho_2=\rho_{E_2}$ be
  two given upper singular cycles. We need to show
  that $\rho_1=\rho_2$. Let $\eta_j(\theta)=\sqrt{2E_j-\rho_j^2}$
  be the corresponding transversal velocity and  $S_j:=\{\theta\in \mathcal
    F_1(E_j)|\;\eta_j(\theta)=0\}$. We consider three cases. 

\noindent {\bf Case I} 
  Suppose $E_1< E_2$ and that 
  \begin{equation*}
    S_1\cap S_2\neq\emptyset.
  \end{equation*}
We introduce $\sigma_j=\rho_j- \sqrt{2E_j}$, and  obtain from  (\ref{eq:peri}) that
  \begin{equation}
    \label{eq:sig}
    \sigma_j'=b+f_{j}(\sigma_j);\;f_{j}(\sigma)=\sqrt{-2\sqrt{2E_j}\sigma-\sigma^2}.
  \end{equation}

We notice the following formulas near any $\theta_0\in S_1\cap S_2$. 
\begin{align}
  \label{eq:asr}
 \sigma_j(\theta)&=\Bigg\{b'(\theta_0)+
 {\sqrt{2E_j}\over 2}\Bigg(-1+\sqrt{1-{4b'(\theta_0)\over
     \sqrt{2E_j}}}\;\Bigg)\Bigg \}{(\theta-\theta_0)^2\over 2}\\
 &+O\big((\theta-\theta_0)^3\big )\text{ for }\theta\downarrow \theta_0,\nonumber
\end{align}
\begin{align}
  \label{eq:asl}
 \sigma_j(\theta)&=\Bigg\{b'(\theta_0)-
 {\sqrt{2E_j}\over 2}\Bigg(1+\sqrt{1-{4b'(\theta_0)\over
     \sqrt{2E_j}}}\;\Bigg)\Bigg \}{(\theta-\theta_0)^2\over 2}\\
 &+O\big ((\theta-\theta_0)^3\big )\text{ for }\theta\uparrow \theta_0.\nonumber
\end{align} 

We deduce  from (\ref{eq:asr}) and (\ref{eq:asl}) that for some
$\epsilon>0$
\begin{equation}\label{eq:sign}
\sigma_2(\theta)
\begin{cases}
  >\sigma_1(\theta)\text { provided
  }0<\theta-\theta_0<\epsilon\\
<\sigma_1(\theta)\text { provided
  }0> \theta-\theta_0>-\epsilon
\end{cases}\;.
\end{equation}

Let us also notice that away from joint zeros of the $\sigma_j$'s,
that is away from the set $S_1\cap S_2$,
\begin{equation}
  \label{eq:difeq}
  (\sigma_2-\sigma_1)'\geq
  -h_2(\sigma_2-\sigma_1);\;h_2=h_{2}(\theta)={2\sqrt{2E_2}+\sigma_2+\sigma_1\over f_{2}(\sigma_1)+f_{2}(\sigma_2)}.
\end{equation}

We learn from (\ref{eq:difeq}) that the function  
\begin{equation*}
  \theta \to \e^{\int_{\bar \theta}^\theta h_2\;d\theta}(\sigma_2-\sigma_1)(\theta)
\end{equation*} is non--decreasing on any interval $[\bar \theta,
\tilde 
\theta]$ disjoint from $S_1\cap S_2$, in particular on some  set of the
form $(\theta_0, \tilde\theta_0)$, where $\theta_0, \tilde\theta_0 \in
S_1\cap S_2$. Here we use the periodicity of the $\eta_j$'s. (If
$S_1\cap S_2\subset \T$ only consists of one  point
$\theta_0$ this amounts to looking at one entire revolution,
$\tilde\theta_0=\theta_0+2\pi$.) By combining this monotonicity with  the upper part of 
(\ref{eq:sign}) at the angle $\theta_0$ and the lower
part at $\theta_0\to \tilde\theta_0$ we arrive at a
contradiction.

\noindent {\bf Case II} Suppose $E_1\leq E_2$ and that 
  \begin{equation*}
    S_1\cap S_2=\emptyset.
  \end{equation*} Introduce again $\sigma_j=\rho_j- \sqrt{2E_j}$. Pick
  arbitrary $\theta_1\in S_1$ and $\theta_2\in S_2$. We may assume  that  
   $\theta_2 <\theta_1$ (otherwise replace $\theta_1\to
  \theta_1+2\pi$). Clearly  $\sigma_2(\theta_2)>\sigma_1(\theta_2)$
   and $\sigma_2(\theta_1)<\sigma_1(\theta_1)$. By using
   \eqref{eq:difeq} to the interval $[\theta_2,\theta_1]$ as before we
   conclude that $\sigma_2(\theta_1)>\sigma_1(\theta_1)$, a contradiction.
 
\noindent {\bf Case III}  Suppose $E_1=E_2$. We modify the arguments
   for uniqueness of outgoing spirals given before Proposition
   \ref{prop:interm}: Let  $\theta_2^-, \theta_2^+\in S_2$ be given such that
   $\theta_2^-< \theta_2^+$ and  $(\theta_2^-,\theta_2^+)\cap
   S_2=\emptyset$ (possibly  $\theta_2^+=\theta_2^-+2\pi$). We solve
   $\theta_2'=\eta_2(\theta_2)$ on $(\theta_2^-,\theta_2^+)$, and  let $r_2({\theta})=\exp
(\int \rho_2/\eta_2\;d\theta)$. As before we look at
$A_2(\theta_2)=r_2(\theta_2)(\rho_1-\rho_2)({\theta_2})$. Since $A_2$
   is non-increasing (same estimate) and $r_2(\theta_2(\tau))\to 0$
   for $\tau\to -\infty$ we
   deduce that $\rho_1\leq \rho_2$ on
   $(\theta_2^-,\theta_2^+)$. By using  this argument for all possible
   intervals of this form, as well as   continuity at $S_2$, we finally see that $\rho_1
   \leq \rho_2$ for all angles, and then by 
symmetry that $\rho_1=\rho_2$.
\end{proof}

\subsubsection{Uniqueness of the lower singular cycle}    
\label{Uniqueness of lower singular cycle}
\begin{prop}
  \label{prop:uniquelower} Suppose $V=0$, Condition \ref{cond:torus} and 
  \eqref{eq:nondege2} for all $E\in I$. Then there is at most one
  energy in $I$ for which a lower singular cycle exists, and
  this solution is uniquely determined.
\end{prop}
\begin{proof} We define $b_1(\theta)=b(-\theta)$. The transformation $\rho(\theta)\to
  -\rho(-\theta)$ turns a lower singular cycle for $b$ to an upper singular
  cycle for $b_1$, cf. \cite[Section 9]{CHS}. Hence  we can invoke Proposition
  \ref{prop:unique} with  $b\to b_1$.
\end{proof} 

\subsection{Incoming spirals}    
\label{Incoming spirals} We
  work in the remaining part of this section under the  conditions of
  Subsection \ref{The set mathcal E} (i.e. \eqref{eq:nondege1}  and 
  \eqref{eq:nondege2} are not used).
Let us denote the set $\mathcal E$ in \eqref{eq: E} by $\mathcal E^+$,
and introduce 
\begin{align*}\label{eq: E-}
  \mathcal E^- =\big \{&E>\max V|\text{ a }\;2\pi-\text{periodic  solution }\rho_E \text { exists},\\&\;2(E-V)>\rho_E^2,\;\int ^{2\pi}_0{\rho_E\over \sqrt
  {2(E-V)-\rho_E^2}}\;d\theta< 0\big \}.
\end{align*} Here $\rho_E$  solves \eqref{eq:peri} (as before).  

The transformation $\rho(\theta)\to
  -\rho(-\theta)$ turns a solution $\rho_E$ corresponding to $E\in
  \mathcal E^-$ to a  solution to \eqref{eq:peri} with  $b(\theta)\to
  b_1(\theta)=b(-\theta)$ and $V(\theta)\to
  V_1(\theta)=V(-\theta)$ 
  but reverses the sign of the integral and  hence $E\in
  \mathcal E^+=\mathcal E^+(b_1,V_1)$, and vice versa. We previously studied
  $\mathcal E^+$; whence  we may deduce for instance that    $\mathcal
  E^-$ is an interval $\mathcal
  E^-=(E_d^-,E_e^-)$ if $V=0$, possibly empty. 

In the case $b<0$ we showed in \cite{CHS} that $\mathcal
  E^+=\mathcal E^-\neq \emptyset$.

 For $a\in (-\sqrt{2E}, \sqrt{2E})$ and $E>0$ we consider the
  initial value problem  
\begin{equation}\begin{cases}
   \label{eq:horia14}
 \rho_E'=b+\eta_E;\;\eta_E=\sqrt {2(E-V)-\rho_E^2}>0\\
\rho_E(\theta=0)=a
\end{cases}
\;.  
 \end{equation} 

If there is a solution on the interval $[0, 2\pi]$ we denote it by
 $\rho=\rho(\theta,a,E)$.  Clearly  $\rho$ extends to a  $2\pi$--periodic
 solution if
 $\rho(2\pi,a,E)=a$.

\begin{lemma} \label{lem:-=+}Assume there exists a periodic solution $\rho_d$ to 
\eqref{eq:horia14} at an energy $E=E_d$, such that 
$$\int ^{2\pi}_0{\rho_d\over \sqrt
  {2(E-V)-\rho_d^2}}\;d\theta = 0.$$
Then   $\mathcal
  E^+$ and $\mathcal
  E^-$ are not empty, and $E_d=E_d^+=E_d^-$. In particular for $V=0$, if the set
  $\mathcal {E}^+\neq \emptyset$ then  $\mathcal
  E^-\neq \emptyset$ (and vice versa). 
\end{lemma}
\begin{proof} 
Let us introduce $a_d=\rho_d(\theta=0)$. We look at the equation
$f(a, E):=\rho(2\pi,a, E)-a=0$ for $| a-a_d|$ and 
$|E-E_d|$
small. By ODE techniques we find that $f$ is smooth  in its
 variables. We compute
\begin{align}
  \label{eq:horia15}
  &\partial_a f(a, E)=\e^{-\int _0^{2\pi}{\rho\over \eta}\;d\theta'}-1;\;\eta=\sqrt
  {2(E-V)-\rho^2},\\\label{eq:horia16}
&\partial_a^2 f(a,E)=-\e^{-\int _0^{2\pi}{\rho\over \eta}d\theta'}\int _0^{2\pi}{2(E-V)\over \eta^3} \e^{-\int
  _0^{\theta}{\rho\over \eta}d\theta'}\;d\theta .
\end{align}

Next we evaluate \eqref{eq:horia15} and \eqref{eq:horia16} at
$(a, E)=(a_d,E_d)$ and  conclude that as a function of $a$, $f$
vanishes to second order at $(a_d,E_d)$.

Another straightforward computation gives:
\begin{align}
  \label{horia13}
  & \partial _{ E}f(a,E) =\e^{-\int _0^{2\pi}{\rho\over
  \eta}\;d\theta'}\int _0^{2\pi}\eta^{-1}\e^{\int
  _0^{\theta}{\rho\over \eta}\;d\theta'}\;d\theta\nonumber\\
&\partial _{ E}f(a_d,E_d)=\int _0^{2\pi}\eta_d^{-1}\e^{\int
  _0^{\theta}{\rho_d\over \eta_d}\;d\theta'}\;d\theta >0.
\end{align}

Using the third order Taylor expansion for $f(\cdot, E)$ in a
neighborhood of $(a_d,E_d)$, we have that $f(a, E)=0$ iff 

\begin{align}\label{horia12}
& f(a,E)-f(a,E_d)\\
&= -\frac{(a-a_d)^2}{2} (\partial_{a}^2 f)(a_d,E_d) -\int_{a_d}^a dx
\int_{a_d}^x dy \int_{a_d}^y dt (\partial_{a}^3 f)(t,E_d).\nonumber 
\end{align}
Recall  that $(\partial_{a}^2 f)(a_d,E_d)<0$, $(\partial_E
f)(a_d,E_d)>0$, and $f(a_d,E)-f(a_d,E_d)>0$ for $E>E_d$ close to
$E_d$. Now choose $\epsilon>0$ and $\delta >0$ small enough and define 
(here $E_d\leq E\leq E_d+\delta$)
\begin{align}\label{horia20}
& F_\pm :[a_d-\epsilon,a_d+\epsilon]\to \R, \\
 F_\pm(a)&=a_d\nonumber\\&\pm 
\sqrt{\frac{f(a,E_d)-f(a,E)}{\frac{1}{2} 
(\partial_{a}^2 f)(a_d,E_d) +\frac{1}{(a-a_d)^2}\int_{a_d}^a dx
\int_{a_d}^x dy \int_{a_d}^y dt (\partial_{a}^3 f)(t,E_d)}}.\nonumber 
\end{align}
If $\delta$ is chosen small enough, we see that $F_\pm$ leave their 
domains invariant, and moreover, they are contractions; more
precisely, one can prove the estimate 
$$\sup_{a\in [a_d-\epsilon,a_d+\epsilon]}|F'_\pm(a)|\leq C \sqrt{E-E_d}.$$

The two fixed
points will in fact be the two possible solutions of  $f(a,E)=0$ near
$(a_d,E_d)$, with the extra condition $E>E_d$. Therefore:
\begin{equation}\label{horia30}
 a_\pm (E)=a_d \pm \sqrt{-(E-E_d) \frac{2(\partial_E
f)(a_d,E_d)}{(\partial_{a}^2 f)(a_d,E_d)}} +\mathcal{O}(E-E_d).
\end{equation}
Using the mean value theorem for $\partial_a f$ we obtain 
\begin{align}\label{horia31}
&(\partial_a f)(a_\pm(E),E)=(\partial_a f)(a_\pm(E),E)-(\partial_a
f)(a_d,E_d)\\
& =\mp \sqrt{-2(E-E_d) (\partial_E
f)(a_d,E_d)\cdot (\partial_{a}^2 f)(a_d,E_d)}
+\mathcal{O}(E-E_d).\nonumber 
\end{align}
Then \eqref{eq:horia15} gives that 
$$\pm \int_0^{2\pi}\frac{\rho_\pm}{\eta_\pm}d\theta >0$$ 
and we are done.
\end{proof} 

%We state it as an open problem to determine
%whether $E_e^+=E_e^-$ in general?

\begin{lemma}
  \label{lem:-<+}Suppose $E^+\in \mathcal E^+$, $E^-\in \mathcal
  E^-$ and $E^+\leq E^-$. Denoting
  the corresponding solutions by $\rho^+_{E^+}$ and $\rho^-_{E^-}$,
  respectively, we have 
  \begin{equation}
    \label{eq::-<+}
   \rho^-_{E^-}(\theta)\leq \rho^+_{E^+}(\theta) \text{ for all }\theta.
  \end{equation}
\end{lemma}
\begin{proof}
  We modify  the last part of the proof of Proposition \ref{prop:unique}
  introducing $A(\theta)=r(\theta)(\rho^+-\rho^-)(\theta)$ where 
  $r=\exp(\int \rho^-/\eta^- \;d\theta)$ and
  $\theta'=\eta^-(\theta)$. As before we verify that $A'\leq
  0$. Whence, since  $r\to 0$ for $\tau \to + \infty$, $\rho^-\leq \rho^+$.
\end{proof}

\subsection{Completeness in $\mathcal E^+$, classical case}
\label{Completeness in interval}
The purpose of this subsection is to outline a proof of asymptotic
completeness  in the set  $\mathcal E=\mathcal E^+$, cf. Remark \ref{rem:e>enedred},  that is generalizable to
quantum mechanics. The one of Proposition \ref{prop:clas-gen} is not generalizable 
since it relies on  topological arguments.

Fix $E\in \mathcal E$. We shall prove classical completeness for orbits with
energy $E$. By a {\it scattering orbit}  we mean below a classical orbit
moving to infinity in configuration space, that is $r(t)=|x(t)| \to \infty$
for $t\to \infty$ $(=+\infty)$. Since the case where $\mathcal F^+(E)=\emptyset$
essentially was treated
in \cite{CHS} let us assume that $\mathcal F^+(E)\neq \emptyset$.

We define $E_{\crt}\in \mathcal E$ as the smallest  energy $E_{\crt}=E'>E$ for which the
equation 
\begin{equation}
  \label{eq:equE}
  \rho_{E'}(\theta)=\sqrt{2(E-V(\theta))}
\end{equation} has a solution $\theta\in \T$. Due to Proposition \ref{prop:intermgen}
 this is a good definition. Let $T_{\crt}$ denote the
set of all solutions  $\theta\in \T$ for this particular energy. From
the definition of $E_{\crt}$ and \eqref{eq:peri} it follows that 
\begin{equation}
  \label{eq:envelope}
  b+\eta_{\crt}=-{V'\over \rho_{E_{\crt}}} \text{ at all angles }\theta\in T_{\crt}.
\end{equation}
 Whence 
\begin{equation}
  \label{eq:upb}
  \epsilon_1 := -\max _{T_{\crt} }\{b+{V'\over \rho_{E_{\crt}}}\}=\min_{\theta \in
   T_{\crt}}\eta_{E_{\crt}}(\theta)>0.
\end{equation} In particular $T_{\crt}\cap
\mathcal F^+_1(E)=\emptyset$.  

Let $T_\kappa=\{\theta| \rho_{E_{\crt}}(\theta)>\sqrt{2(E-V(\theta)}-\kappa\}$; $\kappa>0.$ 
By continuity, for $\kappa$ taken small enough 
\begin{equation}
  \label{eq:2}
  b(\theta)+{V'\over \sqrt{2(E-V(\theta))}}\leq -{\epsilon_1\over 2} \text{ for all }\theta\in T_\kappa.
\end{equation}

\begin{prop}\label{prop:sepa }Let $E, E_{\crt}$ be given as above. Fix
  $E_1\in (E, E_{\crt})$. Then there exists  $\epsilon>0$ small enough
  such that the
  following bounds hold for all scattering orbits with energy $E$:
\begin {enumerate}[\normalfont  1)]
  \item \label{it:int} $\int_1^{\infty} F(\rho-\rho_{E_{\crt}}<\epsilon{})F(\rho-\rho_{E_{1}}>\epsilon{})r^{-1}\;dt<\infty.$
\item\label{it:decay}$F(\rho-\rho_{E_{\crt}}<\epsilon{})F(\rho-\rho_{E_{1}}>\epsilon{})\to
  0 \text{ for }t\to \infty.$
\end {enumerate}
 \end {prop}
 \begin{proof} As in \cite{CHS} we introduce the observable
$B_{E'}=\rho-\rho_{E'}(\theta)$ (for any $E'\in \mathcal E$) and compute
\begin{equation}
  \label{eq:comp1}
 {\bf d}B_{E'}=(\eta-\eta_{E'}){\eta\over r}. 
\end{equation} 

For $E'<E$ in $\mathcal E$ (\ref{eq:comp1}) leads to the
   Mourre estimate
   \begin{equation}
     \label{eq:mourre}
 {\bf d}\{rB_{E'}\} \geq \delta \text { for some } \delta>0.
   \end{equation}

In particular we learn from (\ref{eq:mourre}) that for any given  
$E'<E$ in $\mathcal E$,  $\rho(t)\geq \rho_{E'}(\theta(t))\geq -\sqrt{2(E'-V(\theta(t)))}$ for all $t$ large
enough.

  Using the Mourre estimate and a  maximal velocity bound,
  cf. \eqref{eq:largevelo}, we see that the
  factor $r^{-1}$ in \ref{it:int}) may be replaced by  $t^{-1}$. Hence
  the 
  statement \ref{it:decay}) follows from the subsequence argument,
  cf. the proof of Proposition \ref{lemdoi} (or 
  for example  \cite{CHS}), and
  \ref{it:int}).  
 
If $\eta\neq -\eta_{E'}$ we may rewrite (\ref{eq:comp1}) as 
\begin{equation}
  \label{eq:comp2}
 {\bf d}B_{E'}={2(E-E')\over{\eta+\eta_{E'}}}{\eta\over
   r}-{\rho+\rho_{E'}\over{\eta+\eta_{E'}}}{\eta\over r}B_{E'}=T_1+T_2. 
\end{equation} 

With these preliminaries we can start proving \ref{it:int}).

\noindent {\bf Step I} We shall prove the bound 
\begin{equation}
  \label{eq:negmom}
  \int_1^{\infty} F(B_{E_{\crt}}<\epsilon{})F(-\eta >\epsilon')r^{-1}\;dt<\infty;\;\epsilon'>0.
\end{equation}

For $E'\in (E_d,E_{\crt}]\cap \mathcal E$ consider
  $$\Phi=F(B_{E'}<2\epsilon{})F(-\eta >\epsilon')=F_1F_2.$$  In
  the arguments below
  we  assume that $\epsilon'>0$ is ``small''. We
  compute using (\ref{eq:comp1})
\begin{equation}
  \label{eq:Heineg0}
 {\bf d}\Phi= F'(B_{E'}<2\epsilon{})(\eta-\eta_{E'})F_2{\eta\over r}+F_1F'(-\eta >\epsilon'){(b+\eta)\rho+V'\over r},
\end{equation}
leading to 
\begin{equation}
  \label{eq:Heineg2}
 {\bf d}\Phi\leq  F'(B_{E'}<2\epsilon{})F_2{\epsilon' \min \eta_{E'}\over 2r}.
\end{equation}
Here we used that the second term in \eqref{eq:Heineg0} is non-positive
due to the fact that on the support of $F_1F'(-\eta >\epsilon')$, $\rho$
is positive, and $\eta$ and $b\rho+V'$ are  negative, cf. (\ref{eq:2}).

Clearly (\ref{eq:Heineg2}) yields
\begin{equation}
  \label{eq:negmom2}
  \int_1^{\infty} |F'(B_{E'}<2\epsilon{})|F(-\eta >\epsilon')r^{-1}\;dt<\infty.
\end{equation} 

Taking the freedom of varying $E'\in (E_d,E_{\crt}]\cap \mathcal E$ and
the monotonicity in $\rho_{E'}$ into
account we conclude (\ref{eq:negmom}) from (\ref{eq:negmom2}),
cf the proof of \cite [Proposition 3.5] {CHS}.
 
\noindent {\bf Step II} For any sufficiently small $\epsilon>0$ we shall prove the bound 
\begin{equation}
  \label{eq:posmom}
  \int_1^{\infty} F(B_{E_{\crt}}<\epsilon{})F(B_{E_{1}}>\epsilon{})F(\eta >\epsilon')r^{-1}\;dt<\infty;\;\epsilon'>0.
\end{equation}

Consider for $E'\in [E_1,E_{\crt}]$ the observable 
  $$\Phi=F(B_{E'}<\bar \epsilon)F(\eta >\epsilon')=F_1F_2; \;\bar
  \epsilon \in [\epsilon/2, 2\epsilon].$$ As
    before we may assume that $\epsilon'>0$ is ``small''. We
  compute using (\ref{eq:comp2})
\begin{equation}
  \label{eq:Heineg}
 {\bf d}\Phi= F'(B_{E'}<\bar \epsilon{})(T_1+T_2)F_2-F_1F'(\eta >\epsilon'){(b+\eta)\rho+V'\over r},
\end{equation} Now we may bound the term with $T_1$ from below as 
\begin{equation}
  \label{eq:T1}F'(B_{E'}<\bar \epsilon{})T_1F_2\geq |F'(B_{E'}<\bar \epsilon{})|F_2{c(E'-E)\eta\over r}
\end{equation} for some $c>0$. For the   term with $T_2$ we have  for some $C>0$ 
\begin{equation}
  \label{eq:T2}F'(B_{E'}<\bar \epsilon{})T_2F_2\geq -|F'(B_{E'}<\bar \epsilon{})|F_2{C\bar \epsilon{}\eta\over r}.
\end{equation} 
Finally the last term on the right hand side of (\ref{eq:Heineg}) is
non-negative, cf. (\ref{eq:2})

We conclude that for $\bar \epsilon{}\sim \epsilon<<(E_1-E)$ there is
a $\delta>0$ such that  
\begin{equation}
  \label{eq:Heineg3}
 {\bf d}\Phi\geq |F'(B_{E'}<\bar \epsilon{})|F_2{\delta\epsilon'\over r}.
\end{equation} From this we conclude (\ref{eq:posmom}) by varying $E'\in [E_1,E_{\crt}]$
and $\bar\epsilon \in [\epsilon/2, 2\epsilon]$.

\noindent {\bf Step III} By combining (\ref{eq:negmom}) and
(\ref{eq:posmom}) with energy conservation we obtain that for all
sufficiently small $\bar \epsilon, \epsilon>0$ 
\begin{equation}
  \label{eq:belowcrregion}
  \int_1^{\infty} F(B_{E_{\crt}}+\epsilon+\bar \epsilon <\epsilon{})F(B_{E_{1}}>\epsilon{})r^{-1}\;dt<\infty.
\end{equation}

\noindent {\bf Step IV}  It remains to establish an estimate for the
region $-2\epsilon<B_{E_{\crt}}<\epsilon{}$. For that we introduce the observable
 $$\Phi=\eta
 F(B_{E_{\crt}}<\epsilon)F(B_{E_{\crt}}+2\epsilon>\epsilon)=\eta F_1F_2.$$ 
We
  compute 
\begin{equation}
  \label{eq:Heineg5}
 {\bf d}\Phi= -{(b+\eta)\rho+V'\over r}F_1F_2+\eta
 F_2F'(B_{E_{\crt}}<\epsilon{}){\bf d}B_{E_{\crt}}+\eta F_1{\bf d}F_2.
\end{equation} By Step III the last term is integrable. We insert for
each of the other two terms
$I=F(-\eta >\epsilon') +F(\eta >\epsilon')+R$. The contributions from
$F(-\eta >\epsilon')$ and $F(\eta >\epsilon')$ are integrable due to
Step I and II (with $\epsilon\to 2\epsilon$). But for the contributions
from $R$ we use  (\ref{eq:comp2}) to conclude that for some  positive $\delta$
\begin{align}
  \label{eq:Heinegbou}
 -{(b+\eta)\rho+V'\over r}F_1F_2R&+\eta F'(B_{E_{\crt}}<\epsilon{})(T_1+T_2)F_2R\nonumber\\\geq {\delta\over
   r}F_1F_2R&+0={\delta\over
   r}F_1F_2R;
\end{align} here we used again (\ref{eq:2}) and the arguments in
Step II. By replacing $R\to I$ (by the same arguments as before) we conclude that 
\begin{equation}
  \label{eq:Heineg55}
 {\bf d}\Phi\geq F_1F_2{\delta \over
   r} +\text{ integrable terms}.
\end{equation} 

>From (\ref{eq:Heineg55}) we obtain
\begin{equation}
  \label{eq:crregion}
  \int_1^{\infty} F(B_{E_{\crt}}<\epsilon)F(B_{E_{\crt}}+2\epsilon>\epsilon)r^{-1}\;dt<\infty.
\end{equation}

Clearly \ref{it:int}) follows from the bounds (\ref{eq:belowcrregion}) and (\ref{eq:crregion}).
\end{proof}

\begin{cor}
  \label{cor: completeness} For all scattering orbits with energy
  $E\in \mathcal E$ one of the following two possibilities occur:
  \begin{enumerate}[\normalfont (i)]
  \item \label{it: per0}$\lim_{t\to \infty}|\rho(t)-\rho_E(\theta(t))|=\lim_{t\to \infty}|\eta(t)-\eta_E(\theta(t))|=0.$
\item \label{it: zero0}$\lim_{t\to \infty}\Big\{\eta^2(t)+\Big(b(\theta(t))\rho(t)+V'(\theta(t))\Big)^2\Big\}=0.$
  \end{enumerate}
  \begin{proof} From Proposition \ref{prop:sepa } we learn that either
   \begin{enumerate}[\normalfont 1)]
  \item \label{it: per}$\lim_{t\to \infty}|\rho(t)-\rho_E(\theta(t))|=0,$ or 
\item \label{it: zero}$\liminf_{t\to \infty}\big
  (\rho(t)-\rho_{E_{\crt}}(\theta(t))\big )>0.$
  \end{enumerate}
In the case of \ref{it: per}) we have  (\ref{it: per0}). In the case
of  \ref{it: zero}) we change  $b\to \bar b$ in a small neighborhood of some
$\theta_{\crt}\in T_{\crt}$ such that $\int _0^{2\pi}\bar bd\theta =0$. We
can then proceed as in the proof of Proposition \ref{lemdoi} with $b\to \bar b$ using Proposition \ref{prop:sepa }
to treat errors involving the expression  $b-\bar b$. 
  \end{proof}
\end{cor}

\section{Quantum mechanics in the  general setting} 
\label{Quantum mechanics in the  general setting}

We shall outline how the methods used in the zero flux case to prove
asymptotic completeness may be
modified using now the classical theory of Section \ref{The general
  classical case}.

In addition to Condition  \ref{cond:torus} we shall in this section impose the
following condition.

\begin{cond}[No large oscillation]
\label{cond: no large oscillation} For 
all orbits in the reduced phase space $\T^*_E=\{z\in
\T^* |\;h(z)=E\}$ with energy $E\in I$,
\begin{align}
  \label{eq:rotafree}
  \inf _{\tau_2>\tau_1}\Big \{&\big (\sqrt
  {2(E-V(\theta(\tau_2)))}+\rho(\tau_2)\big )\nonumber\\
&+\big(\sqrt {2(E-V(\theta(\tau_1)))}-\rho(\tau_1)\big)\Big \}>0.
\end{align}
\end{cond}

We remark that Condition \ref{cond: no large oscillation} is fulfilled  
1) for 
$E>E_d^+$ in the case where $\mathcal E^+\neq \emptyset$ (proved by
checking that  $\e^{\int \rho}(\rho-\rho_{E'}^+)$ is increasing in
$\tau$  for $E'<E$ in $\mathcal E^+$), 
2) in general for
all high enough energies or if $b=0$ (proved by using that $\e^{\int \rho}\rho$ is increasing
in $\tau$). 

Clearly  Condition \ref{cond: no large oscillation} excludes
Proposition \ref{prop:clas-gen} \ref{it:cl1}).
Consequently, under the conditions of Lemma \ref{lem:-=+}, the condition \eqref{eq:rotafree}
is false for all orbits with energy $E<E_d^+=E_d^-$ such that $\mathcal
F(E)=\emptyset$. Notice that in this case  indeed Proposition
\ref{prop:clas-gen} \ref{it:cl1}) occurs.  An example is the case, $V=0$ and $b<0$, treated in
\cite{CHS}. We remark that Condition \ref{cond: no large oscillation}
implies Condition \ref{cond:noexcep}. 

Due to the discreteness  of $\mathcal E _{\exc}$, cf. Appendix
\ref{Collapsing orbits}, Condition \ref{cond: no large oscillation}
may be relaxed: For this section it would suffice to  impose the condition that
 Proposition
\ref{prop:clas-gen} \ref{it:cl1}) as well as its negative time version,
cf Remark \ref{remark:negversion}, do not occur.

\begin{cond}[No singular cycles] \label{cond: no  singular cycles} The interval $I$ does not contain an
  energy at which there exists a singular cycle.
\end{cond} 

Notice that by Propositions \ref{prop:unique} and \ref{prop:uniquelower} singular
cycles may occur at most at two energies in the case $V=0$.

\subsection{A partial ordering} 
\label{A partial ordering}

With Conditions \ref{cond:torus}, \ref{cond: no large oscillation} and
 \ref{cond: no  singular cycles}, and \eqref{eq:nondege1} and
 \eqref{eq:nondege2} at all $E\in I$, we can order the set $\mathcal
F^+_{\sa}(E)$ for any such $E$ as follows: We write $z\prec \tilde z$ for $z, \tilde
z \in \mathcal
F^+_{\sa}(E)$ if and only if,  either $z= \tilde z$,  or there exists a chain of orbits $\gamma_1,
\dots,\gamma_n$, $n\in \N$, each orbit solving \eqref{eq:equamotion}
and possessing energy $E$,
such that 
\begin{align}
  \label{eq:cyc+sa}&\lim_{\tau\to
    +\infty}\gamma_1(\tau)=\tilde z,\;\lim_{\tau\to
    -\infty}\gamma_n(\tau)=z,\\
\label{eq:cyc+sa2}&\lim_{\tau\to
    -\infty}\gamma_j(\tau)=\lim_{\tau\to
    +\infty}\gamma_{j+1}(\tau)\in \mathcal
F^+_{\sa}(E);\;1\leq j<n\;(\text{for }1<n).
\end{align}

We claim that this recipe is a partial ordering of $\mathcal
F^+_{\sa}(E)$: The only non-trivial property to check is
anti-symmetry. So suppose $z\prec \tilde z$ and $\tilde z\prec
z$, then we need to verify that $z= \tilde z$, which in turn
amounts to showing that there are no loops. A loop is given  by a
chain  $\gamma_1,
\dots,\gamma_n$, $n\in \N$, such that the two limits in \eqref{eq:cyc+sa}
are replaced by the same value  $z =\tilde z\in \mathcal
F^+_{\sa}(E)$ (while \eqref{eq:cyc+sa2} is left unchanged). Suppose
first $n=1$, so that we 
look at  a homoclinic orbit $\gamma$. Since ${d\over d\tau} (\rho-\tilde
b)=\eta^2$ there exists $m\in \N$ such that 
\begin{equation}
  \label{eq:mult}
 [\theta(\tau)]_{-\infty}^{\infty} =2\pi m.
\end{equation}

Now, due to Conditions  \ref{cond:torus} and \ref{cond: no large oscillation},
$\rho(\tau)> -\sqrt
  {2(E-V(\theta(\tau)))}$ for all $\tau$. This means that
  $\gamma(\tau) \in \T^{*,+}(E)$ (see \eqref{eq:resimpphase} for definition).

The topological structure of $\T^{*,+}(E)$ is an annulus; thus $\gamma$
is a non--self--intersecting closed orbit in an annulus. The equation
of motion for the angle is $\theta'=\eta$. Under these
circumstances  the only
possibility  for angular increment along the orbit is $m=1$ in
\eqref{eq:mult}, cf. \cite [Proposition 5.20]{F},  and  $\eta(\tau) \geq{ 0}$  for all $\tau$. 
(Notice that $\{\eta=0\}$
divides $\T^{*,+}(E)$ into two separate annuli with opposite direction
of flow, and that $\eta(\tau) \leq{ 0}$ is excluded by the
flux-condition $\tilde b(2\pi)\leq 0$.)  The existence of such orbit
is not consistent  with Condition \ref{cond: no 
  singular cycles}.

In the case $n>1$  the  total angular increment
for the loop  is again given by $2\pi m$ for some $m\in \N$,
and we may argue as above. It does not exist.

The partial ordering defines the {\it order} of any given $\tilde
z \in \mathcal
F^+_{\sa}(E)$ as the largest possible $n$ for which there exists $
z \in \mathcal
F^+_{\sa}(E)$ and a chain of orbits $\gamma_1,
\dots,\gamma_n$ such that \eqref{eq:cyc+sa} and \eqref{eq:cyc+sa2}
hold. If $\tilde
z$ is minimal the order is by definition $n=0$.

We order $\mathcal
F^-_{\sa}(E)$ in the same way as done above for $\mathcal
F^+_{\sa}(E)$ (replacing $\mathcal
F^+_{\sa}(E)$ in \eqref{eq:cyc+sa2} by  $\mathcal
F^-_{\sa}(E)$). The arguments for anti-symmetry are similar. Instead of
\eqref{eq:resimpphase} we consider the set $\T^{*,-}(E)$ of \eqref{eq:resimpphase2}.

\subsection{The case $\mathcal E^+\cup \mathcal E^-= \emptyset$} 
\label{The case mathcal E^+= emptyset}

Suppose the conditions of Subsection \ref{A partial ordering}. Using
Proposition \ref{prop:clas-gen} and the   orderings of
Subsection \ref{A partial ordering} 
we shall outline a proof of the limiting absorption principle and
asymptotic completeness 
in the case  where $\mathcal E^+\cup \mathcal E^-=\emptyset$
along the lines of Sections \ref{Propagation of
  singularities}--\ref{Projection=0} and \ref{Asymptotic completeness}. 
Notice that  Subsection
 \ref{Wave front set1} generalizes verbum verbatim, and that
 Proposition \ref{prop:clas-gen} \ref{it:cl2}) (and its negative time
 version, cf. Remark \ref{remark:negversion}) occurs for all orbits with energy $E\in I$.

\subsubsection{Bounds at $\mathcal F^-_{\sa}$}
We shall  modify Section \ref{Saddlepo} using now the more refined
ordering at $\mathcal
F^-_{\sa}(E)$ introduced above. 

So suppose  first that $ z(E) \in \mathcal
F^-_{\sa}(E)$   is minimal  at energy $E=E_0\in I$ in the sense of
Subsection \ref{A partial ordering}. Then we use  the same
contructions as in Section \ref{Saddlepo}: We aim again at proving 
\eqref{eq:bound4} by using the quantizations of the symbols in
\eqref{eq:propobs2}  with $f$ 
supported in an interval $[E_0-\delta,E_0+\delta]$ with
$\delta<<\epsilon$. We define for $\epsilon>0$
\begin{equation*}
  K=\Big \{z\in \T^*_-|\;{\epsilon\over 4}\leq l_{E_0}\leq2\epsilon,\; h=E_0,\;
  \dist (z,M^u_{E_0})\geq{\epsilon\over 2}\Big  \}.
\end{equation*}
If $\epsilon$ is small enough,  then for all  $z\in K$ we cannot have
$\phi_\tau(z)\to \tilde z \in \mathcal F^-_{\sa}$ for $\tau\to
-\infty$, 
due to the fact  that $z(E_0)$ is minimal. Here we  used
Proposition \ref{prop:clas-gen}. In fact, since Proposition
\ref{prop:clas-gen} \ref{it:cl2}) and its negative time version are  valid, 
in the far past $\phi_\tau(z)$  will be close to $\mathcal
F^-_{\so}(E_0)$. Notice that also Condition \ref{cond:noexcep} is
used  here (recall that it is a consequence of Condition  \ref{cond: no large
  oscillation}). We choose
$\tau_0<<-1$ such that we have a good wave front set bound of an open
neighborhood $U_{\tau_0}\subseteq \T^*$ of the
compact set $\phi_{\tau_0}(K)$. Define
$U_1=\phi_{-\tau_0}(U_{\tau_0})$,

\begin{equation*}
  U_2^\delta=\{z\in \T^*_-|\;{\epsilon\over 4}< l_{h}<2\epsilon,\; |h-E_0|<2\delta,\;
  \dist (z,M^u_{h})<{\epsilon}\};\;\delta>0,
\end{equation*} and 
\begin{equation*}
  K^\delta=\{z\in \T^*_-|\;{\epsilon\over 2}\leq l_{h}\leq \epsilon,\; |h-E_0|\leq\delta,\;
  \dist (z,M^u_{h})\geq{\epsilon}\};\;\delta\geq 0.
\end{equation*}

Due to  Lemma \ref{lem:Stability2}
\begin{equation*}
  \tilde K :=\{z\in \T^*_-|\;{\epsilon\over 2}\leq l_{h}\leq\epsilon,\;
  |h-E_0|\leq \delta\}\subseteq U_1\cup  U_2^\delta.
\end{equation*}
Next pick non--negative $\psi_1,\psi_2\in C^\infty (\T^*)$ subordinate to this
covering, that is $\supp \psi_1 \subseteq U_1$, $\supp \psi_2
\subseteq U_2^\delta$ and $\psi_1+\psi_2=1$ on $\tilde K$.

We  decompose 
$F'(l_h<\epsilon)f(h)=(\psi_1+\psi_2)F'(l_h<\epsilon)f(h)$.
The propagation of singularity theorem
applies to the contribution
from the term involving $\psi_1$. Again  the contribution from the term
involving $\psi_2$ has the right sign. We conclude 
\eqref{eq:bound4} if the order of $z(E_0)$ is zero. 

We  repeat the
analysis at   elements $z \in \mathcal
F^-_{\sa}(E)$ of positive order measured at $E=E_0$, proving ``good''
bounds there too. As in Subsection \ref{Saddlepo} this is done by  grouping  the elements according to their order 
and treat recursively the groups one by one in increasing
order. Finally we vary $E_0\in I$ and conclude \eqref{eq:bound4} for
any 
$f\in C_0^\infty(I)$.

\subsubsection{Bounds at $\mathcal F^+_{\sa}$}

We may mimic the analysis of the previous subsubsection at  elements $z \in \mathcal
F^+_{\sa}(E)$ starting again at elements of order zero measured at
$E=E_0$ and treating 
elements of higher order recursively. We use the observable of Subsection \ref{Wave
  front set3},
and we may obtain  \eqref{eq:bound11}, in fact without the last term
to the right. Notice at this point that we do not have 
\eqref{eq:awayfromF}; we use  at each step a stronger local version  derived by
propagation of the established bounds 
at $\mathcal F^-$ and at the saddles in $\mathcal
F^+_{\sa}(E)$ of lower order. 

\subsubsection{Bounds at $\mathcal F^+_{\si}$ and away from $\mathcal F$}

We may obtain  \eqref{eq:bound11} at $\mathcal F^+_{\si}$ by using the
propagation of singularity result. Similarly we may prove
\eqref{eq:bound11} 
 away from $\mathcal F$. 

However we shall also need a stronger type of bound like 
\eqref{eq:awayfromF} for the limiting absorption principle,
cf. \eqref{eq:bound44333}. The idea of the proof is to  look at the construction \eqref{eq:propobs2} with
$t=\tfrac{1}{2}$ rather than $t>\tfrac{1}{2}$
as above, first at points in $\mathcal F^+_{\sa}$ (treated in increasing order) and then at points in $\mathcal
F^+_{\si}$. Whence by repeating  the analysis
above we may obtain $WF^{-1/2}$--bounds in deleted
neighborhoods of $\mathcal F^+$. Using again  the
propagation of singularity result and the (strong) bounds at $\mathcal
F^-$ we consequently obtain the bound
\begin{align}
  \label{eq:awayfromF2}
  \|{\Op}(\chi) u\|^2_{-1/2}&\leq
  C_1\|u\|_{-1}^2+\sigma \|u\|_{-t}^2
  +C_2\sigma^{-1}\|v\|^2_t;\\&\supp \chi \subset \T^*\setminus
  \mathcal F^+,\;\sigma>0,\;t\in (1/2,1).\nonumber
\end{align}

\subsubsection{Limiting absorption principle} We may now mimic
Section \ref{LAP} to show the analogue of Theorem \ref{thm:LAP}.

\subsubsection{Preliminary estimates}
 
Using \eqref{eq:lapfinal} and \eqref{eq:awayfromF2} we obtain the
following bound away
from eigenvalues:
\begin{equation}
  \label{eq:awaycrit}
 WF^{-{1\over
    2}}(R(E+i0)v)\subseteq\mathcal F^+\text{ for all }v\in
    L^{2,t};\;t>1/2.
\end{equation}

Given \eqref{eq:awaycrit} we may obtain (by the Fourier transform) a
version of \eqref{eq:bounde}; to the left we have the expectation of
${\Op}(\chi) $ with $\chi$ as in \eqref{eq:awayfromF2}, however $\|\psi\|$
to the right needs to be replaced by  $\|\psi\|_t$ with
$t>\tfrac{1}{2}$. Since
\eqref{eq:bounde} is used to derive the integral estimates
\eqref{eq:boundrho} and \eqref{eq:boundrovertmin} the right hand side
of those needs to be changed similarly.
Whence we have a slightly weaker version of Lemma \ref{lemma:1}. We
may verify    Lemma 
\ref{lemma:2} as before for $\psi\in L^{2,t}$ and hence for $\psi\in L^{2}$. 
\subsubsection{Decomposition into channels formula}\label{Decomposition into channels formula}
As for Section
\ref{Projections} we need to proceed differently; it is not possible
to show the existence of the projections $P_j$ along the same
line. This is due to the fact that we only have  integral bounds  with $\|\psi\|_t$
appearing to the right. 

Instead we may start by defining $P_j$ at the
stable fixed points. We define in this case, in terms of the operator
$L_1(t)=\Op(l_1(t))$ with $l_1(t)$ specified in \eqref{eq:l_111}, 
\begin{equation}
  \label{eq:Pnew}
  P_jf(H)\psi =\lim_{t\to \infty}\e^{itH}L_1(t)\e^{-itH}f(H)\psi, 
\end{equation}
for $\psi\in L^{2,t};\;t>\tfrac{1}{2}$. Here we impose
\eqref{eq:epsilons}; consequently due to
Lemma  \ref{lemma:2} and the proof of Lemma  \ref{lemma:extrabounds}
the right hand side is independent of the
$\epsilon$'s By definition $P_j$ is the smallest orthogonal
projection obeying \eqref{eq:Pnew}. We notice that indeed the limit
exists,  cf. \eqref{eq:poisF_4} and 
\eqref{eq:poisF_1}--\eqref{eq:boundb12} (notice that $\|\psi\|$
appears in the integral bounds). Whence   (by  Lemma
\ref{lemma:2} and the proof of Lemma  \ref{lemma:extrabounds}) $P_j$
is well-defined.

We are left with proving the formula 
\begin{equation}
    \label{eq:completejssmod}
  1_{I}(H)=P_I:=\sum_{z_j(\cdot)\in \mathcal F_{\si}^+}P_j.  
  \end{equation}

This may by done by  looking at  the evolution of states of the form 
\begin{equation*}
 \psi_{\exc}=(1_{I}(H)-P_I)f(H)\psi;\;\psi\in
L^{2,t},\; t>\tfrac{1}{2}.
\end{equation*} We may decompose
\begin{equation}\label{eq:sadsta}
 \psi_{\exc}(t)=\e^{-itH}\psi_{\exc}\approx \sum_{z_j(\cdot)\in \mathcal
 F_{\sa}^+}\Op (\chi_j)\e^{-itH}f(H)\psi,
\end{equation} where $\chi_j$ is a symbol localized near
 $z_j(\cdot)$.  

At this stage  we can show that indeed 
\begin{equation*}
 \lim
_{t\to \infty} \|\e^{itH}\Op (\chi_j)\e^{-itH}f(H)\psi\|=0 
\end{equation*} by  
invoking 
\cite[Theorem 1.2]{HS2}. 

\subsubsection{Completeness}\label{Completeness}
Given    \eqref{eq:completejssmod} 
we may proceed as in the case of zero flux in Section                         
\ref{Asymptotic completeness},                in particular we may obtain
 analogues of Theorems \ref{thm:-11} and
\ref{thm:-111}. 

%Notice the following consequence of
%\eqref{eq:awaycrit} in combination with 
%\cite{HS2}. 
%or all $\chi$ supported away from  $\mathcal F^+_{\si}$ 
%begin{equation}\label{eq:awaysad}
%\|\Op(\chi) \e^{-itH}f(H)v\| \to 0\text{ for } t\to \infty.
%end{equation}

\subsection{The case $\mathcal E^+\neq \emptyset$} 
\label{The case mathcal E^+not= emptyset}

Suppose the conditions of Subsection \ref{A partial ordering}. 
 In the case $\mathcal E^+\neq \emptyset$ we shall discuss  quantum
completeness above $E_d^+$. The limiting absorption principle follows
from quantizing \eqref{eq:mourre}, cf. \cite {CHS}. For completeness
in $\mathcal E^+$ we refer to Subsection \ref{Completeness in
  interval} and \cite {CHS}. Notice that  the arguments given in Subsection \ref{Completeness in
  interval}  generalize to quantum mechanics. Consequently the outcome is analogues of Theorems \ref{thm:-11} and
\ref{thm:-111} for the part of the wave packets at fixed    points and
spiraling  behaviour as in \cite{CHS} for another part,
 cf. \eqref{eq:acomega2} and \eqref{eq:chan000}.

So we focus on the case where $I\cap \overline{\mathcal
  E^+}=\emptyset$. (Notice  that this amounts to looking at
  $I\subseteq (E_e^+, \infty)$ if $V=0$.)

Suppose first that also $I\cap {\mathcal E^-}=\emptyset$. Then again
Proposition \ref{prop:clas-gen} \ref{it:cl2}) and its negative time
version  hold, and we obtain the wave front set bound
\eqref{eq:awaycrit} and completeness as before.

If $I\cap {\mathcal E^-}\neq \emptyset$ the incoming spiral may seem
to cause problems in establishing \eqref{eq:awaycrit}. However 
 the Mourre estimate gives a good bound of
$Au$ where $A=\Op (F(r>1)f(h)F(\rho-\rho_E^-<\epsilon))$ for some $\epsilon>0$ and $u=R(E+i0)v$, due to the (classical) inequalities 
\begin{equation}
  \label{eq:rhosin}
  \rho-\epsilon\geq\rho^+_{E'}\geq  \rho_E^-;\;E'<E\text{ in
  }\mathcal E^+\text{ and } \epsilon=\epsilon(E')>0.
\end{equation} Here we assume that $\rho$ belongs to an orbit with
energy $E\in I$. The first  inequality  is due to the Mourre estimate
while 
the second follows from Lemma \ref{lem:-<+}. By a ``good'' bound we mean
the statement $WF^{-s}(Au)=\emptyset$
for  some $s<{1\over
    2}$. Technically the implementation of \eqref{eq:rhosin} for this
  purpose may
be done by considering the  symbol
$$F(r>1)f(h)\big(r(3\epsilon-\rho+ \rho^+_{E'})\big )^{1-2s}F(\rho-
\rho^+_{E'}<2\epsilon);$$
 we skip the details. See \cite{GIS} for a somewhat similar
 construction and procedure. 

The outcome is again analogues of Theorems \ref{thm:-11} and
\ref{thm:-111}.

\begin{remark}
\label{rem: belowma} Quantum mechanics for  energies below $\max V$ is
easier to deal with, cf. Remark \ref{remark:belowener},  and will not be
discussed.
\end{remark}
\appendix

\section{Collapsing and exceptional classical orbits}
\label{Collapsing orbits}

Under  Condition \ref{cond:torus}  we shall study the collapsing
orbits with energy $E\in I$ in the ``torus'' $\T^*_E=\{z\in
\T^* |\;h(z)=E\}$. By
definition a 
{\it collapsing  orbit} is an integral curve of  the system \eqref{eq:equamotion} for which the quantity $r=\exp (\int_0^\tau \rho\;d\tau)\to 0$
for $\tau\to +\infty$. We encountered     such 
an 
example
     in Remark \ref{rem:excep}. The set of $z's$ in $\T^*_E$ in the
     range of a collapsing  orbit is denoted by 
     $A_{\col}(E)$.

We introduce a new  independent variable $\psi$ by
\begin{equation}
  \label{eq:psi} (\cos \psi,\sin \psi)={(\eta,\rho)\over \sqrt{2(E-V)}},
\end{equation}
and compute
\begin{equation}
  \label{eq:eqnpsi}
\psi'=  \sqrt{2(E-V)}\cos \psi+b+{V'\over \sqrt{2(E-V)}}\sin \psi=:F_2.
\end{equation}

Combining \eqref{eq:eqnpsi} and 
\begin{equation}
   \label{eq:eqnpfi}
\theta'=  \sqrt{2(E-V)}\cos \psi  =:F_1
\end{equation}
leads to the following formula for the Jacobian of the flow
$\phi_\tau=(\theta(\tau),\psi(\theta))$ for the
vector field $F=(F_1,F_2)$ on $\T^*_E$,
\begin{equation}
  \label{eq:comp55}
  \exp\Big (\int _0^\tau\nabla \cdot F\;d\tau\Big )=\exp\Big (\int_0^\tau -\rho\;d\tau\Big )=r^{-1}.
\end{equation}

Hence for any measurable set $A\subseteq \T^*_E$
\begin{equation}
  \label{eq:change}
  \int_{\phi_\tau (A)}d\theta d\psi=\int_A r(\tau;\theta,\psi)^{-1}d\theta d\psi.
\end{equation}

\begin{lemma}\label{lem:col}
  The set $A_{\col}(E)\subseteq \T^*_E$ has (relative)
   measure zero. 
\end{lemma}
\begin{proof}
  We combine \eqref{eq:change} and  the Fatou Lemma. 
\end{proof}

Under more conditions the structure of  $A_{\col}(E)$ can be specified
further. For example under the conditions of Proposition
\ref{lemdoi}, $A_{\col}(E)$ is the union of stable manifolds/curves at
$\mathcal F^-(E)$. A less obvious example is provided by the following
result.

\begin{lemma}\label{lem:col2}
  Suppose $\int_0^{2\pi}b\;d\theta<0$ and $\mathcal E^-\neq
  \emptyset$. Then for any $E\in \mathcal E^-$ the associated orbit
  $\theta^-\to \gamma^-(\theta^-)=(\theta^-, \eta^-(\theta^-),\rho^-(\theta^-))$ is collapsing, and if
  $\gamma=(\theta, \eta, \rho)$ is another collapsing orbit
  with energy $E$ then 
  \begin{equation}
    \label{eq:inincom}\rho(\tau)\leq \rho^-(\theta(\tau))).
  \end{equation}
Suppose in addition that $\mathcal F^-(E)=\emptyset$ then
\begin{equation}
  \label{eq:singl}
  A_{\col}(E)=\{\gamma^-(\theta^-)|\theta^-\in \R\}.
\end{equation}
\end{lemma}
\begin{proof}
 As for $\gamma^-$ we have $r=\exp (\int_0^\tau \rho\;d\tau)= \exp
 (\int_0^\tau {\rho^-\over \eta^-}\;d\theta^-)\to 0$ for $ \tau \to +\infty$. 

Next we look at $A(\tau)=A=r(\rho^-(\theta)-\rho)$  defined in terms
of $\rho^-$ and the given collapsing orbit $\gamma$. We compute
$A'\leq 0$, and  since $A\to 0$ for $\tau\to +\infty$ we conclude
that $A\geq 0$. Whence \eqref{eq:inincom} holds.

Finally under the additional condition $\mathcal F^-(E)=\emptyset$ the
function $f(\theta)=f=b\sqrt{2(E-V)}-V'$ has a definite sign. As in
the proof of Proposition \ref{prop:clas-gen} we see that
$\theta(\tau)\to +\infty$. We distinguish between the cases 1) $\eta(\tau_n)=0$ along a sequence
$\tau_n\to +\infty$,  2) $\eta>0$
for all $\tau$ large enough.

We show that 1) does not occur: Since $\eta'=-b\rho-V'$ at any
crossing given by   $\eta=0$  and the sign of $\eta'$ must alternate for two
consequtive crossings, we conclude from the fact that $f$ has a
definite sign that the sign of $\rho$ cannot be negative for two
consequtive crossings. In particular $\rho=\sqrt{2(E-V)}$ occurs,
which contradicts  \eqref{eq:inincom}.

In the case 2) we consider  $A(\tau)=A=\{\exp(\int{
\rho^-\over \eta^-}\;d\theta)(\rho-\rho^-)\}(\theta^-)$ where ${\theta^-}'=\eta^-(\theta^-)$
and $\rho$ is considered as a function $\theta$. The latter is
legitimate  since $\theta=\int \eta\;d\tau$ increases monotonely to
$\infty$; in particular  we may consider $\tau$ as a function of the angle. Again we
 have $A'\leq 0$ and $A\to 0$ for $\tau\to +\infty$, whence  $A\geq
 0$. In combination with \eqref{eq:inincom} we conclude that 
\begin{equation}
    \label{eq:inincom2}\rho(\tau)=    \rho^-(\theta(\tau)))\text{ for
      all large } \tau.
  \end{equation}
 From \eqref{eq:inincom2} we  obtain \eqref{eq:singl} by noting that a
 similar relation holds
 for $\eta$ and $\eta^-$ (since $\eta>0$ eventually).
\end{proof} 

 We complete the appendix by showing discreteness of the exceptional
 set of energies $\mathcal E_{\exc}$ of Condition \ref{cond:noexcep}
 (under Condition \ref{cond:torus}), cf. Remark \ref{rem:excep}. 

\begin{lemma}\label{lem:col3} The set $\mathcal E_{\exc}$ is discrete
 in $I$. 
\end{lemma}
\begin{proof} Suppose $E_1\in \mathcal E_{\exc}$. Let $\gamma$ be a
  corresponding heteroclinic orbit and $z_1=\gamma(0)$. All that needs
 to be checked is that  $\gamma$ splits at $z_1$ under perturbation
 with respect to variation of the energy. More precisely we may
 introduce a (fixed) transversal curve at $z_1$ and look at its  intersection
 with the unstable orbit from one of the saddles (as a function of the
 energy-parameter) and similarly look at its  intersection
 with the stable orbit from the other saddle. The homoclinic orbit
 splits if for all energies in a small deleted  neighborhood of $E_1$
 the two points of intersection are different. There is a criterion
 for splitting due to Melnikov \cite {Meln}. Here we refer the reader
 to \cite [Section 6.1]{C}. It suffices to check that the ``Melnikov
 integral'' \cite [(6.12)]{C} (with $\lambda_j=E-E_1$) is nonzero. Due
 to (\ref{eq:comp55}) this amounts in
 our case to checking that
 \begin{equation*}
   \int_{-\infty}^{\infty} r\big (2V'\omega^{-2}\cos \psi \sin
   \psi +b\omega^{-1} \cos \psi \big )d\tau\neq 0;
 \end{equation*} here $\omega=\sqrt {2(E-V)}$ and the integrand is
 evaluated at the unperturbed heteroclinic orbit $\tau\to
 \gamma(\tau)$.

This integral may be computed as follows: We express the last term as 
\begin{equation*}
 b\omega^{-1} \cos \psi =\omega^{-2} \frac{d}{d\tau{}}\rho{}-\omega^{-2} \eta^2,  
\end{equation*}
and treat the first term to the right after substitution by integrating  by parts. Since $r\to 0$ for
$|\tau|\to \infty$ boundary terms disappear. After a cancellation the
integrand simplifies yielding the following expression
 \begin{equation*}
   -\int_{-\infty}^{\infty} r\omega^{-2}\big (\eta^2+\rho^2 \big )d\tau=-\int_{-\infty}^{\infty} rd\tau
 \end{equation*}   
 for the above integral. Obviously it is negative.
\end{proof} 

\begin{remark} For the case $V=0$ one can construct examples of $b$'s
  for which 
  $\mathcal E_{\exc}\neq \emptyset$ as follows: Fix any non-constant
  $b$ with $\int ^{2\pi}_0b\;d\theta{}=0$, and consider
  $b_\kappa=b+\kappa$ with $\kappa\in ( -2\max |b|,0)$. Claim:  There 
  exist $\kappa_0\in (-2\max |b|,0)$, $E_{0}>0$,  and angles $\theta_1$ and
  $\theta_2$ such
  that  
  upon replacing  $b\to b_{\kappa_0}$ and
  $E\to E_{0}$ in \eqref{eq:peri} indeed the equation has a periodic solution $\rho=\rho_{0}$ obeying   $\rho(\theta_1)=\sqrt{2E_{0}}$  and 
$\rho(\theta_2)=-\sqrt{2E_{0}}$. Given this claim we  solve
  $\frac{d}{d\tau{}}\theta{}=\sqrt {2E_0-\rho^2}=:\eta$ to obtain 
 an exceptional orbit. 

To outline a proof of  the claim  we introduce $f(E,\kappa)=\int ^{2\pi}_0{\rho_{E,\kappa}\over
  \eta_{E,\kappa}}\;d\theta$ for  $\kappa<-\max |b|$ and sufficiently large
  $E$'s. We look at the condition 
   $f(E,\kappa)=1$, for example, starting from  a fixed solution
  $\rho_{E,\kappa}$ and
  then 
trace the
 dependence of $\kappa$ using the implicit function
 theorem. Notice here the properties Proposition \ref{prop:interm} 
  \ref{it:133}), \ref{it:1344}) and \ref{it:135}. We may in fact
  compute using Proposition \ref{prop:interm} \ref{it:1344}), 
  $E'={d\over d\kappa}E=-\partial_\kappa f/\partial_Ef<0$.
 Let us denote by  $\kappa_0$ the right
  endpoint  of the
  maximal $\kappa$--interval legitimate for this procedure. We
  may use  the Arzel{a}--Ascoli
theorem to define a limiting solution $\rho=\lim_{\kappa\to \kappa_0^-}
  \rho_{E(\kappa),\kappa}$.  If  $\kappa_0=0$ we integrate \eqref{eq:peri}
  yielding  $\int ^{2\pi}_0\eta\;d\theta{}=0$, which
  means that $\rho$ is constant, and therefore in turn 
  $b=0$. By assumption this cannot be Consequently, indeed
  $\kappa_0<0$.

It is an open problem whether $\mathcal E_{\exc}\neq \emptyset$  may
occur  for the zero flux case.
 
\end{remark}
\section{A similar model}
\label{A similar model}

\label{exam:riema33} Consider the symbol $h$ on $(\mathbf{R}^{2}\setminus{\left\{
0\right\}})\times \mathbf{R}^{2}$
  \begin{equation*}
 h=h(x,\xi)=\tfrac
 {1}{2}g^{-1} \xi^2,   
  \end{equation*} where the conformal (inverse) metric factor is
 specified in polar coordinates 
 $x=(r\cos \theta, r\sin \theta)$ as
 $g^{-1}=e^f;\;f=f(\theta-c\ln r)$. We assume $f$ is a given smooth 
 non-constant $2\pi$--periodic function and that $c>0$. 
We introduce $v=(x_1-cx_2)\partial_{x_1}+(cx_1+x_2)\partial_{x_2}-c\xi
_2\partial_{\xi_1}+c\xi_1 \partial_{\xi_2}$. Computations show that  $v$
  and the Hamitonian vector field 
  $v_h$ fulfill the conditions of \cite[Appendix A]{HS2} along the positive
  orbit of $v$ originating at  $(r_0,0;\rho_0,c\rho_0)$; here $\rho_0=\sqrt
  {2E(1+c^2)^{-1}e^{-f_0}}$ where    
  $f_0 =f(\theta_0)$ is given in terms
  of any $r_0>0$ satisfying  the
  equation
  \begin{equation}
    \label{eq:237}
 -f'(\theta_0) =2c(1+c^2)^{-1};\;\theta_0=-c\ln r_0, 
  \end{equation}  and $E=h>0$ is arbitrary.  (Notice that there 
  are at least two solutions to \eqref{eq:237} for  
  all small as well as for  all large values of $c$.) The $x$--space part of the orbit (a geodesic) is the logarithmic spiral
 given by the equation $\theta-c\ln r= \theta_0$. 
 We  take  $S\subset \{(x,\xi)|\,x_2=0\}$, cf. \cite[Appendix A]{HS2},
  and compute the eigenvalues
  for the linearized reduced flow to be given by
\begin{equation} \label{eq:238}
 -\rho_0 \tfrac
 {1}{2} \Big \{1\pm \sqrt{1-2(1+c^2)^2f''_0}\Big \};\;f''_0=f''(\theta_0).
 \end{equation}  For $f''_0<0$ the
family of fixed points consists of  saddles. There are no resonances for ``generic''
values of $c$, and we also notice that  taking  $c\to 0$ in
\eqref{eq:237} and \eqref{eq:238} yields the formulas for the
corresponding homogeneous model (here the equations are  considered to be
 equations  in $c$ and $\theta_0$).

Finally, using the new angle $\tilde \theta= \theta-c\ln r$ one may 
conjugate to a homogeneous model. 
More precisely the relevant  symplectic change of variables is induced
(expressed here in terms of  rectangular coordinates) by the map $x\to \tilde x=(x_1g_1+x_2g_2,x_2g_1-x_1g_2)$, where
$g_1=\cos (c\ln |x|)$ and $g_2=\sin (c\ln |x|)$. One may check that  
$v\to \tilde v
:=\sum x_j\partial _{x_j}$, and that $h\to \tilde h$ given by
\begin{equation*}
\tilde h =\tfrac
 {1}{2}e^{f(\theta)}\Big ( \big \{(c\sin \theta +\cos
 \theta)\xi_1+(\sin \theta-c\cos \theta)\xi _2\big  \}^2+\{-\sin
 \theta  \xi _1 +\cos \theta  \xi _2\}^2 \Big );
\end{equation*}
 we changed notation back to the old one, $x=(r\cos \theta, r\sin
 \theta)$ for position and $\xi$ for momentum. 

Introducing  dual polar variables $\rho$ and $\l$ for $r$ and
$\theta$, respectively, 
the expression for this Hamiltonian simplifies as
\begin{equation*}
h=\tfrac
 {1}{2}e^{f(\theta)}\big ( (\rho -c\l /r)^2+ (\l /r)^2 \big )
\end{equation*}

Let us introduce a new angle $\psi$ by the relations 
\begin{align*}
  & \rho =\sqrt {2E(1+c^2)} e^{-f(\theta)/2}\cos \psi,\\
& (1+c^2)\l /r -c\rho =\sqrt {2E(1+c^2)} e^{-f(\theta)/2}\sin \psi.
\end{align*}

The equations of motion  are reduced in the variables $\theta$ and $\psi$:
\begin{equation}\label{eq:equamotion3}
\begin{cases}
\frac{d}{d\tau{}}\theta{}=\sin \psi\\
\frac{d}{d\tau{}}\psi{}{}=-\tfrac {1}{2} g(\theta)\cos \psi- \tfrac {1}{1+c^2}\sin \psi 
\end{cases}\;;
\end{equation} here $g(\theta)=f'(\theta)+2c/(1+c^2)$ and $\tau$
represents a ``new time''. 

Notice that the fixed points are given by $g(\theta)=0$ and $\psi \in
\pi \Z$, in agreement with \eqref{eq:237}.

To study \eqref{eq:equamotion3} it is useful to observe that the
observable $a_1=\exp\big ({-\tfrac {1}{2} \int_0^\theta g d\theta}\big )\cos \psi$
obeys
\begin{equation}
  \label{eq:apost}
  {d\over d\tau}a_1=\exp\Big ({-\tfrac {1}{2} \int_0^\theta g
  d\theta}\Big ) \dfrac {\sin ^2\psi }{1+c^2}.
\end{equation}

If $\sin \psi \neq 0$ for all $\tau$ we may consider $\psi $ as a
function of $\theta$ and look at 
\begin{equation}
  \label{eq:psired}
 {d\over d\theta}\psi= -\tfrac {1}{2} g(\theta)\cot \psi- \tfrac {1}{1+c^2}. 
\end{equation}

Equipped with \eqref{eq:apost} and \eqref{eq:psired} we can prove  the
following analogue of Proposition \ref{prop:clas-gen}. If
$\psi(\theta)$ is a $2\pi$-periodic solution to \eqref{eq:psired}   with $\sin
\psi(\theta)\neq 0$ for all $\theta$ then we say $\psi=\psi_p$ is
{\it regular}. If $\psi(\theta)$ is a continuous $2\pi$-periodic  function
solving  \eqref{eq:psired} away from the zero set of $g$ and obeying
 1) $\sin \psi(\theta)= 0$ for at least one zero $\theta$ of $g$, and
 2) either 
$\sin
\psi(\theta)\geq 0$ or $\sin
\psi(\theta)\leq 0$ (for all $\theta$), then we call  $\psi=\psi_p$ {\it singular}. We
suppose that $g$ has at most a finite number of zeros, all of which
are non-degenerate.

\begin{prop} \label{prop:clas-gen3}Let $\gamma=(\theta,\psi)$  be  an
   arbitrary solution of \eqref{eq:equamotion3}. Then  one of the following 
  cases occurs:
\begin {enumerate}[\normalfont  i)]
\item \label{it:cl21}The set of fixed points $\mathcal F$ for
  \eqref{eq:equamotion3} is non-empty, and there exists $z\in \mathcal F$ such that
  $\gamma(\tau)\to z$ for $\tau \to +\infty$. 
\item \label{it:cl32} There exists a regular solution
  $\psi_p=\psi_r(\theta)$ to the equation \eqref{eq:psired},
  such that  
  \begin{equation}
    \label{eq:rhor222}
    \lim_{\tau\to + \infty}|\psi(\tau)-\psi_r(\theta(\tau))|=0.
  \end{equation}
\item \label{it:cl43}There exists a singular solution
  $\psi_p=\psi_s(\theta)$ to the equation \eqref{eq:psired}, such that
  \begin{equation}
    \label{eq:rhos22}
 \lim_{\tau\to + \infty}|\psi(\tau)-\psi_s(\theta(\tau))|=0.   
\end{equation}
\end {enumerate}
\end {prop}
\begin{proof} Using \eqref{eq:apost} we have two possibilities, either 1) $\theta(\tau)$ stays bounded
  near $+\infty$, or 2)  $\theta(\tau)$ is not bounded
  near $+\infty$. In the case of 1) we introduce the observable 
\begin{equation}
  \label{eq:apost22}
  {d\over d\tau}a_2=Ca_1- \exp\Big ({-\tfrac {1}{2} \int_0^\theta g
  d\theta}\Big ) g(\theta)\sin \psi;\;C>>1.
\end{equation} We may show that
\begin{equation}
  \label{eq:apost23}
  {d\over d\tau}a_2\geq  \tfrac {1}{4} \exp\Big ({-\tfrac {1}{2} \int_0^\theta g
  d\theta}\Big ) (\sin ^2 \psi+g(\theta)^2).
\end{equation} 
 Since $a_2(\tau)$ by assumption is bounded we obtain an
 integral estimate  which in turn, as in the proof of Proposition
 \ref{lemdoi},  may be used to conclude that $\sin ^2
 \psi+g(\theta)^2\to 0$ yielding Proposition \ref{prop:clas-gen3} \ref{it:cl21}).

As for the case 2) we notice that if $\cos \psi(\tau_0)\geq 0$ for
  some $\tau_0$ then the same relation holds for all $\tau\geq
  \tau_0$, cf. \eqref{eq:apost}. Consequently, either $\cos
  \psi(\tau)< 0$ for all $\tau$'s  or $\cos \psi(\tau)\geq 0$ for all
  sufficiently large values of  $\tau$. Let us introduce the sets
  $\T^{*,\pm}=\T^2\setminus \{\cos \psi \neq \mp 1\}$. Then, 
  either   $\gamma(\tau)\in \T^{*,+}$ or $\gamma(\tau)\in \T^{*,-}$
  for all
  sufficiently large $\tau$'s. Let us for convenience in the following
   assume
  $\gamma(\tau)\in T^{*,+}$. We claim that eventually either $\sin
  \psi >0$ or $\sin
  \psi <0$. This statement follows from the assumption 2) and
  \eqref{eq:equamotion3}  using  the topological
  structure of $\T^{*,+}$ (it is an annulus). Let us  for convenience in the following
   assume $\sin
  \psi <0$. We may write $\psi=\psi(\theta)$ and use 
  \eqref{eq:psired}. Next we consider the quantity
  $\triangle (\theta)=\psi(\theta-2\pi)-\psi(\theta)$. If $\triangle
  (\theta)=0$ we have a regular solution to \eqref{eq:psired}. If not
  we can construct a solution as
  \begin{equation}
    \label{eq:rhope22}
    \psi_p(\theta)= \lim_{n\to \infty}\psi (\theta-2\pi n);
  \end{equation}
 notice that this sequence is either monotone increasing for all
 $\theta$ or  monotone decreasing for all
 $\theta$ (depending on the sign of $\triangle (\theta)$). If the
 sequence is decreasing we have  that $\psi_p(\theta)\geq -\pi/2$ and
 therefore that $\psi_p$ is regular (this is due to the fact that
 $a_1$ is increasing). So let us  look at the case where the sequence is
 increasing. It is readily seen (using this monotonicity) that if $\psi_p(\theta_0)=0$
 at some $\theta_0$ then $g(\theta_0)=0$. We can now prove that
 $\psi_p$ is continuous,  in fact absolutely continuous:  Let
 $I=[\theta_1,\theta_2]$ be such that  $g(\theta)\neq  0$ on either
 $[\theta_1,\theta_2)$ or $(\theta_1,\theta_2]$. Let us only consider
 the first case. We write
 \begin{align}
   \label{eq:psig0}&\psi(\theta_2-2\pi n)-\psi(\theta_1-2\pi
 n)\nonumber \\&=-\int
 _{\theta_1}^{\theta_2}\{\tfrac {1}{2} g(\theta)\cot
 \psi( \theta- 2\pi n) +(1+c^2)^{-1}\}d\theta.
 \end{align} 

Using the monotone convergence theorem we may take $n\to \infty$ in
 \eqref{eq:psig0} yielding 
\begin{equation}
   \label{eq:psig00}\psi_p(\theta_2)-\psi_p(\theta_1)=-\int
 _{\theta_1}^{\theta_2}h(\theta)d\theta,
 \end{equation} where $h(\theta)=\tfrac {1}{2} g(\theta)\cot
 \psi_p(\theta) +(1+c^2)^{-1}$ for $\theta
 \neq \theta_2$. In particular $h$ is integrable. By ``gluing'' these
 formulas together we obtain a representation of the form
 \eqref{eq:psig00} without any restriction on the interval $I$. This
 shows   absolute continuity of $\psi_p$. Furthermore we obtain from
 these arguments that the limit
 \eqref{eq:rhope22} is attain locally uniformly in $\theta$. Using
 these facts  we deduce 
 the conclusion \ref{it:cl32}) or \ref{it:cl43}).

 For the other cases (assuming 2)) we may argue similarly.
\end{proof}
\begin{prop} \label{prop:clas-gen33} For any solution $\psi_p$ as
  considered in Proposition \ref{prop:clas-gen3} its range is an 
  interval  of length  $<\pi/2$.
\end{prop}
 \begin{proof} As in the previous proof we concentrate on the case  $\sin
  \psi (\theta)\leq 0$. It suffices to show that $\cos \psi >0$. For that
  we introduce $a_1(\theta)$ by substituting $ \psi (\theta)$ in the
  argument for $\psi $ in  the defining expression for
  $a_1$. Differentiating using \eqref{eq:psired} yields 
\begin{equation}
  \label{eq:apost55}
  {d\over d\theta}a_1=\exp\Big ({-\tfrac {1}{2} \int_0^\theta g
  d\theta}\Big ) \dfrac {\sin \psi }{1+c^2}< 0.
\end{equation} 

Upon integrating \eqref{eq:apost55} (splitting the interval of
  integration at possible zeros of $g$ to deal with singular
  solutions) we see that
$a_1(\theta)<a_1(\theta-2\pi)$,
but this means that 
\begin{equation*}
\Big (\exp\Big ({-\tfrac {1}{2} \int_{\theta-2\pi}^\theta g
  d\theta}\Big )-1\Big )\cos \psi(\theta)<0,  
\end{equation*} from which we see, using 
  $g(\theta)=f'(\theta)+2c/(1+c^2)$,  that indeed $\cos \psi(\theta) >0$.
\end{proof}

\begin{cor} \label{cor:99}Suppose  $g(\theta)= f'(\theta) + 2c(1+c^2)^{-1}\leq 0$ on an interval of
length $(1+c^2)\pi /2$,  then only \ref{it:cl21}) of Proposition
\ref{prop:clas-gen3} occurs.
\end{cor}
\begin{proof} We need to exclude the existence of solutions  $\psi_p$ as
  considered in Proposition \ref{prop:clas-gen3}. Suppose
  $I=[\theta_1,\theta_2]$ is an  interval of
length $(1+c^2)\pi /2$ on which $g(\theta)\leq 0$. Again we consider
  only the case $\sin
  \psi (\theta)\leq 0$. Upon integrating \eqref{eq:psired} we learn from
  the proof of 
  Proposition \ref{prop:clas-gen33} that
  $\psi(\theta_2)-\psi(\theta_1)\leq -|I|(1+c^2)^{-1}=-\pi /2$, which
  contradicts Proposition \ref{prop:clas-gen33}. Here we have again split the interval of
  integration at possible zeros of $g$ in $I$ (to deal with singular solutions).
\end{proof}

For the limiting case, $c=0$, we  have the same conclusion as in Corollary
\ref{cor:99} (seen by using \eqref{eq:apost23}; this is similar to
 the zero flux case).  
We remark that \ref{it:cl21}) of Proposition
\ref{prop:clas-gen3} obviously does not occur if $g$ is strictly positive. For
some models  both of \ref{it:cl21}) and \ref{it:cl32}) may occur. Presumable
\ref{it:cl43}) may only occur at a finite numbers of $c$'s. If
\ref{it:cl43}) does not occur then presumable
LAP would follow by the methods of this paper. The case of Corollary
\ref{cor:99} 
would be ``easy''. The quantum channel corresponding to \ref{it:cl32})
seems to be similar to the one considered in \cite {CHS}.

\end{document}